\documentclass[10pt, letterpaper]{IEEEtran}
\IEEEoverridecommandlockouts
\usepackage{cite}
\usepackage{amsmath,amssymb,amsfonts}
\usepackage{graphicx}
\usepackage{textcomp}
\usepackage{xcolor}
\def\BibTeX{{\rm B\kern-.05em{\sc i\kern-.025em b}\kern-.08em
    T\kern-.1667em\lower.7ex\hbox{E}\kern-.125emX}}
    
\usepackage{amsthm,amssymb,graphicx,multirow,amsmath,color,amsfonts}
\usepackage[update,prepend]{epstopdf}
\usepackage[latin1]{inputenc}
\usepackage{tikz}
\usepackage{bbm} 
\usepackage{pdfpages}
\usepackage{tabulary}
\usepackage{multirow}
\usepackage{comment}
\usepackage{graphicx}
\usepackage{amsthm}
\usepackage{caption}
\usepackage{subcaption}
\usepackage{mathtools}
\usepackage{epstopdf}
\usepackage{comment}
\usepackage{adjustbox}
\usepackage{subcaption}
\usepackage{caption}
\usepackage{algorithm}
\usepackage{algpseudocode}
\usepackage[most]{tcolorbox}
\tcbuselibrary{skins, breakable}

\usepackage{tabularx}
\usepackage{booktabs}
\usepackage{pifont}

\usepackage{xcolor}
\newcommand{\cmark}{\textcolor{green!60!black}{\ding{51}}} 
\newcommand{\tmark}{\textcolor{orange}{\ding{115}}}        
\newcommand{\xmark}{\textcolor{red}{\ding{55}}}   

\usepackage[noframe]{showframe} 
\usepackage{array,ragged2e}
\newcolumntype{P}[1]{>{\RaggedRight\arraybackslash}p{#1}}

\include{notation}
\allowdisplaybreaks
\begin{document}

\newcommand{\subparagraph}{}
\captionsetup{belowskip=0pt}


\newcommand{\ts}{\textsuperscript}
\setlength{\belowcaptionskip}{-10pt}
\long\def\/*#1*/{}

\def\chr#1{{\color{red}#1}}
\def\chb#1{{\color{blue} #1}}


\title{A Survey on Cloud-Based 6G Deployments: Current Solutions, Future Directions and Open Challenges}

\author{\IEEEauthorblockN{Tolga O. Atalay\IEEEauthorrefmark{1},
Alireza Famili\IEEEauthorrefmark{3}
Amirreza Ghafoori\IEEEauthorrefmark{2}
Angelos Stavrou\IEEEauthorrefmark{1}\IEEEauthorrefmark{3}} \\
\IEEEauthorblockA{
\IEEEauthorrefmark{1}A2 Labs LLC, Arlington VA, USA\\
\IEEEauthorrefmark{2}Department of Electrical and Computer Engineering, Virginia Tech, USA\\
\IEEEauthorrefmark{3}WayWave, Inc, Arlington, VA, USA\\
Email: tatalay@a2labs.com,
afamili@wayvave.com,
ghafoori@vt.edu,
angelos@a2labs.com,
}}


\maketitle


\begin{abstract}

The next generation of cellular networks is designed to provide ubiquitous connectivity to a wide range of devices. As Telecommunication Service Providers (TSPs) increasingly collaborate with public cloud providers to deploy 5G and beyond networks, a fundamental shift is underway, from hardware-bound Physical Network Functions (PNFs) to cloud-native, containerized deployments managed through platforms like Kubernetes. While this transition promises greater scalability, flexibility, and cost efficiency, it also introduces a complex set of technical and operational challenges that must be thoroughly understood before large-scale cellular deployments can take place in cloud environments. In this survey, we present a structured taxonomy that categorizes the design space of cloud-based cellular deployments across four dimensions: deployment architecture, resource management and orchestration, multi-tenancy and isolation, and economic and ownership models. Using this taxonomy as a foundation, we critically analyze six key investigation areas, security and privacy, scalability and elasticity, performance and latency, cost optimization, resilience and fault management, and compliance and sovereignty, examining each through a cloud-native lens. To benchmark the state of industry adoption, we examine the deployment strategies of leading Infrastructure-as-a-Service (IaaS) providers, namely Amazon Web Services (AWS), Microsoft Azure, and Google Cloud Platform (GCP). Finally, we identify emerging trends such as AI-driven orchestration, quantum-safe protocols for virtualized network functions, and serverless networking for 6G, while articulating the open challenges that remain in realizing robust, scalable cloud-based cellular networks.

\end{abstract}


\section{Introduction} \label{sec:intro}
The next generation of mobile networks is designed to accommodate a wide range of vertical use cases, each with distinct Quality of Service (QoS) requirements. These networks must adapt to diverse performance, mobility, and security needs without compromising service fidelity. With a significant leap in service range compared to legacy Long Term Evolution (LTE) networks, Fifth Generation (5G) cellular networks can provide the necessary connectivity for a broader array of use cases. This enhanced versatility is enabled by adopting key technologies such as Network Functions Virtualization (NFV)\cite{abdelwahab2016network, basu2020softwarized, zhang2018network}, Software-Defined Networking (SDN)~\cite{tadros2020software, long2022software}, and Mobile Edge Computing (MEC)\cite{taleb2017multi, pham2020survey}. Adopting these fundamental building blocks in the design process enables the construction of more software-based cellular networks that can adapt to the diverse connectivity requirements promised by 5G. Mobile network functionality in LTE is bound to proprietary hardware in the form of Physical Network Functions (PNFs). Leveraging NFV, the 5G core network design revolves around Virtual Network Functions (VNFs) that can be deployed on Commercial-off-the-Shelf (COTS) servers. With SDN, this virtual mobile core network can be centrally managed as decoupled sets of control and user planes. In addition to a virtual core network, the 5G Radio Access Network (RAN) can also be virtualized to enable more flexible deployments. 

\begin{figure}[t]
    \centering
    \includegraphics[width=\linewidth,trim={5cm 0.1cm 4.2cm 1.2cm},clip]{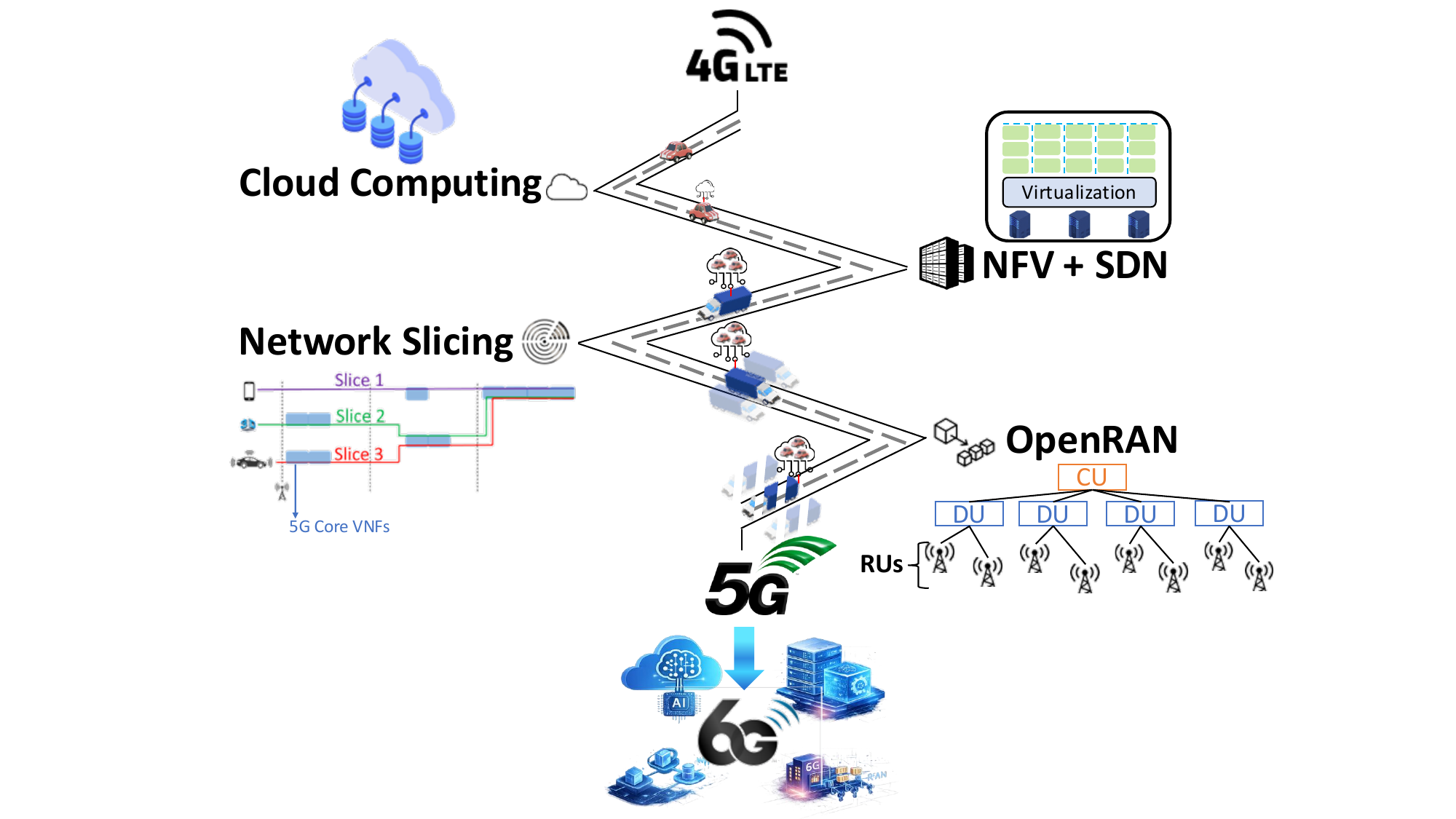}
    \caption{Transitioning milestones from LTE to 5G with cloud computing; adoption of NFV and SDN; network slicing and OpenRAN}
    \label{fig:transition_lte_to_5G}
\end{figure}

The recent emergence of OpenRAN~\cite{polese2023understanding}, led by the O-RAN Alliance~\cite{ORANALLI61online}, advocates for the disaggregation of the RAN, allowing different functional splits~\cite{brik2024explainable, marinova2024intelligent, alam2024comprehensive} to optimize services for specific use cases. OpenRAN standardizes the communication interfaces between RAN components, such as the Radio Unit (RU), Distributed Unit (DU), and Central Unit (CU). This standardization fosters a vendor-agnostic ecosystem, encouraging collaborative growth within the cellular network community. OpenRAN, not only facilitates interoperability and reduces dependency on single vendors but also accelerates innovation by enabling different vendors to contribute specialized solutions, thereby driving the evolution of cellular networks. Integrated with MEC, the virtualized RAN and core deployments can achieve enhanced scalability and flexibility. This approach allows networks to be fine-tuned for more granular performance requirements, ensuring that they can meet the diverse needs of modern connectivity landscapes. 

The collective features of NFV, SDN, and MEC provide a fertile ground for mobile network deployment. Leveraging these fundamental building blocks, services in 5G and beyond are currently offered in the form of end-to-end, logically isolated network fragments denoted as ``network slices''~\cite{zhang2019overview, li2017network, wu2022ai}. Within the scope of network slicing, the Third Generation Partnership Project (3GPP) defines a set of umbrella Slice Service Types (SSTs) such as enhanced Mobile Broadband (eMBB), Ultra-Reliable Low Latency Communication (URLLC), massive Internet of Things (mIoT), Vehicular to Everything (V2X), High-Performance Machine-Type Communications (HMTC), and High Data Rate and Low Latency Communications (HDLLC)~\cite{3gpp23501}. Mobile Virtual Network Operators (MVNOs) can provide services through various other SSTs or customize the delivery further with specific network slicing designs. This step towards virtualization creates enterprise opportunities for service providers to enter the mobile network ecosystem. 
Small-to-Medium Enterprises (SMEs) can build Software-as-a-Service (SaaS), Platform-as-a-Service (PaaS), and Network-Slice-as-a-Service~\cite{gsm2021official,atalay_globecom2022,zhou2016network} business models to create custom 5G offerings for end users.

Figure~\ref{fig:transition_lte_to_5G} highlights the major milestones in the transition from LTE to 5G, leading to 6G. The journey begins with the emergence of cloud computing, the primary facilitator. The adoption of NFV and SDN followed, paving the way for network slicing, which allows the network to accommodate various use cases with different QoS requirements. To create a more configurable and vendor-agnostic RAN with diverse functional split options, OpenRAN was introduced. While other foundational principles such as microservice architecture, 5G New Radio (NR), and MEC also play crucial roles, Figure~\ref{fig:transition_lte_to_5G} provides a concise overview of the critical milestones in the LTE to 5G evolution. Moving forward, Artificial Intelligence (AI), along with the flexibility and scalability and flexibility of cloud-based deployments, will pave the way for the 6G

\begin{figure}[t]
    \centering
    \includegraphics[width=\columnwidth,trim={9cm 5.3cm 12.2cm 6.5cm},clip]{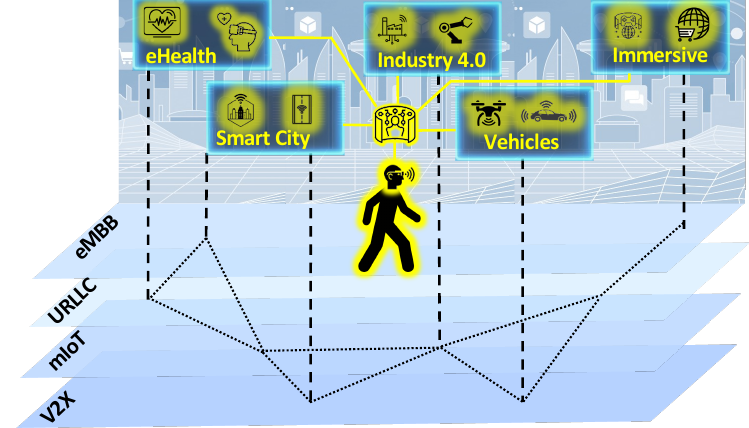}
    \caption{Coexistence of network slices supporting different use cases with different SSTs}
    \label{fig:5gnetworkov}
\end{figure}

Figure~\ref{fig:5gnetworkov} illustrates the coexistence of network slices supporting different use cases within a 5G network. The figure shows how various verticals such as eHealth, Smart City, Industry 4.0, Vehicles, and Immersive applications are enabled by distinct network slices, each tailored to specific requirements. These slices are categorized into eMBB, URLLC, mIoT, and V2X. The layered structure indicates that each use case leverages the appropriate network slice to ensure optimal performance, reliability, and scalability. Thus, Figure~\ref{fig:5gnetworkov} emphasizes the flexibility and adaptability of 5G networks in meeting diverse application demands through network slicing.

As the landscape of devices and use cases in cellular networks continues to expand, deployments must scale accordingly to meet these growing demands. Accommodating the diverse QoS requirements necessitates not only highly flexible deployments but also an infrastructure capable of matching this flexibility. While major Telecommunication Service Providers (TSPs) such as Verizon, AT\&T, Deutsche Telekom, Vodafone, and NTT DOCOMO have a foundational RAN to build upon, they lack the necessary compute infrastructure for deploying virtual RAN (vRAN) and core network resources, which are essential for a comprehensive 5G and beyond deployment. To fill this vacuum, TSPs have started to create collaborations with Infrastructure-as-a-Service (IaaS) providers such as Amazon Web Services (AWS), Microsoft Azure, and Google Cloud Platform (GCP). 

AWS has already entered into a partnership with Telefonica to host the 5G Standalone (SA) core network built entirely in the cloud~\cite{O2Telefo59online}. This represents the first instance where a telecom provider is migrating their existing network to a 5G deployment running on AWS. Furthermore, NTT DOCOMO has announced that they will be leveraging the AWS infrastructure to deploy a nationwide Open Radio Access Network (OpenRAN) within Japan~\cite{NTTDOCOM16online}. Similarly, AWS and DISH have a long-standing partnership to create a cloud-native OpenRAN deployment in the United States~\cite{DISHandA13online}. 

Similarly to AWS, Microsoft Azure has begun offering private 5G deployments~\cite{AzurePri65online} to provide enterprise-grade connectivity to MVNOs. With the Azure Private 5G core, businesses can deploy a fully self-contained 5G core at the network edge, supporting both 4G and 5G RANs. This development is a significant step towards enabling 5G connectivity for SMEs without the need to own or operate proprietary equipment, relying solely on cloud resources instead.

\begin{figure*}[t]
    \centering
    \includegraphics[width=\linewidth,trim={0cm 0cm 0cm 0.5cm},clip]{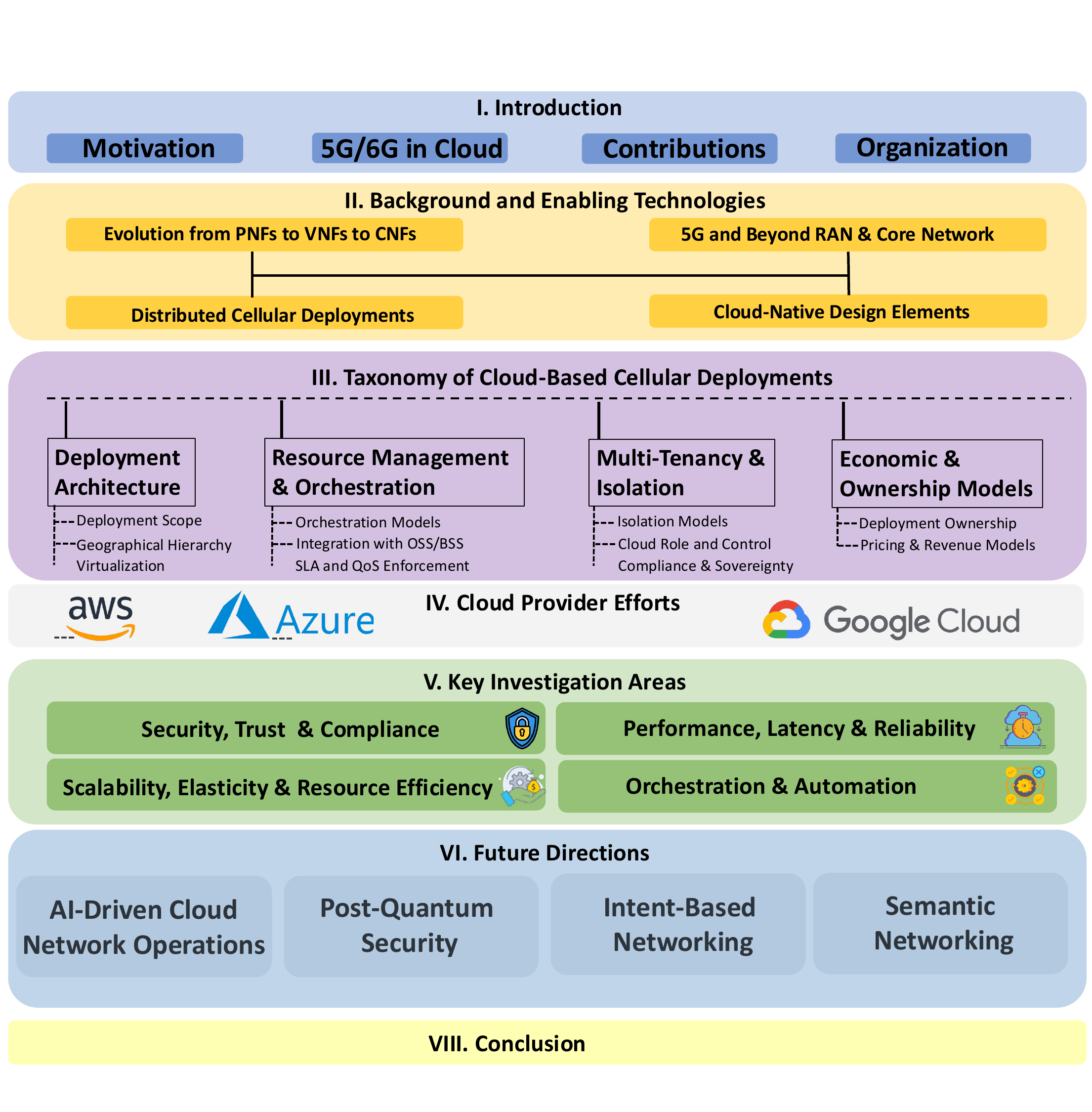}
    \caption{Organization of this survey consisting of background and enabling technologies; a taxonomy for cloud-based cellular deployments; key technical investigation areas; the deployments trends across major cloud providers; future trends and research directions and the open challenges}
    \label{fig:surveysummary}
\end{figure*}

Last but not least, T-Mobile and GCP announced their collaboration to enhance 5G capabilities through 5G Advanced Network Solutions (ANS) and Google Distributed Cloud Edge (GDC Edge)~\cite{TMobilea76online}. This partnership aims to provide enterprises and government organizations with the tools to drive digital transformation across various industries, including retail, manufacturing, logistics, and smart cities. Integrating T-Mobile's 5G networks with Google's edge computing technology, businesses can leverage low latency, high speeds, and reliable connectivity for innovative applications such as augmented reality (AR) and computer vision. An example of this collaboration is the magic mirror proof of concept, which uses cloud-based processing to create interactive retail experiences. The partnership promises to expand the use of AR and VR technologies from limited applications to large-scale adoption, significantly benefiting businesses across the country.

With all these serial partnerships forming between IaaS providers and major TSPs, cloud providers have started tailoring services for finer-grained TSP requirements. This comes in the form of diversification for resource allocation and in-house tools to facilitate mobile network deployment, security, monitoring, and management. As a result of all these moving parts, deploying 5G and beyond networks in cloud environments has become a complex task requiring significant planning. 

\begin{table*}[t]
\caption{Cloud-centric comparative analysis of key technical and deployment dimensions across existing 5G/6G cellular network survey papers}
\label{tbl:survey_comparison_compact}
\centering
\small
\renewcommand{\arraystretch}{1.2}
\begin{tabularx}{\textwidth}{lcccccccc}
\toprule
\textbf{Survey} & \rotatebox{50}{\textbf{E2E Coverage}} & \rotatebox{50}{\textbf{Cloud-Native Arch.}} & \rotatebox{50}{\textbf{Zero-Touch Ops}} & \rotatebox{50}{\textbf{Edge Integration}} & \rotatebox{50}{\textbf{Network Slicing}} & \rotatebox{50}{\textbf{Security}} & \rotatebox{50}{\textbf{Public Cloud}} & \rotatebox{50}{\textbf{Business Models}} \\
\midrule
~\cite{scalise2024systematic}      & \cmark   & \tmark  & \tmark  & \cmark   & \cmark   & \cmark & \xmark   & \xmark \\
~\cite{de2023survey}               & \cmark   & \xmark  & \xmark  & \xmark   & \cmark   & \cmark & \xmark   & \xmark \\
~\cite{zieba2024cloud}             & \tmark   & \cmark  & \tmark  & \cmark   & \tmark   & \tmark & \xmark   & \cmark \\
~\cite{lee2024federated}           & \cmark   & \cmark  & \cmark  & \cmark   & \xmark   & \cmark & \xmark   & \xmark \\
~\cite{taleb2017multi}             & \tmark   & \cmark  & \xmark  & \cmark   & \xmark   & \xmark & \xmark   & \xmark \\
~\cite{cruz2022edge}               & \cmark   & \tmark  & \xmark  & \cmark   & \xmark   & \xmark & \xmark   & \xmark \\
~\cite{eswaran2023private}         & \tmark   & \cmark  & \xmark  & \xmark   & \cmark   & \tmark & \xmark   & \cmark \\
~\cite{spinelli2020toward}         & \tmark   & \cmark  & \xmark  & \cmark   & \xmark   & \tmark & \xmark   & \tmark \\
~\cite{coronado2022zero}           & \cmark   & \tmark  & \cmark  & \tmark   & \tmark   & \cmark & \xmark   & \xmark \\
\textbf{This Work}  & \cmark   & \cmark  & \cmark  & \cmark   & \cmark   & \cmark & \cmark   & \cmark \\
\bottomrule
\end{tabularx}
\vspace{1mm}
\footnotesize{\textbf{Legend:} \cmark = Covered, \tmark = Partially Covered, \xmark = Not Covered.}
\end{table*}

\textbf{Contributions. }
In our effort to represent the uncovered gaps in cloud-based deployments, we have compiled the comparison in Table~\ref{tbl:survey_comparison_compact}. This table presents key contribution areas relevant to the cellular networking community focused on cloud-based deployments. In evaluating each category, we specifically examined whether the corresponding survey addressed the topic from a cloud-native deployment perspective. To that end, Table~\ref{tbl:survey_comparison_compact} highlights our effort to construct a cohesive survey that offers comprehensive coverage of well-established areas such as edge integration, network slicing, and security, which are topics that have already received notable attention within the community. However, our work extends beyond prior surveys by analyzing each of these domains through a cloud-native lens and further incorporating emerging considerations related to the transition toward public cloud deployments and the issues that accompany such a major deployment shift. 

\textbf{Organization.} In this survey, we present a systematic analysis of the ongoing transition of cellular networks to cloud-based deployments. The survey summary is presented in Figure~\ref{fig:surveysummary}. To create a smooth narrative, we identify the enabling technologies and discuss their key features in greater detail in Section~\ref{sec:background_enabling}. This includes the NFV and SDN-enabled transition to VNFs and Containerized Network Functions (CNFs) from PNFs, followed by an overview of 5G and beyond RAN and core network. We then provide a proof-of-concept approach to how a distributed cloud-based cellular deployment can occur. Last but not least, we identify key cloud-based deployment trends and the emerging cloud-native orchestration frameworks used by the industry to manage hyper-scale deployments.

After establishing the necessary background, a taxonomy is presented in Section~\ref{sec:taxonomy} to define and categorize the design space within which cloud-based cellular deployments can have varying properties. This taxonomy will serve as a map of the landscape to identify what dimensions exist, what options are available, and how different pieces fit together. After the taxonomy, Section~\ref{sec:bigthree} presents the current deployment ecosystem and trends embraced by the big three public IaaS providers (i.e., AWS, Azure, GCP).

Using the taxonomy, Section~\ref{sec:key_investigation} presents the key investigation areas to critically analyze the specific challenges, limitations, and research opportunities within the design space. This survey distinguishes between structural classification and technical analysis as exemplified in Box~\ref{box:taxonomy_vs_analysis} and echoed throughout Section~\ref{sec:taxonomy} and Section~\ref{sec:key_investigation}.

\begin{tcolorbox}[colback=gray!5!white,
                 colframe=black!50,
                 title=Box~1: Declarative vs. Investigative Framing,
                 fonttitle=\bfseries,
                 breakable,
                 enhanced,
                 sharp corners,
                 boxrule=0.5pt]
\label{box:taxonomy_vs_analysis}

\textbf{Section III = The Taxonomy - Design Space Blueprint Definition}

"There are VM-based, CNF-based, and bare-metal deployments; some are centralized, others are edge-native."

\vspace{1em}

\textbf{Section V = Investigation Area - Technical Analysis}

"However, CNF-based deployments at the edge face significant challenges due to orchestration complexity, performance jitter, and limited fault isolation mechanisms."

\end{tcolorbox}

This is followed by the open challenges and future directions in Section~\ref{sec:future_trends}. Finally, Section~\ref{sec:conc} concludes the survey.

\section{Background and Enabling Technologies} \label{sec:background_enabling}
\subsection{Evolution from PNFs to VNFs and CNFs} \label{sec:evolpnfcnf}

\begin{figure}[t]
    \centering
    \includegraphics[width=\columnwidth,trim={1cm 5cm 11cm 3cm},clip]{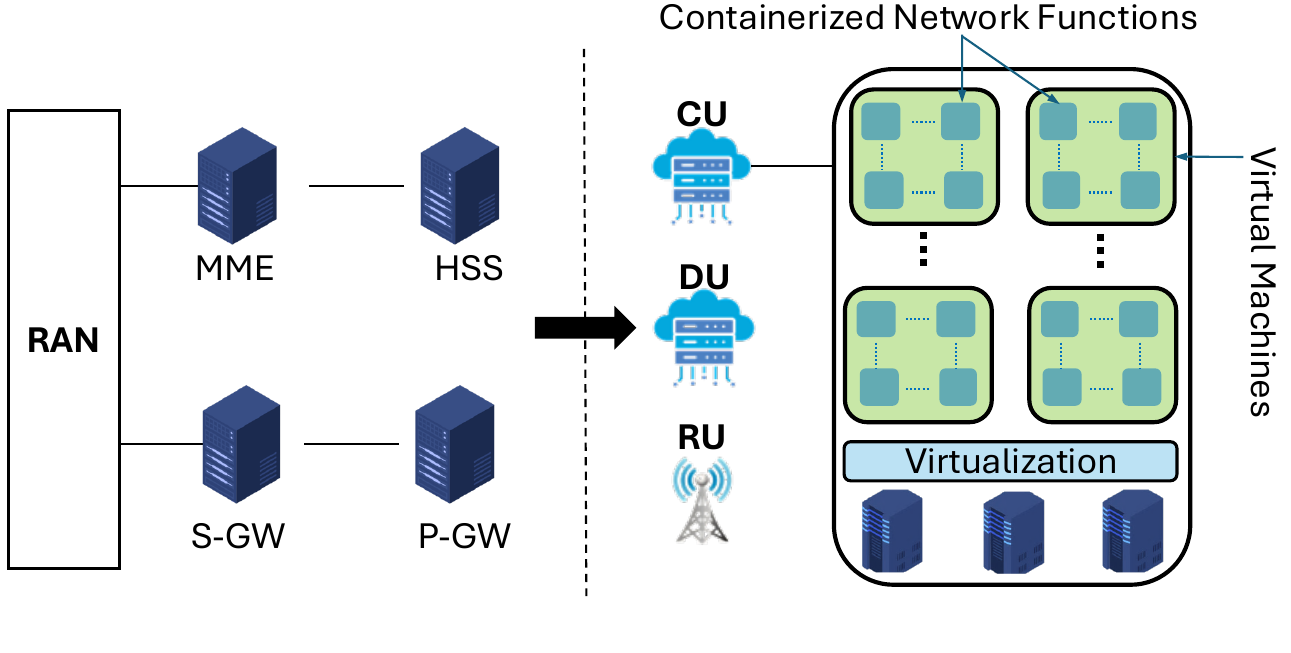}
    \caption{Transition from physical network functions to virtual and containerized network functions as a result of embracing NFV in cellular networks}
    \label{fig:transitiontovnfs}
\end{figure}

The design of 5G and future cellular systems is fundamentally shaped by transitioning from hardware-centric architectures to highly virtualized, software-defined infrastructures. NFV decouples network functions from proprietary hardware appliances and instead deploys them as VNFs and Containerized Network Functions (CNFs) on COTS servers~\cite{yi2018comprehensive, han2015network}. As a result of embracing this virtualization, a greater degree of resource elasticity and flexibility is enabled for multi-tenant and multi-domain deployments. This architectural shift is illustrated in Figure~\ref{fig:transitiontovnfs}, where monolithic PNFs predominant in LTE deployments~\cite{} are now packaged as VNFs and CNFs for 5G and beyond. On the left side, the LTE Evolved Packet Core (EPC) is shown with hardware-bound elements such as the Mobility Management Entity (MME), Serving Gateway (S-GW), Packet Gateway (P-GW), and the Home Subscriber Server (HSS), all tightly coupled to physical infrastructure~\cite{nguyen2016sdn}. This legacy architecture represents PNFs, which are difficult to scale, maintain, or deploy flexibly. As the architecture transitions rightward, the RAN components are disaggregated into the RU, DU, and CU, as defined by O-RAN~\cite{polese2023understanding}. These disaggregated elements support more flexible functional splits and allow selective placement across edge and core locations depending on latency and performance requirements. Furthermore, the entire set of core network functions can now be hosted within virtual cloud environments as microservice-based deployments and be controlled by well-defined, lightweight orchestration frameworks. Overall, these directions represent a significant shift towards a more scalable and flexible cellular architecture.

\subsection{5G RAN \& Core Network} \label{sec:rancoreoverview}

\begin{figure}[t]
    \centering
    \includegraphics[width=\linewidth,trim={11.5cm 7cm 15cm 8cm},clip]{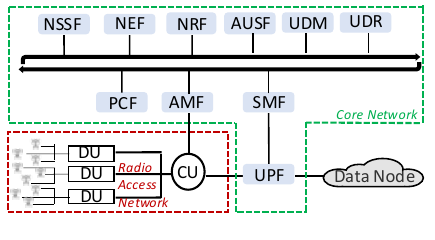}
    \caption{5G Core and RAN overview with a service-based-architecture and disaggregated gNodeB}
    \label{fig:5gcoreranov}
\end{figure}

To accurately follow the discussions and analysis presented in this survey, the reader needs to have a working understanding of mobile networking architecture. Thus, in this subsection, we discuss the fundamentals of the 5G and beyond RAN and core network, as well as the fundamental interactions that take place across different entities for the system to function. Figure~\ref{fig:5gcoreranov} presents a concise overview of the RAN and core. In this case, the RAN is represented by a disaggregated base station comprising the CU, DU, and RU, while the core network consists of a set of VNFs that communicate over service-based interfaces.

Within the RAN, the RU handles the RF front-end, transmitting and receiving radio signals to and from the UE. Typically co-located with the antenna, the RU is responsible for analog-digital conversion, amplification, and beamforming. It connects to the DU over the fronthaul interface to support high-throughput, low-latency communication. The DU sits between the RU and the core, managing Layer 1 and Layer 2 processing, including PHY, MAC, and RLC functions~\cite{3gpp38801}. It handles time-sensitive operations such as scheduling, HARQ, and uplink timing alignment, and is often deployed at the network edge to meet strict latency requirements. A single DU may coordinate multiple RUs, enabling centralized baseband processing for improved efficiency. The CU, positioned upstream of the DU, is split into CU-CP and CU-UP components. CU-CP terminates RRC and forwards NAS signaling to the AMF over the N2 interface, while CU-UP handles user-plane traffic delivery to the UPF over the N3 interface. This disaggregated architecture provides deployment flexibility, allowing operators to balance latency, scalability, and resource efficiency across different RAN scenarios.

In the 5G core, network functions interact using standardized RESTful APIs defined by the Common API Framework (CAPIF)~\cite{3gpp23222}, as outlined by 3GPP. Figure~\ref{fig:5gcoreranov} illustrates the 5G Service-Based Architecture (SBA), with core functions grouped into those belonging to the Serving Network (SN) and the Home Network (HN). At the heart of the SN is the Access and Mobility Management Function (AMF), which serves as the main signaling hub that manages registration, mobility, and NAS message routing between the RAN, the UE, and other core functions.

Authentication procedures are handled by the Authentication Server Function (AUSF) in coordination with the Unified Data Management (UDM) and Unified Data Repository (UDR), both typically hosted in the HN. Together, these functions participate in the 5G Authentication and Key Agreement (AKA) process, ensuring secure access~\cite{3gpp23501}.

Once authentication is complete, the Session Management Function (SMF) takes over to establish and manage the user's data sessions. It selects and configures the appropriate User Plane Function (UPF), which forwards user traffic toward external applications or data networks, often deployed in the form of containerized DNNs. The SMF also works closely with the Policy Control Function (PCF), which provides real-time policy decisions on QoS, charging, and access rules based on subscription profiles and service requirements~\cite{3gpp23501}.

To coordinate service discovery and enable dynamic interaction among VNFs, the Network Repository Function (NRF) maintains a registry of available network functions and their capabilities. Additionally, the Network Exposure Function (NEF) provides a secure gateway for external applications to access core network services and policies, enabling functions like QoS requests or event notifications. Finally, the Network Slice Selection Function (NSSF) ensures that UEs are assigned to the appropriate network slice based on subscription, service type, or network conditions, supporting the multi-tenancy and SLA guarantees central to 5G~\cite{3gpp23501}.

\subsection{Distributed Cellular Deployments} \label{sec:distrocell}

In Section~\ref{sec:evolpnfcnf} we summarized the transition from PNFs to VNFs to eventually CNFs, which have a significant impact on how 5G and beyond deployments are packaged. Furthermore, Section~\ref{sec:rancoreoverview} provided an overview of the distributed and decentralized mobile network architecture. Together, these developments create a foundation that is well-suited for deployment in distributed cloud environments, where scalability, flexibility, and dynamic resource allocation can be more effectively realized. Furthermore, with the adoption of network slicing, 5G offers fine-grained isolation and management capabilities for a large variety of vertical use cases. 

\begin{figure}[t]
    \centering
    \includegraphics[width=\columnwidth,trim={5.4cm 4cm 5.4cm 3.8cm},clip]{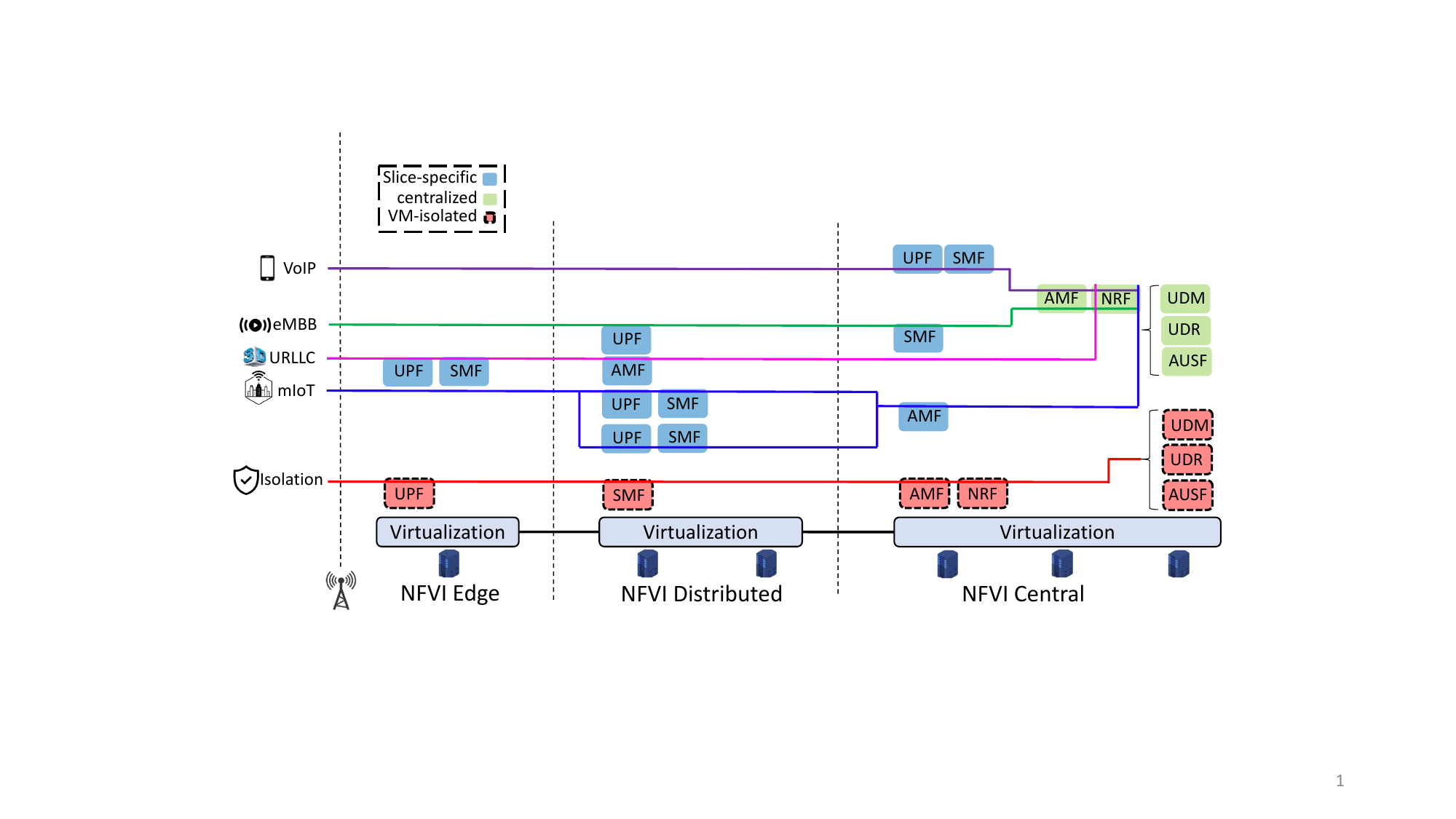}
    \caption{Network slice deployment taking place in a distributed cloud hierarchy formed of edge, distributed, and central Network Functional Virtualization Infrastructure (NFVI) hierarchies}
    \label{fig:sliceclouddep}
\end{figure}

Figure~\ref{fig:sliceclouddep} depicts the deployment of network slices within a distributed cloud hierarchy, consisting of edge, distributed, and central Network Functional Virtualization Infrastructure (NFVI) tiers. The diagram illustrates how different service types, eMBB, URLLC, and mIoT, are allocated across these tiers. Each slice utilizes various network functions, such as the UPF and SMF, distributed across the NFVI Edge, NFVI Distributed, and NFVI Central layers. The color-coded paths indicate the distinct network slices, with solid lines representing slice-specific deployments, dashed lines for centralized functions, and dotted lines for VM-isolated functions. The figure emphasizes the role of virtualization in enabling flexible, scalable, and isolated network slice deployments to meet the diverse requirements of modern 5G use cases.

\subsection{Cloud-Native Design Elements}

With the transition to cloud-based infrastructure, it becomes possible for mobile network operators to leverage cloud-native design frameworks and elements to create more flexible and scalable cellular deployments. In this subsection, we present an overview of some cloud-native design concepts and frameworks to prepare the reader for the subsequent analysis regarding cloud-based cellular deployments.

\subsubsection{Kubernetes-Based Deployments}
The containerization of mobile network functions as described in Section~\ref{sec:evolpnfcnf}, enables the adoption of new deployment and management frameworks. One framework that has gained significant traction over the past few years is Kubernetes~\cite{Kubernet17online}, an orchestration platform for the deployment and management of containerized applications. Kubernetes was originally developed by Google and is now maintained by the Cloud Native Computing Foundation (CNCF)~\cite{CloudNat29online}. It provides a declarative infrastructure model that abstracts physical and virtual resources into a unified control plane.

Within Kubernetes, the smallest unit of deployment is defined as a pod. A single pod can host multiple containers within the same network namespace and storage environment. In Kubernetes, the system is structured around several high-level object types that represent the core building blocks for deploying and managing applications. This includes a variety of workload types tailored to the state requirements of the application, supported by a range of auxiliary Kubernetes objects, such as \texttt{Services} for abstracting network access and \texttt{Volumes} for persistent data storage, that operate alongside the primary workloads. These support constructs play a critical role in enabling decentralized, modular deployment and runtime behavior within cloud-native environments.

Kubernetes manages pods across a cluster of nodes, scheduling workloads based on resource availability, policy constraints, and desired state declarations. Built-in mechanisms such as self-healing, auto-scaling, service discovery, and rolling updates enable resilient and highly available service delivery, making Kubernetes particularly well-suited for managing distributed, cloud-native 5G core functions~\cite{atalay_wcnc2022, larrea2023corekube, 5g_map_mobicom2025}.

Last but not least, Kubernetes allows the orchestrator to define custom resources~\cite{CustomRe63online} depending on the specific application environment. This is especially useful given the modular nature of 5G and beyond networks that utilize network slices to organize VNF service chains. Such a capability presents opportunities for synergy when deploying 5G networks with Kubernetes

\subsubsection{Side Car Proxies, Service Meshes \& eBPF}
To support the complexity of modern cloud-native deployments, architectural components such as sidecar proxies, service meshes, and eBPF have emerged as complementary mechanisms for managing traffic, enforcing policies, and ensuring observability across distributed network functions.

\begin{figure}[t]
    \centering
    \includegraphics[width=\columnwidth,trim={6.1cm 5.8cm 12.5cm 6.6cm},clip]{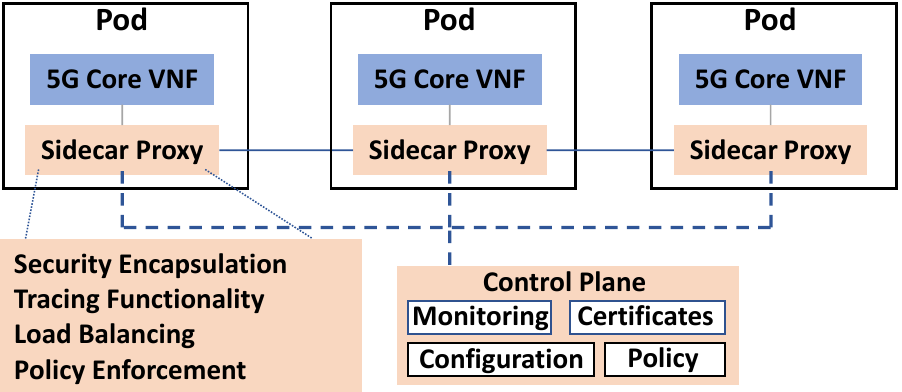}
     \caption{Overview of SCP and service mesh deployments}
     \label{fig:backscp}
\end{figure}

Sidecar proxies have become foundational components in cloud-native systems, particularly within microservice-oriented architectures. As illustrated in Figure~\ref{fig:backscp}, the SCP is a secondary application container that sits adjacent to the primary application container within the same Kubernetes pod. By offloading platform-level concerns, such as authentication, traffic management, logging, and observability, from the core application, sidecars enable a clean separation of functional and non-functional responsibilities~\cite{li2021automatic, yuan2022sidecar, xrf_infocom2023}. This separation allows developers to deploy and manage complex applications more flexibly and securely. In a typical Kubernetes deployment, sidecar proxies are deployed as containers within the same pod as the main application, intercepting all ingress and egress traffic. This co-location model facilitates consistent enforcement of security policies, performance monitoring, and protocol translation, without modifying application code.

When integrated into a service mesh, sidecar proxies can be centrally managed to apply global communication policies, automate service discovery, enforce mutual authentication, and provide telemetry at scale~\cite{saokar2023servicerouter, IstioThe37online, ashok2021leveraging}. Service meshes build a logical overlay network, often using control planes to configure and coordinate sidecars across the mesh. This enables scalable management of distributed network functions, making service meshes particularly suitable for cloud-native 5G core environments where inter-function communication patterns are highly dynamic and performance-sensitive. The 5G-MAP platform, for example, adopts this architecture to inject sidecars dynamically into 5G core pods, facilitating platform-agnostic deployment across both private and public Kubernetes clusters.

\begin{figure}[t]
    \centering
    \includegraphics[width=\columnwidth,trim={0cm 7cm 16.8cm 0cm},clip]{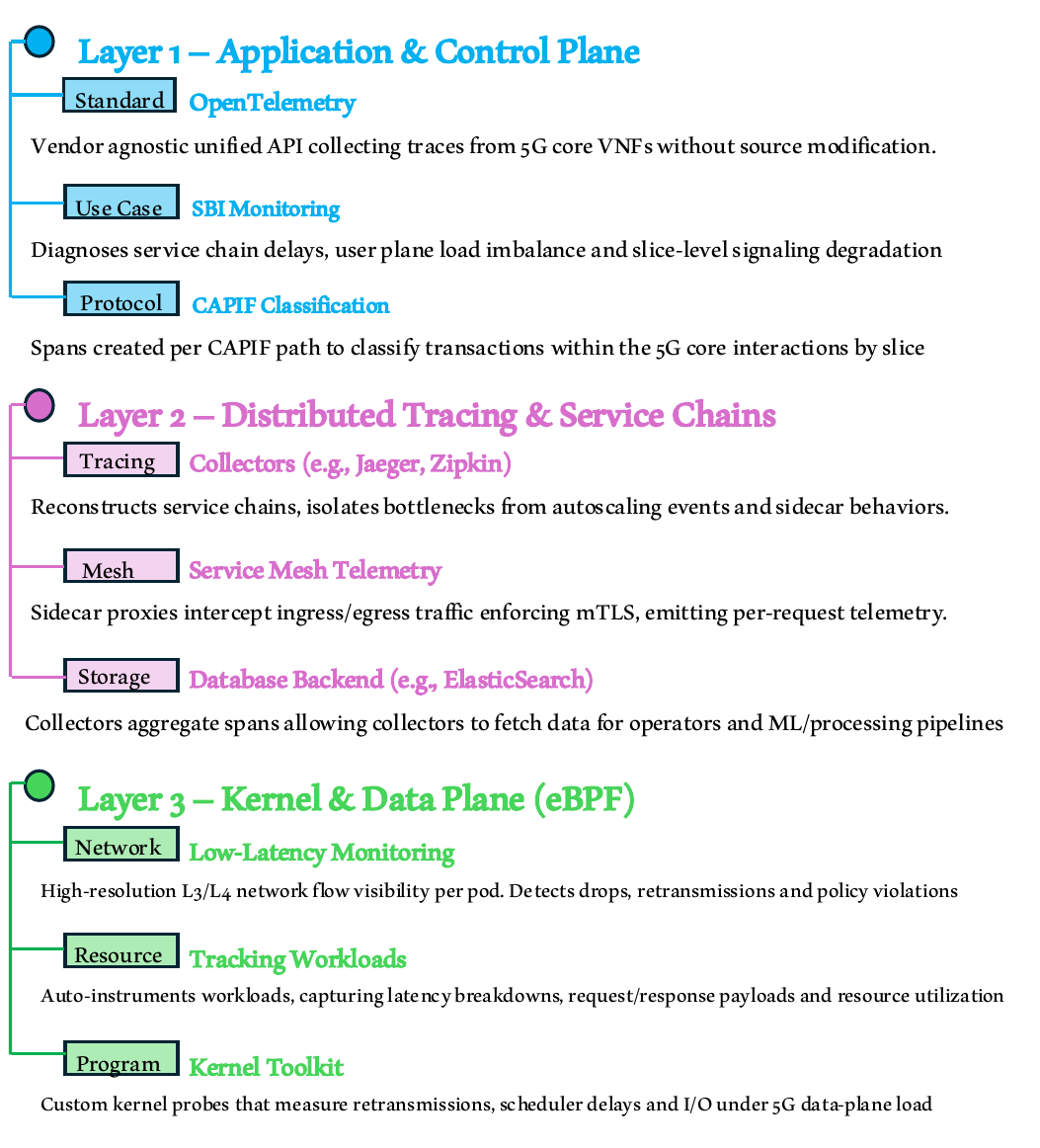}
     \caption{Overview of observability stack used for cellular network deployment monitoring}
     \label{fig:obsstack}
\end{figure}

Complementing these user-space mechanisms, the extended Berkeley Packet Filter (eBPF) offers a high-performance alternative for inspecting and controlling traffic directly within the Linux kernel~\cite{huang2023accelerating, loureiro2024enhancing, 5GVisibi38online}. eBPF enables safe, sandboxed execution of custom programs at various kernel hook points, allowing for low-overhead packet filtering, traffic classification, performance profiling, and event tracing, all without context switching to user space. Because eBPF operates at the kernel level, it introduces minimal latency and is highly efficient for tasks such as flow monitoring, load balancing, and telemetry collection.

However, eBPF's strengths in performance and visibility come with trade-offs in programmability and protocol awareness. Its limited support for complex application-layer logic makes it less suitable for use cases requiring detailed header manipulation, authentication handshakes, or multi-step request routing. Service meshes, by contrast, excel in those areas but typically incur higher overhead due to user-space execution and proxy chaining. For this reason, platforms combine both approaches: eBPF for lightweight kernel-space interception and telemetry, and service meshes with sidecar proxies for fine-grained control and rich application-layer functionality. This hybrid model balances performance with flexibility, offering a scalable and secure foundation for next-generation mobile core deployments.

\subsubsection{Orchestration and Operations}

The shift toward cloud-native cellular networks fundamentally redefines how mobile infrastructure is deployed, scaled, and managed. Modern operational workflows rely on declarative orchestration systems that continuously reconcile desired state with observed state, enabling fully automated lifecycle management of network functions. At the center of this paradigm is Kubernetes, which provides primitives such as \texttt{Deployments}, \texttt{StatefulSets}, \texttt{DaemonSets}, and \texttt{Jobs} that allow mobile operators to deploy VNFs and CNFs in a deterministic and modular manner~\cite{Workload80online}.

On top of Kubernetes, specialized orchestration frameworks, such as ONAP, OSM, and vendor-specific solutions, extend operational capabilities with telecom-tailored mechanisms for network service composition, policy-driven scaling, lifecycle automation, and multi-domain coordination~\cite{dalgitsis2024cloud}. These frameworks bridge the gap between telecom requirements and general purpose cloud-native tooling by incorporating orchestration logic for service chaining, slice instantiation, SLA-driven autoscaling, and lifecycle management of distributed network functions.

A core requirement for mobile networks is the ability to dynamically instantiate VNFs/CNFs across heterogeneous compute pools, including edge nodes, regional data centers, and centralized clouds. This is particularly important for latency-sensitive RAN workloads and scale-sensitive core workloads. By using Infrastructure-as-Code (IaC) tools such as Terraform, Pulumi, or Ansible, operators can manage the underlying compute, networking, and storage resources in a repeatable manner~\cite{dalvi2022cloud}. Combined with GitOps practices, operational pipelines become versioned, auditable, and reproducible, enabling rapid rollouts and controlled rollbacks.

Finally, orchestration in 5G networks extends beyond the life cycle of individual VNFs/CNFs. It includes continuous management of slice admission, slice updates, cross-slice policy enforcement, and dynamic placement of UPFs and latency-sensitive functions. As 5G advances toward multi-access and multi-cloud deployments, orchestration increasingly requires abstractions that span clouds, Kubernetes clusters, and diverse edge domains. Such capabilities establish the operational foundation for next-generation programmable mobile networks.

\subsubsection{Observability and Tracing}

As cellular networks adopt microservice architectures and distributed cloud deployments, ensuring end-to-end observability becomes essential for maintaining performance, reliability, and security. Unlike monolithic PNFs, where most packet processing occurred within vertically integrated appliances, cloud-native 5G cores distribute functionality across numerous microservices that interact over service-based interfaces. This distributed topology requires new mechanisms to trace interactions, correlate events, and diagnose faults across heterogeneous execution environments~\cite{inaolaji2024analyzing}. To that end, Figure~\ref{fig:obsstack} presents the multi-faceted approach to observability in cloud-based cellular deployments with select examples.

OpenTelemetry has emerged as a de facto standard for instrumentation, enabling collection of logs, metrics, and distributed traces through a unified, vendor-neutral API~\cite{OpenTele37online}. When applied to cellular networks, OpenTelemetry allows operators to monitor SBi interactions, interface latencies (e.g., N2, N3, N6), and control-plane events across AMF, SMF, UPF, NRF, and other VNFs/CNFs. This visibility is crucial for diagnosing issues such as delayed PDU session establishment, load imbalance across UPFs, or slice-level performance degradation.

Distributed tracing, typically implemented through Jaeger or Zipkin, provides fine-grained visibility into the request flows between microservices. For example, in a typical 5G registration procedure, a single UE attach operation leads to a cascade of interactions among the RAN, AMF, AUSF, UDM, and PCF. Tracing allows these interactions to be reconstructed, enabling operators to identify bottlenecks, misconfigurations, or anomalous behaviors. This is especially important in cloud-native deployments where container restarts, autoscaling events, or sidecar proxy behaviors may introduce unexpected latencies.

Complementing application-level tracing, eBPF-based observability tools, such as Cilium Hubble, Pixie, and BCC-enable high-resolution kernel-level monitoring without significant overhead~\cite{soldani2023ebpf}. These tools can capture packet flows, measure TCP retransmissions, track per-pod latencies, and identify congestion or failures within the data plane. Combined with service mesh telemetry, they form a multilayer observability stack spanning kernel, networking, and application layers.

Overall, cloud-native observability is essential for maintaining SLA guarantees, enforcing slice isolation, and enabling reliable AI/ML-driven analytics. It provides the operational transparency required to manage distributed, multi-access cellular deployments.

\subsubsection{Resilience}

Resilience is a foundational property of 5G and beyond systems, particularly given the scale, heterogeneity, and mission-critical nature of contemporary cellular deployments. Cloud-native architectures enhance resilience through redundancy, automated recovery, self-healing workflows, and adaptive traffic steering. These behaviors contrast with PNF-era systems, where failures often required manual intervention and could lead to prolonged service outages.

Kubernetes natively provides mechanisms for failure detection and remediation. Liveness and readiness probes ensure that malfunctioning VNFs/CNFs are automatically restarted or removed from service endpoints~\cite{sannareddy2024autonomous}. Autoscaling mechanisms, i.e., Horizontal Pod Autoscalers (HPA), Vertical Pod Autoscalers (VPA), and Cluster Autoscalers enable elastic adaptation to fluctuating subscriber load, mobility patterns, or application traffic. These capabilities help maintain UE experience continuity even during flash-crowd events or localized node failures.

Another critical enabler of resilient mobile networks is~\cite{5g_stream_dsn2025}. Within the 5G core, the Network Repository Function (NRF) provides dynamic registration and discovery of VNFs via the service-based architecture~\cite{5g_stream_dsn2025, tran2025open5glos}. This allows AMF, SMF, PCF, and other functions to query available service instances at runtime, adapt to scaling events, and seamlessly integrate new CNF instances into operational workflows~\cite{3gpp23501}. Through combining NRF-driven discovery with Kubernetes-native DNS and service abstractions, modern cellular deployments achieve high agility and avoid the rigid configuration paradigms of previous generations~\cite{larrea2023corekube}.

Resilience also extends to the user plane. Multi-UPF architectures, redundant fronthaul/backhaul paths, and service mesh-based load balancing ensure continued connectivity in the presence of failures or congestion~\cite{su2024multi}. eBPF-based failover logic can be used to reroute traffic at kernel level with minimal latency impact, enabling graceful degradation instead of catastrophic failure.

Finally, strong resilience is increasingly tied to autonomous behaviors. AI-driven control loops can detect anomalies, such as RAN degradation, slice SLA violations, or control-plane congestion, and trigger corrective actions. These may include dynamic VNF placement, UPF relocation, slice reconfiguration, or prioritization of mission-critical traffic. Such autonomous resilience mechanisms represent a necessary evolution toward the reliable, adaptive, and self-managing networks envisioned for 6G.
\section{Taxonomy of Cloud-Based Cellular Deployments} \label{sec:taxonomy}
\subsection{Deployment Architecture}
\label{subsec:deployment_architecture}

\begin{figure}[t]
    \centering
    \includegraphics[width=\columnwidth,trim={0.2cm 7cm 22cm 0cm},clip] {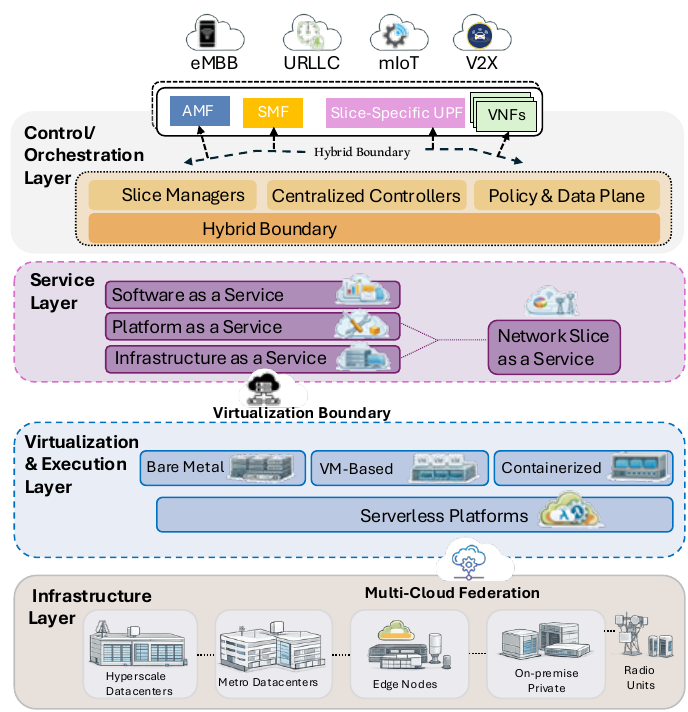}
    \caption{Layered deployment architecture for cloud-based cellular systems, spanning infrastructure, execution substrates, service models, and control/orchestration across federated cloud domains.}
    \label{fig:deployment_architecture}
\end{figure}

Deployment architecture defines the structural blueprint of cloud-based cellular systems and forms the first dimension of the taxonomy introduced in Section~\ref{sec:taxonomy}. As illustrated in Figure~\ref{fig:deployment_architecture}, the architecture is organized into four vertically aligned layers: infrastructure, virtualization and execution, service realization, and control and orchestration. Horizontally, these layers span a federated multi-cloud environment comprising hyperscale regions, metro data centers, edge nodes, and on-premise or private domains. 

In contrast to legacy LTE systems, where network functions were tightly coupled to proprietary hardware appliances, 5G and emerging 6G architectures operate within a programmable cloud fabric that decouples functional logic from physical topology. This shift is aligned with the cloud-native principles formalized in 3GPP Release 17 and beyond and further emphasized in ongoing 6G architectural studies~\cite{saad2019vision, 3gpp23501}. Deployment architecture, therefore, no longer represents a fixed topology, but rather a structural envelope within which network functions, slices, and services may be dynamically instantiated and migrated.

\subsubsection{Infrastructure Layer and Multi-Cloud Federation}

The infrastructure layer forms the physical substrate of the cellular cloud. As depicted in Figure~\ref{fig:deployment_architecture}, it encompasses hyperscale data centers, metro or regional facilities, edge nodes, on-premise private deployments, and radio units. These domains are interconnected through a multi-cloud federation fabric that enables cross-domain orchestration and service continuity.

Hyperscale regions provide elastic compute and storage capacity suitable for subscriber databases, policy control functions, analytics engines, and global service registries. Metro data centers reduce latency while preserving substantial resource pools, enabling regional control-plane clustering and aggregation. Edge nodes, co-located with radio infrastructure, host latency-sensitive user-plane functions and multi-access edge computing workloads. On-premise or private domains satisfy enterprise sovereignty and deterministic performance requirements, increasingly relevant for industrial 5G and 6G verticals.

Federation across these domains introduces challenges in identity propagation, certificate management, service discovery, and telemetry aggregation. Recent studies emphasize the necessity of interoperable orchestration layers and cross-domain slice management to support multi-provider 6G ecosystems~\cite{ismail2024towards}. Deployment architecture therefore evolves from a hierarchical tree into a graph of cooperating administrative domains.

\subsubsection{Virtualization and Execution Layer}

Above the infrastructure substrate resides the virtualization and execution layer, which determines how network functions are instantiated and isolated. Figure~\ref{fig:deployment_architecture} distinguishes four execution paradigms: bare-metal, virtual machine-based VNFs, containerized network functions, and serverless platforms.

Bare-metal execution provides deterministic performance and direct hardware acceleration, which remains valuable for specific radio access workloads and high-throughput user-plane processing. However, limited elasticity and coarse lifecycle management constrain large-scale automation~\cite{aniruddh2021comparison, chiueh2005survey}.

Virtual machine-based virtual network functions represent the classical NFV model. Hypervisor-enforced isolation and mature operational tooling provide strong tenant separation and compliance capabilities. Nevertheless, scaling granularity and resource overhead remain less efficient than container-native deployments.

Containerized network functions have become the dominant realization of cloud-native 5G cores and increasingly of virtualized RAN components. Kubernetes-based orchestration enables declarative lifecycle management, rolling upgrades, horizontal autoscaling, and microservice decomposition. Integration with service meshes enhances observability, traffic control, and security enforcement. These characteristics align with cloud-native transformation roadmaps outlined by major operators and industry bodies~\cite{alliance2021cloud}.

Serverless platforms represent an emerging execution model in which event-driven functions supplement stateful control-plane services. Although not yet pervasive in production-grade mobile cores, serverless paradigms are being investigated for telemetry processing, slice-specific policy evaluation, and edge-triggered service adaptation in 6G research.

Hybrid execution environments frequently coexist within a single deployment. For example, stateful data repositories may remain VM-based, control-plane functions may be containerized, and performance-critical data-plane components may leverage bare-metal acceleration. Deployment architecture must therefore accommodate heterogeneous execution substrates across geographical tiers.

\subsubsection{Service Layer and Network-Slice-as-a-Service}

The service layer abstracts execution substrates into cloud consumption models. As shown in Fig.~\ref{fig:deployment_architecture}, Infrastructure-as-a-Service (IaaS), Platform-as-a-Service (PaaS), and Software-as-a-Service (SaaS) form the foundational cloud stack upon which Network-Slice-as-a-Service (NSaaS) is realized.

IaaS provisions compute, storage, and networking primitives. PaaS offers managed Kubernetes clusters, databases, and messaging frameworks that support network function deployment. SaaS exposes higher-level operational and analytics services. NSaaS overlays these abstractions to provide logically isolated cellular networks tailored to specific service-level objectives.

Slice instantiation may allocate dedicated user-plane functions, policy control chains, and resource quotas while sharing physical infrastructure. Ultra-reliable low-latency slices may anchor user-plane functions at the edge, whereas enhanced mobile broadband slices may aggregate traffic within regional domains. Massive machine-type communication slices may prioritize scalability over strict latency constraints. This workload-aware placement~\cite{liang2024resource} reflects the service differentiation targets of IMT-2030 and 6G visions~\cite{hossain20256g}.

Logical isolation mechanisms, including namespace segmentation, role-based access control, and slice-aware resource scheduling, ensure differentiated quality-of-service enforcement even within shared clusters. The service layer thus operationalizes the economic and multi-tenancy dimensions of the taxonomy introduced in Section~III.

\subsubsection{Control and Orchestration Layer}

The uppermost layer in Fig.~\ref{fig:deployment_architecture} comprises centralized controllers, slice managers, and policy and data-plane coordination components. This layer spans a hybrid boundary that separates logical control from underlying execution domains.

Core control-plane functions such as AMF and SMF may operate within regional or central clusters, while slice-specific user-plane functions are distributed according to latency objectives. Orchestration frameworks coordinate lifecycle management, scaling policies, and placement decisions across heterogeneous clusters. Policy engines enforce quality-of-service constraints and security rules that traverse cloud boundaries.

Hybrid boundaries arise when control logic is centralized, but execution is distributed, or when enterprise slices maintain local autonomy while federating with public cloud cores. Recent research highlights the importance of intent-based orchestration and AI-assisted operations to manage this complexity in 6G environments~\cite{alliance2021cloud}. 

The control and orchestration layer, therefore, binds infrastructure, execution substrates, and service abstractions into a coherent cellular fabric. Its design determines scalability, fault containment, and compliance capabilities across federated domains.

\subsection{Resource Management and Orchestration}
\label{subsubsec:cloud_resource_control}

\begin{figure}[t]
    \centering
    \includegraphics[width=\columnwidth]{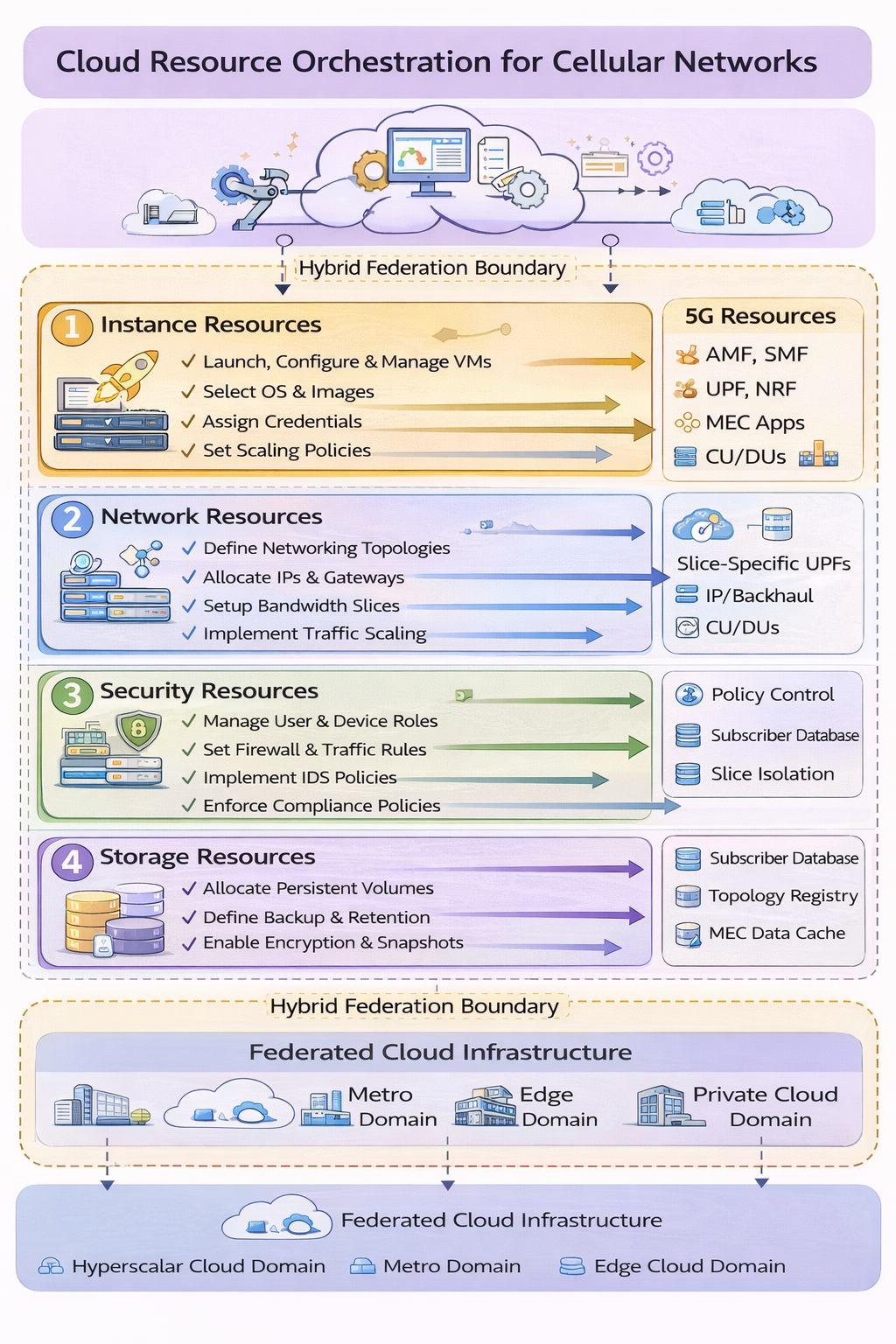}
    \caption{Cloud resource categories and their mapping to 5G functions and slice-specific components across federated domains.}
    \label{fig:resource_slice_mapping}
\end{figure}

Figure~\ref{fig:resource_slice_mapping} refines the orchestration view by decomposing resource management into four primary cloud-level resource domains, instance, network, storage, and security resources,~\cite{ranjan2015cloud, CloudSer16online}, and explicitly mapping them to 5G system components and slice constructs. This abstraction avoids provider-specific terminology and instead emphasizes the generic control primitives required to operationalize cloud-native cellular systems across federated infrastructures.

\paragraph*{Instance Resources}

Instance resources encompass compute primitives used to instantiate network functions and supporting services. These include virtual machines, container runtimes, CPU and memory allocations, hardware acceleration capabilities, and placement constraints across central, metro, or edge domains. Lifecycle operations such as image selection, bootstrap configuration, scaling policies, restart behavior, and affinity rules form part of this control surface.

In 5G systems, instance resources directly determine the realization of control-plane and user-plane functions, including AMF, SMF, UPF, NRF, PCF, and slice-specific application servers. For virtualized RAN deployments, compute allocation governs the instantiation of CU and DU functions and their scaling under load. For edge-resident slices, instance placement defines whether user-plane breakout and MEC workloads satisfy latency constraints. Consequently, instance orchestration becomes a primary mechanism for enforcing slice-specific performance guarantees.

\paragraph*{Network Resources}

Network resources define the logical and physical connectivity fabric within and across cloud domains. These include virtual network topologies, address spaces, routing domains, gateway configuration, bandwidth reservations, traffic steering policies, and inter-domain peering links. At a higher abstraction level, orchestration frameworks manage service chaining and topology constraints associated with slice realization.

Within 5G architectures, network resource control governs N2 and N3 connectivity, fronthaul and midhaul transport for CU/DU deployments, and backhaul paths toward regional or central cores. Slice-specific user-plane functions rely on network resource partitioning to enforce isolation and differentiated throughput guarantees. Traffic classification, policy-based routing, and bandwidth allocation mechanisms enable the realization of enhanced mobile broadband, ultra-reliable low-latency communication, and massive IoT slices over shared infrastructure. Network resource orchestration, therefore, acts as the connective tissue linking distributed compute placements into coherent end-to-end service graphs.

\paragraph*{Storage Resources}

Storage resources support both stateful network functions and operational data pipelines. These include persistent volumes, distributed databases, object storage backends, snapshot and backup policies, encryption controls, and data lifecycle management rules. Orchestration logic determines replication scope, retention intervals, and locality placement across federated domains.

In 5G systems, storage underpins subscriber databases, policy repositories, slice descriptors, topology registries, and telemetry archives. For edge-deployed slices, localized storage enables low-latency caching and application state persistence, while central domains maintain authoritative subscriber and analytics repositories. Storage isolation and encryption policies directly influence compliance and sovereignty requirements, particularly for enterprise or private network slices. Thus, storage orchestration is inseparable from both reliability engineering and regulatory adherence.

\paragraph*{Security Resources}

Security resources encompass identity management, credential distribution, policy enforcement, firewalling, encryption frameworks, and compliance controls. These resources operate horizontally across compute, networking, and storage domains. Orchestration platforms define trust boundaries, rotate credentials, enforce role-based access control, and ensure secure communication channels between distributed components.

In a 5G context, security resource management maps to policy control functions, subscriber authentication mechanisms, slice-level isolation constructs, and inter-domain trust establishment. Certificate propagation and key management become particularly critical in federated multi-cloud scenarios where control-plane elements span administrative boundaries. Isolation policies must ensure that slice-specific UPFs, control-plane instances, and application workloads cannot interfere with co-resident tenants. Security orchestration, therefore, provides the enforcement substrate that maintains integrity and confidentiality across shared cloud fabrics.

\paragraph*{Slice-Level Synthesis Across Federated Domains}

The lower portion of Figure~\ref{fig:resource_slice_mapping} situates these resource domains within a federated cloud infrastructure spanning hyperscale, metro, edge, and private environments. The hybrid federation boundary indicates that orchestration decisions may be centralized in logical terms, whereas execution occurs across distributed domains. 

For a given network slice, the orchestrator translates service objectives into coordinated actions across all four resource categories. Instance policies define where core or RAN functions are instantiated. Network configurations establish slice-specific connectivity and traffic prioritization. Storage rules ensure state durability and alignment with locality. Security controls enforce tenant isolation and compliance constraints. The slice, therefore, emerges as a composite allocation of heterogeneous cloud resources bound together by declarative intent and closed-loop control.

This resource-centric view clarifies that slice management is not a singular control primitive but rather a coordinated orchestration of compute, network, storage, and security abstractions. Effective resource management frameworks must therefore expose programmable interfaces across all four domains while maintaining consistency across federated cloud boundaries. Such cross-domain coordination represents a prerequisite for scalable, multi-tenant 5G and future 6G cloud-native deployments.

\subsection{Multi-Tenancy \& Isolation}

A central premise of cloud-based cellular deployments is the ability to support multiple independent tenants on a shared infrastructure while preserving strict operational, performance, and security boundaries~\cite{kotulski2017end}. Unlike traditional mobile networks, which were typically deployed as vertically integrated systems operated by a single entity, cloud-native cellular architectures enable infrastructure providers, mobile operators, enterprises, and service providers to coexist on a common compute and networking substrate. This shift toward shared infrastructure introduces new challenges in ensuring that workloads belonging to different tenants do not interfere with one another in terms of resource consumption, performance behavior, security exposure, or management control.

\begin{figure*}[t]
    \centering
    \includegraphics[width=\textwidth,trim={0.2cm 12.5cm 2cm 0cm},clip] {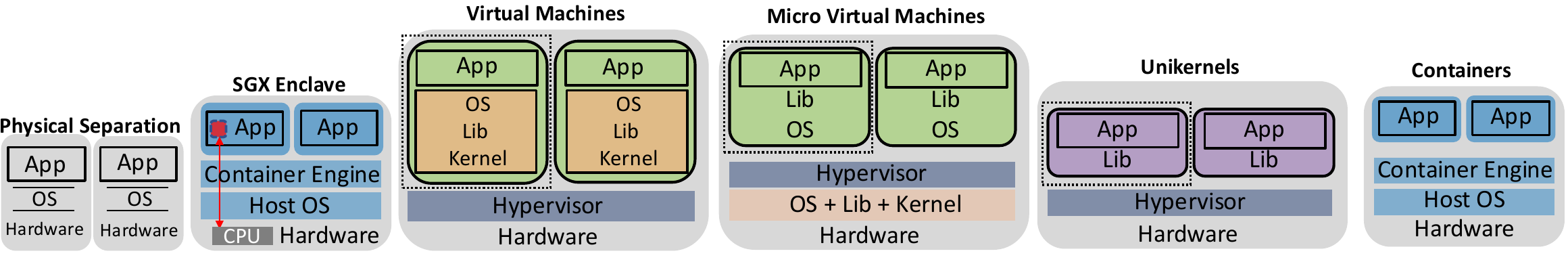}
    \caption{Layered deployment architecture for cloud-based cellular systems, spanning infrastructure, execution substrates, service models, and control/orchestration across federated cloud domains.}
    \label{fig:isolation_models}
\end{figure*}

Multi-tenancy in cellular systems extends beyond simple infrastructure sharing. In modern 5G and emerging 6G architectures, tenants may operate logically independent network slices, deploy custom network functions, and maintain distinct operational policies while utilizing a common cloud platform. Achieving this level of coexistence requires isolation mechanisms that span the entire system stack, including compute environments, operating systems, virtualization substrates, networking resources, and orchestration frameworks. The strength, flexibility, and overhead of these isolation mechanisms vary depending on the underlying execution model. Figure~\ref{fig:isolation_models} illustrates several representative isolation strategies that have emerged in cloud-native computing and are increasingly applied to cellular network deployments.

\subsubsection{Isolation Models}

Isolation mechanisms in cloud-based cellular systems span a spectrum from dedicated physical infrastructure to lightweight virtualization techniques. Each approach provides a different balance between security isolation, resource efficiency, and operational flexibility. As illustrated in Figure~\ref{fig:isolation_models}, these models differ primarily in how applications interact with operating system components and hardware resources~\cite{shu2016study}.

At one end of the spectrum, physical separation provides the strongest form of isolation by dedicating hardware resources to individual workloads. In this model, applications execute directly on their own operating system instance running on exclusive hardware. Traditional telecommunications infrastructure frequently relied on this approach, deploying network functions on dedicated appliances or specialized servers. Although this model offers strong isolation guarantees and predictable performance, it lacks the elasticity and resource efficiency required for large-scale cloud deployments.

Trusted execution environments represent a hybrid model that introduces hardware-enforced isolation within shared infrastructure. Technologies such as secure enclaves enable sensitive application components to execute inside protected processor regions that remain inaccessible to the host operating system and other workloads~\cite{costan2017secure}. This approach is particularly relevant for protecting security-critical components of cellular networks, such as subscriber identity processing, key management functions, or sensitive control-plane logic. By leveraging hardware-supported confidentiality and integrity guarantees, trusted enclaves enable operators to run sensitive workloads even within shared multi-tenant environments~\cite{maene2017hardware}.

Virtual machine virtualization provides isolation through a hypervisor that partitions hardware resources into multiple logically independent guest environments. Each virtual machine includes its own operating system kernel and runtime environment, allowing applications to operate as if they were running on dedicated hardware~\cite{iqbal2009overview}. This model became the foundation of NFV, enabling telecom operators to replace proprietary appliances with software-based network functions deployed on commodity servers. While virtual machines provide strong isolation and compatibility with legacy software stacks, they incur higher resource overhead due to the duplication of operating system components across tenants.

Micro virtual machines attempt to reduce this overhead by minimizing the software stack required for each workload. In these architectures, the guest environment contains a simplified operating system layer that integrates application libraries with essential kernel components~\cite{mavridis2019lightweight}. The resulting execution model preserves the hardware isolation benefits of virtualization while significantly reducing startup latency and resource consumption. MicroVM-based approaches are increasingly being explored in multi-tenant cloud environments, where strong isolation is required without sacrificing the efficiency and elasticity of containerized deployments.

Unikernel-based architectures take this concept further by compiling application logic together with only the minimal operating system functionality required for execution. The resulting specialized machine image contains a single address space that integrates application code and operating system libraries~\cite{talbot2020security}. By eliminating unnecessary system components, unikernels significantly reduce memory footprint and attack surface while maintaining compatibility with hypervisor-based execution environments. These characteristics make unikernels attractive for latency-sensitive edge deployments and lightweight network functions.

Container-based virtualization represents the most lightweight execution model among the approaches shown in Figure~\ref{fig:deployment_architecture}. Containers share a common host operating system kernel while isolating applications through kernel namespaces and resource control mechanisms. This model drastically reduces deployment overhead and startup latency, enabling highly elastic cloud-native environments. As a result, containerized network functions have become the dominant deployment model for modern 5G core networks and edge computing platforms.

In practice, large-scale cloud-native cellular systems often combine multiple isolation mechanisms simultaneously. Containerized network functions may execute within virtual machines to strengthen security boundaries, while hardware-assisted trusted execution environments protect sensitive operations. The combination of these techniques allows operators to balance isolation strength, performance efficiency, and operational flexibility when deploying multi-tenant cellular services across shared cloud infrastructure.

\subsection{Cloud Role and Control}

Cloud-based cellular deployments introduce a layered operational model in which responsibility for system deployment, management, and control is distributed across multiple actors. Unlike traditional mobile networks where a single operator typically owns and operates the entire infrastructure stack, cloud-native cellular systems rely on a combination of cloud infrastructure providers, telecom operators, service brokers, and enterprise tenants. Each actor maintains control over a different portion of the system stack, resulting in a shared responsibility model that governs how infrastructure resources are provisioned, secured, and operated. Understanding the distribution of these responsibilities is essential for analyzing the operational, security, and governance implications of cloud-hosted cellular networks.

\begin{figure}[t]
\centering
\includegraphics[width=\columnwidth,trim={6cm 2.5cm 6.5cm 3cm},clip]{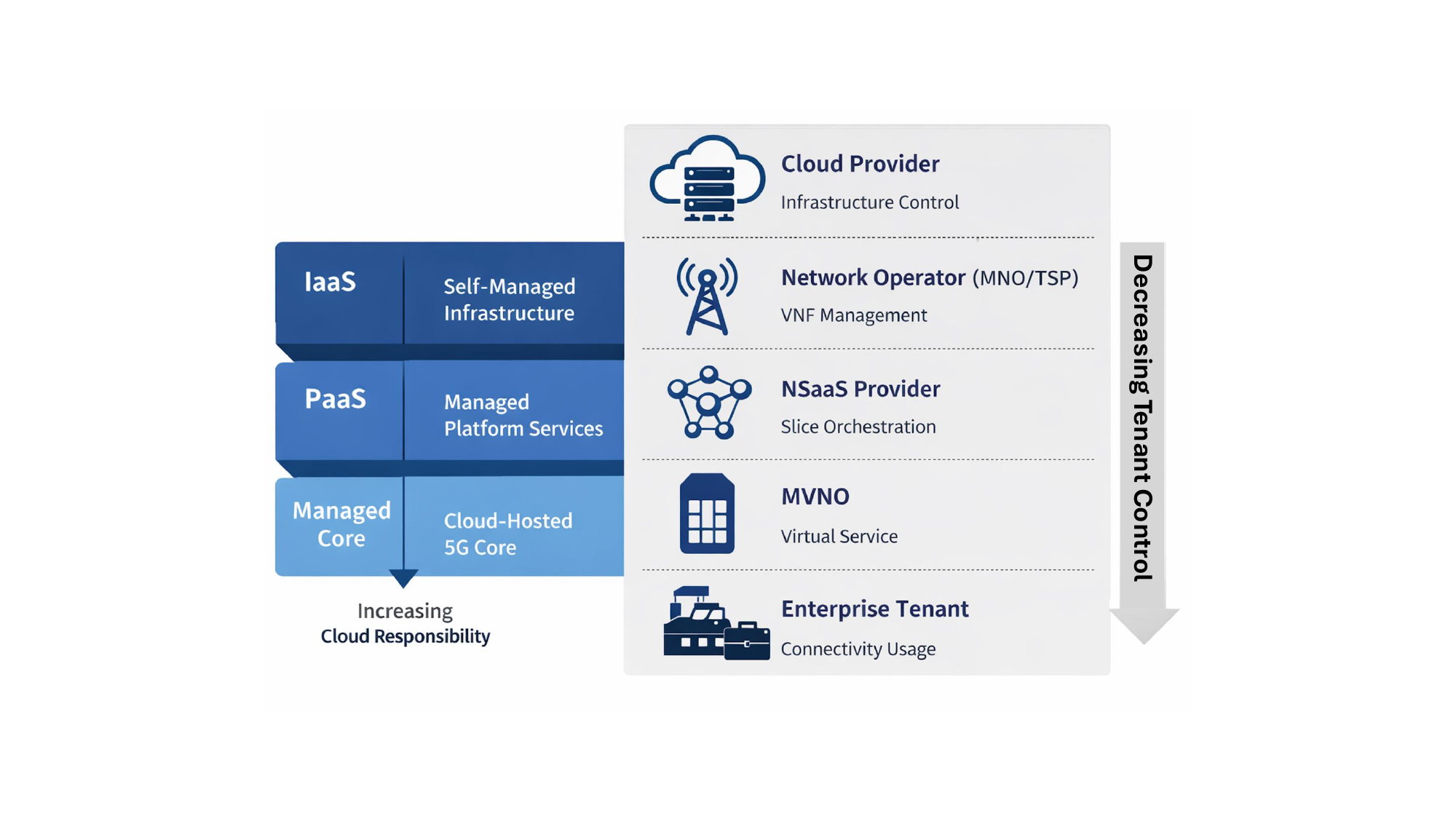}
\caption{Cloud responsibility continuum and telecom role hierarchy in cloud-based cellular deployments. The left axis illustrates increasing infrastructure responsibility assumed by the cloud provider, while the right axis shows the hierarchical distribution of control across telecom stakeholders.}
\label{fig:cloud_roles_control}
\end{figure}

Two complementary dimensions characterize this distribution of responsibility. The first dimension concerns the \emph{cloud responsibility continuum}, which determines how much of the infrastructure stack is managed by the cloud provider versus the telecom operator. The second dimension concerns the \emph{telecom role hierarchy}, which describes how network control is delegated across stakeholders that consume or operate connectivity services. Figure~\ref{fig:cloud_roles_control} illustrates these two dimensions. The left axis of the figure represents the progression of cloud service models from infrastructure provisioning to fully managed cellular platforms, while the right axis shows the hierarchical distribution of control among telecom stakeholders.

\subsubsection{Shared Responsibility in Cloud-Based Cellular Systems}

The deployment of cellular network functions on public cloud infrastructure inherits the shared responsibility paradigm widely used in cloud computing platforms. In this model, responsibility for system operation is divided between the cloud provider and the telecom operator depending on the level of abstraction at which infrastructure services are consumed. At the infrastructure layer, cloud providers expose programmable compute, storage, and networking resources that serve as the substrate for network function deployment. Telecom operators deploy virtual network functions (VNFs) or cloud-native network functions (CNFs) on top of these resources while maintaining control over orchestration frameworks, network configuration, and operational management.

When cellular systems are deployed directly on infrastructure services, the telecom operator retains significant control over the system architecture. In this configuration, VNFs or CNFs are deployed on operator-managed container orchestration environments such as Kubernetes clusters running on cloud compute instances. The operator is responsible for cluster lifecycle management, networking overlays, security policies, service mesh deployment, and monitoring infrastructure. Several research platforms and experimental 5G deployments adopt this architecture because it allows operators to modify control plane behavior and experiment with new network management techniques \cite{taleb2021cloudnative, foukas2017network}.

As cloud platforms introduce increasingly sophisticated managed services, operators may choose to delegate portions of the infrastructure stack to the cloud provider. Managed container platforms, managed databases, and integrated observability services reduce the operational burden associated with maintaining distributed infrastructure at scale. In such deployments, operators focus primarily on telecom-specific service logic and network orchestration policies while the cloud provider assumes responsibility for cluster availability, infrastructure scaling, and platform-level security controls. The resulting architecture preserves operator control over network behavior while leveraging cloud automation for infrastructure management \cite{aws5gwhitepaper, azure5gcore}.

At the highest level of abstraction, some cloud providers offer fully managed cellular core services in which the cloud platform assumes responsibility for the majority of the network infrastructure. In these deployments, telecom operators interact with the cellular system primarily through service configuration interfaces rather than direct infrastructure management. While this model simplifies deployment and reduces operational complexity, it also shifts significant operational authority to the cloud provider and introduces new trust dependencies between telecom operators and the underlying infrastructure.

\subsubsection{Telecom Role Hierarchy and Control Boundaries}

In addition to the infrastructure responsibility continuum, cloud-based cellular systems exhibit a hierarchical structure of operational roles that determine how network control is distributed across stakeholders. The right side of Figure~\ref{fig:cloud_roles_control} illustrates this hierarchy. At the base of the hierarchy resides the cloud provider, which operates the physical infrastructure including data center facilities, compute hardware, storage systems, and programmable networking fabrics. This infrastructure forms the execution substrate upon which higher-layer cellular services are deployed.

Above the infrastructure layer resides the telecom operator or mobile network operator (MNO), which deploys and manages the cellular control plane and user plane functions. The operator maintains authority over network architecture design, subscriber management systems, authentication infrastructure, and network slicing policies. Even when cellular network functions are hosted in public cloud environments, operators typically retain control over the logical architecture of the cellular system and the orchestration policies that govern network behavior.

Additional abstraction layers may emerge when connectivity services are offered to downstream tenants. Network service providers may operate intermediate platforms that expose network slices or programmable connectivity services to external organizations. These entities function as brokers that abstract infrastructure complexity while enabling tenants to request network resources through higher-level service interfaces. Enterprises and application providers occupy the upper layers of the hierarchy, consuming connectivity services without direct visibility into the underlying infrastructure.

Each transition in this hierarchy introduces a control boundary in which operational authority and infrastructure visibility are partially delegated. Lower layers retain responsibility for infrastructure provisioning and platform availability, while higher layers maintain authority over service configuration and application-level policies. These boundaries directly influence how service-level agreements, operational accountability, and fault management are enforced across cloud-hosted cellular systems.

\subsubsection{Trust Implications of Cloud-Managed Cellular Infrastructure}

The distribution of operational responsibility across multiple actors introduces important trust considerations for cloud-based cellular deployments. When telecom operators retain direct control over infrastructure and orchestration platforms, they maintain strong authority over network behavior but assume the operational complexity associated with maintaining large-scale distributed systems. Conversely, when infrastructure management is delegated to cloud providers, operational complexity decreases while trust dependencies increase.

This shift in trust boundaries is particularly significant for sensitive components of the cellular control plane. Functions such as authentication servers, subscriber databases, and policy management systems process highly sensitive subscriber information that historically resided within operator-controlled facilities. Hosting these components within shared cloud infrastructure requires mechanisms that preserve confidentiality and isolation even when the underlying infrastructure is not fully controlled by the telecom operator.

Recent research has explored several techniques for strengthening isolation guarantees in such environments, including hardware-backed trusted execution environments, secure enclaves, and micro-virtualization approaches that protect sensitive network functions from the surrounding infrastructure \cite{costan2016intel, gu2022hardware}. These mechanisms allow telecom operators to maintain strong security guarantees even when cellular network functions are deployed on infrastructure managed by external cloud providers.

As cloud-native cellular architectures continue to evolve toward 6G-era deployments, the relationship between infrastructure ownership, operational control, and trust will remain a defining design consideration. The shared responsibility continuum and telecom role hierarchy illustrated in Figure~\ref{fig:cloud_roles_control} provide a conceptual framework for understanding how operational control is distributed across actors in cloud-hosted cellular systems and how these boundaries influence the security, governance, and reliability properties of the resulting network.


\subsection{Economic \& Ownership Models}

\begin{figure}[t]
\centering
\includegraphics[width=\columnwidth,trim={5cm 5cm 9cm 3.3cm},clip]{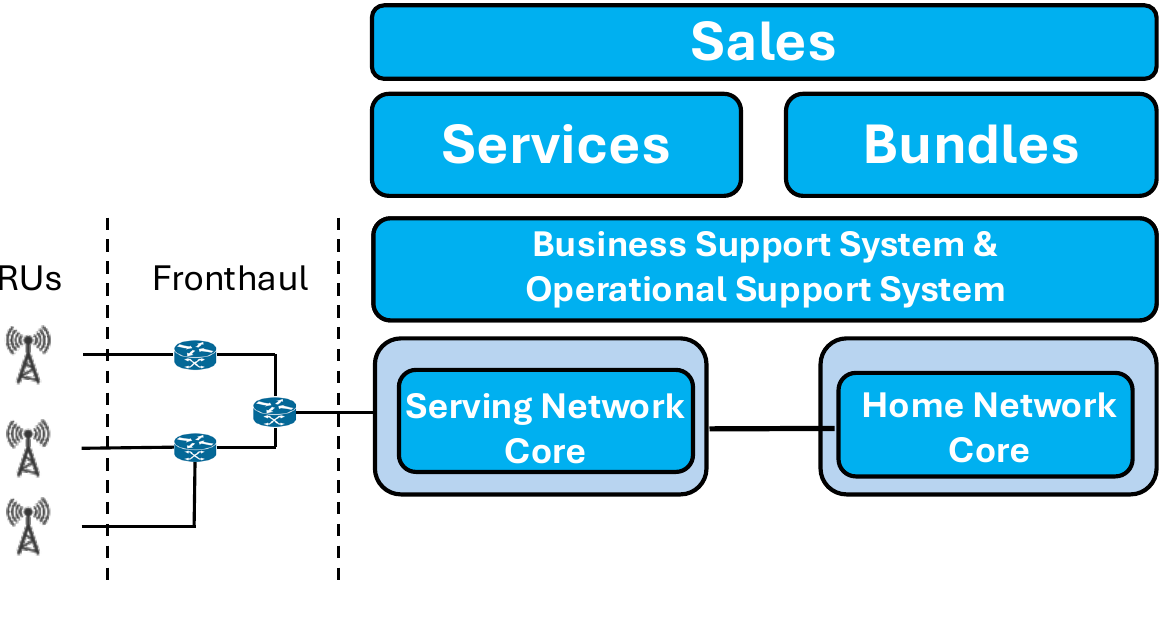}
\caption{Traditional vertically integrated mobile network operator architecture including radio access infrastructure, core network functions, and OSS/BSS operational systems that support subscriber management and service provisioning.}
\label{fig:mno_architecture}
\end{figure}

The migration of cellular networks toward cloud-native architectures fundamentally alters the economic and ownership structures that govern mobile network deployments. Traditional cellular systems have historically been vertically integrated, with mobile network operators (MNOs) owning and operating the majority of the infrastructure required to deliver connectivity services. This includes radio access networks, transport infrastructure, core network elements, and the operational support systems that manage subscribers and services. While this model provides operators with strong control over network performance and service provisioning, it also requires substantial capital investment and operational overhead.

Cloud-based cellular deployments decouple infrastructure ownership from network service provisioning. Infrastructure resources may be owned by hyperscale cloud providers, while telecom operators deploy network functions on top of these resources to deliver connectivity services. At the same time, virtual operators and enterprise tenants may consume connectivity services without directly managing underlying infrastructure. This layered ownership structure resembles the service-oriented economic models common in cloud computing ecosystems and introduces new actors into the cellular value chain. Understanding how these actors interact economically is essential for analyzing how future 5G and 6G networks will be deployed and monetized.

\subsubsection{Vertically Integrated Operator Model}

Historically, mobile network operators have operated under a vertically integrated ownership model in which the same entity owns the majority of the infrastructure required to deliver cellular connectivity. As illustrated in Figure~\ref{fig:mno_architecture}, this model includes radio infrastructure, core network systems, and operational platforms such as business support systems (BSS) and operational support systems (OSS). These systems collectively manage subscriber provisioning, billing, network monitoring, and service delivery.

The vertically integrated structure provides operators with end-to-end control over network performance and security policies. However, this architecture requires significant capital expenditures associated with spectrum licenses, radio infrastructure deployment, transport networks, and data center facilities. As network architectures evolve toward virtualized and cloud-native deployments, operators increasingly seek to reduce infrastructure costs by leveraging shared cloud platforms and virtualized network functions \cite{taleb2021cloudnative, foukas2017network}.

\subsubsection{Mobile Virtual Network Operator Models}

One of the earliest examples of economic separation between infrastructure ownership and service provisioning is the mobile virtual network operator (MVNO) model. MVNOs provide cellular services to subscribers without owning the underlying radio infrastructure. Instead, they lease capacity from host MNOs while providing differentiated services such as pricing models, branding, or specialized service offerings.

\begin{figure}[t]
\centering
\includegraphics[width=\columnwidth,trim={2.2cm 4cm 2cm 3cm},clip]{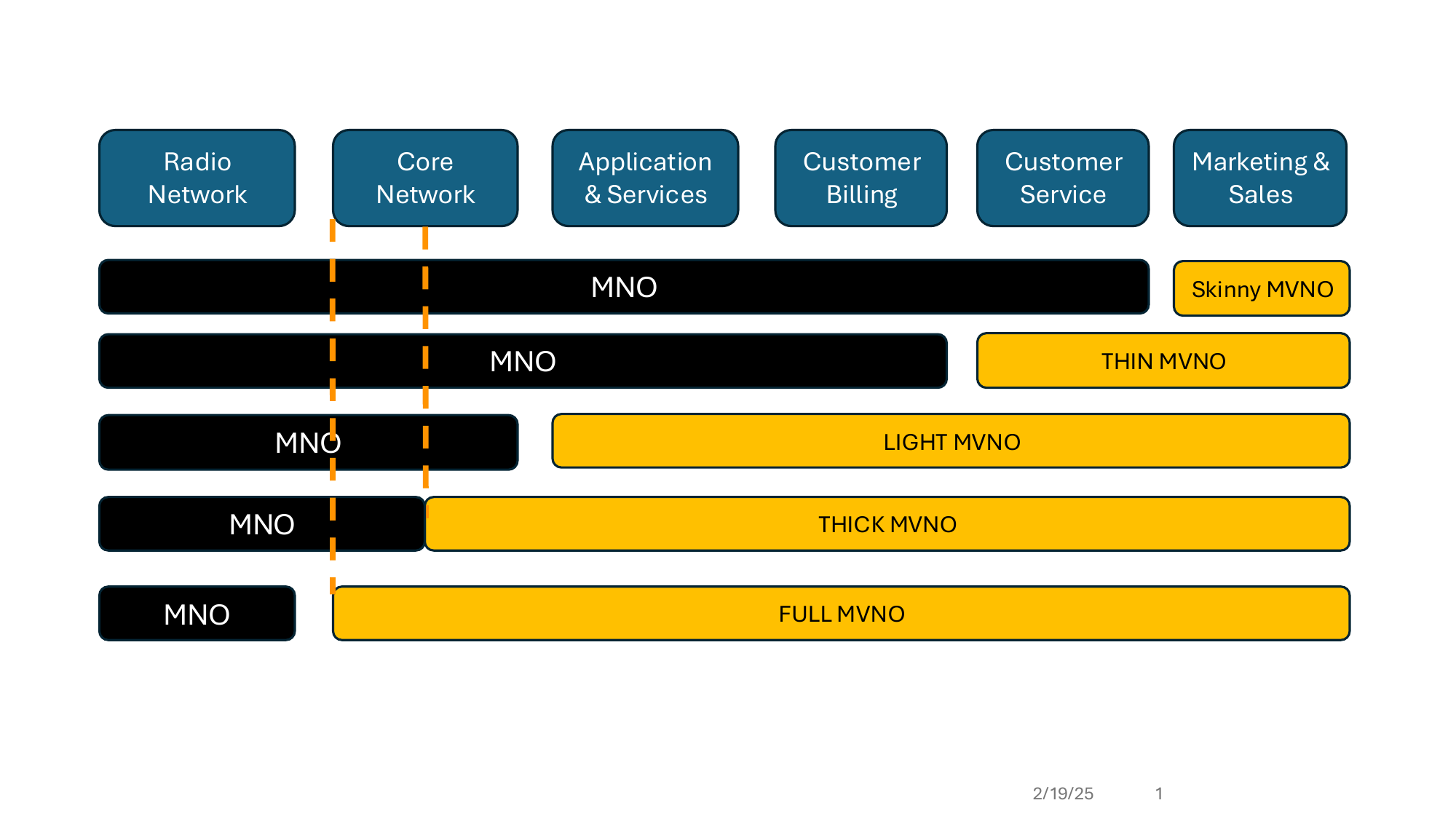}
\caption{Spectrum of mobile virtual network operator (MVNO) models illustrating increasing operational independence from the host MNO. As MVNOs move from branded or skinny models toward full MVNO deployments, they assume greater control over core network and service infrastructure.}
\label{fig:mvno_models}
\end{figure}

Figure~\ref{fig:mvno_models} illustrates the spectrum of MVNO configurations that exist within cellular ecosystems. At one extreme are \emph{skinny} or \emph{branded} MVNOs, which rely heavily on the host MNO for network operation and infrastructure management. These operators typically focus on marketing, customer relationship management, and billing functions while outsourcing most network operations to the host operator.

As the degree of independence increases, \emph{thin} or \emph{light} MVNOs may operate portions of the service infrastructure such as subscriber databases or service platforms while still relying on the host MNO for core network connectivity. At the highest level of independence are \emph{full MVNOs}, which operate substantial portions of the home network infrastructure including authentication servers, subscriber management systems, and service platforms while leasing only radio access capacity from the host network \cite{ghaznavi2019mvno}.

The MVNO model demonstrates how cellular services can be economically separated from infrastructure ownership. This separation becomes even more pronounced in cloud-native cellular architectures where infrastructure resources can be dynamically provisioned from shared cloud platforms.

\subsubsection{Cloud-Native Network Service Providers}

The emergence of cloud-native network architectures introduces new actors that extend beyond the traditional MNO-MVNO relationship. In particular, some platforms act as intermediaries that provide \emph{network-as-a-service} (NaaS) or network slicing services built on top of shared cloud infrastructure. In this model, the underlying infrastructure may be owned by hyperscale cloud providers, while telecom operators deploy virtualized network functions that expose programmable connectivity services to downstream tenants.

This layered service structure allows connectivity to be offered as a programmable resource. Enterprises or application providers may request network slices that provide guaranteed performance characteristics such as latency bounds, bandwidth allocations, or reliability guarantees. Operators allocate network resources dynamically to satisfy these service requirements while enforcing service-level agreements across tenants \cite{foukas2017network, rost2017network}. Such service abstractions allow connectivity to be monetized in ways that resemble cloud computing resource models rather than traditional subscription-based telecom services.

\subsubsection{Enterprise and Private Cellular Networks}

Another emerging ownership model involves enterprise-operated private cellular networks. Industrial enterprises, campuses, and government organizations increasingly deploy private 5G networks to support specialized connectivity requirements such as industrial automation, robotics, and mission-critical communications. In these deployments, the enterprise may own portions of the infrastructure stack including radio access equipment and local edge computing resources.

Private networks often adopt hybrid architectures in which radio infrastructure is deployed locally while portions of the network core are hosted in cloud environments. This approach allows enterprises to maintain control over sensitive data and latency-sensitive applications while leveraging the scalability and operational automation of cloud platforms \cite{etalebprivate5g}. As a result, the boundary between telecom operators and enterprise network operators becomes increasingly fluid, with enterprises assuming greater responsibility for connectivity infrastructure.

\subsubsection{Revenue Models and Connectivity Monetization}

Cloud-native cellular architectures also enable new revenue models for connectivity services. Traditional mobile networks rely primarily on subscription-based pricing in which users pay fixed monthly fees for connectivity services. By contrast, cloud-based deployments enable more flexible economic models that resemble cloud computing pricing structures.

Connectivity services may be priced based on usage metrics such as network slice allocation, data throughput, quality-of-service guarantees, or edge computing resources associated with application workloads. Infrastructure providers may charge operators for the compute, storage, and networking resources required to host cellular network functions, while operators monetize connectivity services offered to downstream tenants.

This layered pricing model creates a multi-sided connectivity marketplace in which infrastructure providers, telecom operators, service brokers, and enterprise tenants each capture value at different layers of the ecosystem. As cellular networks evolve toward fully cloud-native architectures, the economic structure of connectivity services will increasingly resemble the service-oriented ecosystems that characterize modern cloud computing platforms.
\section{The Big Three and 5G} \label{sec:bigthree}

\subsection{Amazon Web Services}
\label{subsec:aws}

Among the three major hyperscale cloud providers, AWS has emerged as the most prolific platform for cloud-based cellular deployments. Through a combination of purpose-built infrastructure services and strategic partnerships with global TSPs, AWS has positioned itself as the cloud backbone for next-generation mobile networks. In this subsection, we examine the AWS infrastructure tailored for 5G, its flagship operator partnerships, and the services that collectively form the AWS telecom ecosystem.

\subsubsection{AWS Infrastructure for 5G Deployments}
\label{subsubsec:aws_infra}

\begin{figure}[t]
    \centering
    \begin{subfigure}[t]{0.24\columnwidth}
        \centering
        \includegraphics[width=\textwidth,trim={14.5cm 6.1cm 13.7cm 3cm},clip]{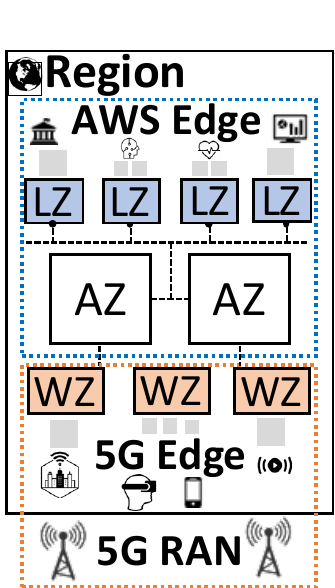}
        \caption[]%
        {{\small }}    
        \label{fig:awsinfra}
    \end{subfigure}
    \hfill
    \begin{subfigure}[t]{0.74\columnwidth}  
        \centering 
        \includegraphics[width=\textwidth,trim={7.5cm 4.6cm 8.8cm 4cm},clip]{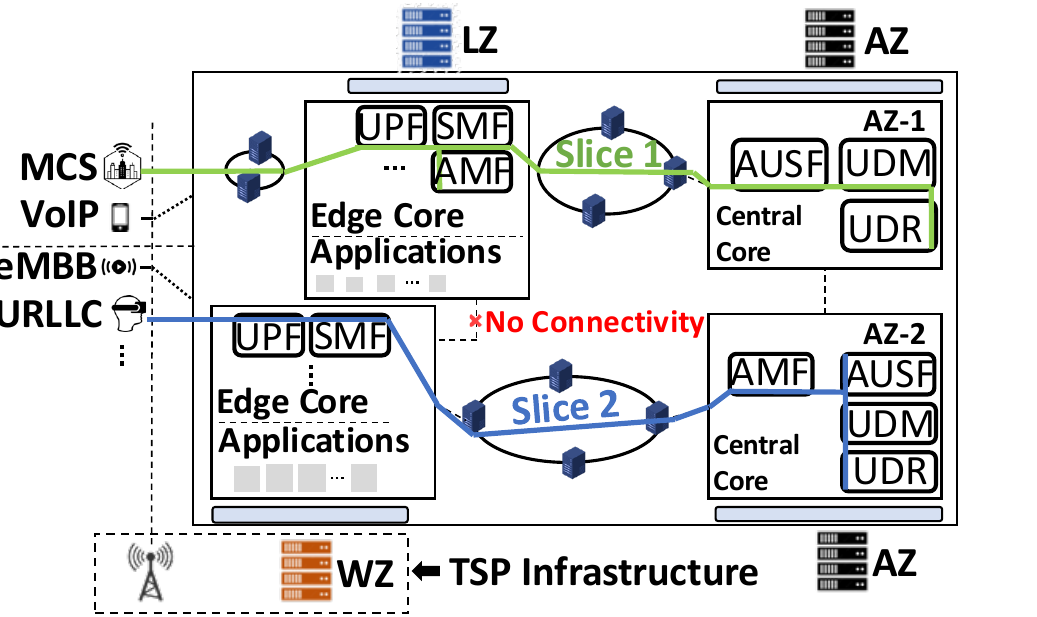}
        \caption[]%
        {{\small }}    
        \label{fig:aws5gdep}
    \end{subfigure}
    \vspace{0.1in}
    \caption{(a) Breakdown of an AWS region with different infrastructure zones; (b) QoS-sensitive 5G deployment on top of AWS across AZs, LZs, and WZs}
    \label{fig:awsow}
\end{figure}

The AWS global infrastructure is organized into geographical segments known as \textit{regions} (e.g., US-East). As illustrated in Figure~\ref{fig:awsinfra}, each region consists of multiple \textit{Availability Zones} (AZs) and, for selected regions, finer-grained \textit{Local Zones} (LZs)~\cite{aws_localzones}. While an AZ serves as a general-purpose deployment zone with full access to the AWS service portfolio, LZs are positioned closer to end-users to enable low-latency applications at the edge of the AWS network. In addition to AZs and LZs, AWS has introduced the \textit{Wavelength Zone} (WZ)~\cite{aws_wavelength_faq} to embed AWS compute and storage services directly within TSP infrastructure. Compared to LZs, WZs deploy applications on top of the carrier's own 5G network edge, in close proximity to the Radio Access Network (RAN), thereby delivering an integrated cloud environment capable of supporting URLLC applications.

As depicted in Figure~\ref{fig:aws5gdep}, these three zone types create a tiered deployment hierarchy that is particularly well-suited for QoS-sensitive 5G core architectures. The figure illustrates two network slices deployed across a hybrid AZ--LZ--WZ layout. For Mission Critical Services (MCS) such as Slice~1, where managing network slice setup and transfer times is paramount for high reliability, the edge core VNFs (UPF, SMF, AMF) are deployed across an LZ--AZ connection, leveraging the lower latency of the LZ--AZ path compared to the WZ--AZ alternative. Conversely, for latency-sensitive applications such as eMBB and URLLC (Slice~2), the data session VNFs are deployed closer to the 5G edge in WZs for improved user plane QoS. As discussed in~\cite{5g_map_mobicom2025}, the lack of direct LZ--WZ connectivity limits hybrid deployment options; establishing such inter-connectivity would unlock richer topologies where VNFs and application servers can be distributed across both edge zone types simultaneously.

As of 2025, AWS Wavelength Zones operate in over 30 cities globally through partnerships with seven TSPs: Verizon in the United States~\cite{aws_wavelength_faq}, Vodafone in the United Kingdom and Germany, KDDI in Japan, SK Telecom in South Korea, Bell in Canada, British Telecom in the United Kingdom, and Orange/Sonatel in Africa~\cite{aws_wavelength_locations}. Notable recent expansions include Africa's first Wavelength Zones in Casablanca (January 2025) and Dakar (April 2025) through the Orange partnership~\cite{aws_wz_casablanca, aws_wz_dakar}. In~\cite{5g_map_mobicom2025}, the authors conducted a global-scale measurement campaign across 18 edge zones in seven countries, evaluating both the control and user plane performance of cloud-based 5G core deployments on AWS. Their findings demonstrate that the strategic placement of VNFs across WZs, LZs, and AZs can reduce inter-VNF latency by up to five times, highlighting the critical role of zone selection in achieving optimal 5G core performance.

\subsubsection{AWS Services for Telecom}
\label{subsubsec:aws_services}

To complement its infrastructure zones, AWS has developed a portfolio of services specifically targeting telecom workloads. The \textit{AWS Telco Network Builder} (TNB)~\cite{aws_tnb} is a fully managed service for automating the deployment and lifecycle management of telecom networks. TNB adheres to the ETSI NFV MANO standards (SOL001--SOL005 in TOSCA format), enabling operators to define network function packages and deploy them across AWS Regions, Local Zones, and Outposts. As of 2025, TNB is available in ten AWS Regions and has been demonstrated in conjunction with partners such as Fujitsu for automated end-to-end 5G RAN and core deployment~\cite{aws_fujitsu_tnb}.

\textit{AWS Outposts} represents a critical bridge between public cloud and on-premises telecom infrastructure. At Mobile World Congress (MWC) 2025, AWS unveiled second-generation Outposts racks specifically designed for telecom workloads~\cite{aws_outposts_telco_mwc2025}. These racks feature bare-metal instances with up to 800~Gbps accelerated networking and support Layer~2 networking capabilities (VLANs, multicast, hardware PTP) that are essential for 5G Core User Plane Function (UPF) and RAN Centralized Unit (CU) workloads~\cite{aws_outposts_2ndgen_blog}. The significance of Outposts for telecom was further underscored in March 2026, when O2~Telef\'{o}nica deployed production-scale 5G core VNFs on AWS Outposts racks within its own data center~\cite{telefonica_outposts_2026}, marking the first such deployment worldwide.

Notably, AWS retired its \textit{Private 5G} service on May~20, 2025~\cite{aws_private5g_retired}. Originally launched in 2022 as an integrated offering combining small cell radios, Outposts servers, and 5G core/RAN software, the service was discontinued in favor of the \textit{Integrated Private Wireless on AWS} program~\cite{aws_ipw_program}. This strategic pivot signaled AWS's recognition that partnering with established carriers---rather than competing with them in providing radio infrastructure---is a more sustainable approach to the private 5G market.

For container orchestration, Amazon Elastic Kubernetes Service (EKS) and its on-premises variant EKS Anywhere have become the de facto platform for deploying cloud-native 5G core VNFs on AWS. Multiple operator deployments, including NTT DOCOMO~\cite{docomo_aws_2024} and O2~Telef\'{o}nica~\cite{telefonica_aws_2024}, rely on EKS for managing Kubernetes-based 5G core workloads. In~\cite{5g_map_mobicom2025}, the authors utilize the Rancher Kubernetes Engine (RKE) to deploy self-managed production-grade clusters across 15 EC2 instances per testing location, demonstrating how operators can leverage AWS compute instances (e.g., t3.xlarge) across AZs and edge zones for fine-grained VNF placement experimentation.

\subsubsection{Major Operator Partnerships}
\label{subsubsec:aws_partnerships}

The maturation of AWS as a telecom cloud platform is best illustrated through its flagship operator partnerships, which have progressed from initial proofs-of-concept to production-scale commercial deployments.

\textbf{NTT DOCOMO.} The DOCOMO--AWS partnership represents the most thoroughly documented telecom cloud migration to date, spanning four years from proof-of-concept to commercial launch. In March 2022, DOCOMO and NEC began testing NEC's 5G SA Core on AWS using Graviton2 processors, achieving a 72\% average power consumption reduction compared to x86 alternatives~\cite{docomo_graviton2_2022}. By February 2023, the two companies completed a carrier-grade hybrid cloud redundancy design~\cite{docomo_redundancy_2023}. The formal selection of AWS was announced on February~26, 2024, when DOCOMO---then serving over 89 million subscribers---chose AWS to deploy a nationwide 5G Open RAN in Japan, also joining DOCOMO's OREX initiative to promote Open RAN globally~\cite{docomo_aws_2024}. The culmination arrived on March~2, 2026, when DOCOMO and NEC launched Japan's first commercial 5G core network on AWS~\cite{docomo_nec_5gcore_2026, docomo_5gcore_pr_2026}. The deployment leverages Agentic AI (Amazon Bedrock AgentCore combined with GitOps) to automate 5G core design and construction on AWS, reportedly reducing construction time by approximately 80\% compared to conventional methods. The system runs on Graviton3 processors with a confirmed approximately 70\% power consumption reduction, and DOCOMO has also deployed an agentic AI system for network operations analyzing data from over one million network devices~\cite{docomo_nec_5gcore_2026}.

\textbf{O2 Telef\'{o}nica.} In May 2024, O2~Telef\'{o}nica became the first existing telecom operator to migrate its production 5G core to a hyperscaler's public cloud~\cite{telefonica_aws_2024, telefonica_nokia_aws_2024}. Using Nokia's cloud-native 5G core architecture orchestrated via Amazon EKS on AWS, the initial phase supported approximately one million customers on the 5G Standalone network. In a significant follow-up announced at MWC 2026, O2~Telef\'{o}nica deployed production-scale 5G core VNFs on AWS Outposts within its own data center~\cite{telefonica_outposts_2026}, making it the first operator globally to move production 5G core functions to on-premises AWS infrastructure. Serving over 35 million mobile connections, the operator is also integrating Amazon Bedrock AgentCore for AI-driven autonomous network operations.

\textbf{DISH Network.} The DISH--AWS strategic collaboration, announced in April 2021~\cite{dish_aws_2021}, was a landmark achievement as the first cloud-native 5G network built entirely on public cloud infrastructure. DISH deployed a standalone 5G Open RAN using AWS Outposts, Local Zones, EKS, and Graviton2 instances, with Nokia providing the 5G SA Core~\cite{dish_aws_architecture}. The hierarchical architecture spanned national, regional, and breakout edge data centers, all managed through Infrastructure-as-Code (IaC) with AWS CDK and CloudFormation. However, following the DISH--EchoStar merger (completed January 2024), financial pressures led to a dramatic reversal. In late 2025, EchoStar announced approximately \$42 billion in spectrum sales to AT\&T and SpaceX~\cite{echostar_att_spectrum, echostar_spacex_spectrum}, effectively ending the standalone network build. While Boost Mobile continues operating its cloud-native 5G core (reportedly still on AWS) as a hybrid MVNO connected to AT\&T's physical RAN, the DISH saga offers a cautionary narrative about the gap between technical innovation and financial sustainability in greenfield cloud-native network deployments.

\textbf{Swisscom and Deutsche Telekom.} Beyond the flagship deployments, AWS has cultivated partnerships with European operators targeting specific use cases. Swisscom, Ericsson, and AWS announced a proof-of-concept trial at MWC 2023 to deploy Ericsson's 5G Core on AWS in a hybrid cloud configuration, exploring scenarios such as cloud offloading during maintenance windows or traffic peaks~\cite{swisscom_ericsson_aws}. Deutsche Telekom partnered with AWS under the Integrated Private Wireless program to offer 5G campus networks for enterprises~\cite{dt_aws_campus}, leveraging DT's portfolio of over 30 local mobile networks in Germany and Austria.

\textbf{Citymesh.} In February 2026, Belgium's Citymesh launched the world's first commercial mobile service on 5G Core Software-as-a-Service (SaaS), powered by Nokia and AWS~\cite{citymesh_5g_saas}. This subscription-based model transforms the 5G core into a flexible, on-demand service targeting enterprise verticals such as aviation, healthcare, and large-scale events. The deployment represents a new paradigm where smaller operators can access carrier-grade 5G core capabilities without the capital expenditure traditionally associated with core network ownership.

\subsubsection{Emerging Directions: Agentic AI and Autonomous Networks}
\label{subsubsec:aws_ai}

AWS's telecom strategy in 2025--2026 has evolved beyond infrastructure provisioning toward \textit{agentic AI for autonomous network operations}. Amazon Bedrock AgentCore, which reached general availability in October 2025~\cite{bedrock_agentcore_ga}, has become a cornerstone of this proposition. Two vendor integrations exemplify this direction. First, Ericsson's \textit{Agentic rApp as a Service}~\cite{ericsson_agentic_rapp_aws} is built on AWS with Amazon SageMaker AI and Bedrock AgentCore for autonomous RAN optimization, targeting TM Forum Autonomous Network Level~4 by 2030. Second, Nokia's \textit{intent-based network slicing}~\cite{nokia_bedrock_slicing} leverages Amazon Bedrock to deploy specialized AI agents for real-time 5G network slice management, communicating via the Agent-to-Agent (A2A) protocol. At re:Invent 2025, AWS's telecom leadership discussed the concept of ``network language models''---small language models custom-trained on individual operator networks and compact enough for edge deployment~\cite{aws_reinvent2025_telco}. These developments, combined with DOCOMO's production deployment of agentic AI for network operations~\cite{docomo_nec_5gcore_2026}, suggest that AI-driven automation is becoming an integral component of cloud-based 5G deployments on AWS.


\subsection{Microsoft Azure}
\label{subsec:azure}

Microsoft Azure pursued the most aggressive entry into the telecom cloud market among the three hyperscalers, acquiring two cloud-native network function vendors in 2020 and securing AT\&T as its anchor tenant in 2021. However, the Azure 5G story is also one of strategic recalibration, as Microsoft restructured its telecom-focused team in 2024 and retired key services by 2025 to redirect resources toward AI. In this subsection, we examine Azure's architecture for 5G, its operator partnerships, and the implications of its strategic pivot.

\subsubsection{Azure Infrastructure for 5G Deployments}
\label{subsubsec:azure_infra}

\begin{figure}[t]
    \centering
    \begin{subfigure}[t]{0.49\columnwidth}
        \centering
        \includegraphics[width=\textwidth,trim={0cm 9.5cm 23cm 0cm},clip]{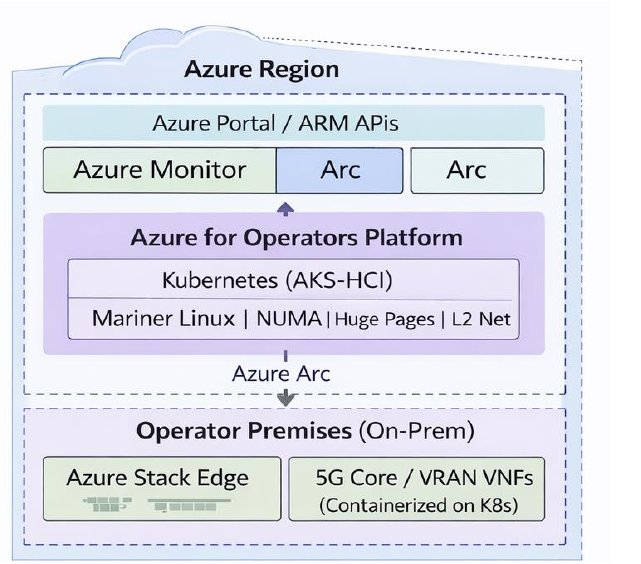}
        \caption[]%
        {{\small }}    
        \label{fig:azure_arch}
    \end{subfigure}
    \hfill
    \begin{subfigure}[t]{0.49\columnwidth}  
        \centering 
        \includegraphics[width=\textwidth,trim={0cm 9.5cm 23cm 0cm},clip]{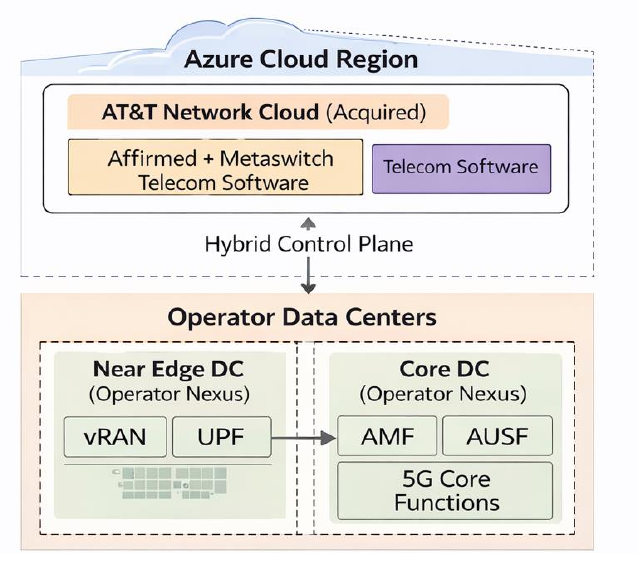}
        \caption[]%
        {{\small }}    
        \label{fig:azureop}
    \end{subfigure}
    \vspace{0.1in}
    \caption{Azure hybrid cloud architecture for 5G deployments. (a) Azure telecom platform architecture showing the separation between the Azure cloud control plane and operator-hosted network infrastructure. (b) Example hybrid deployment in which Azure manages lifecycle and orchestration while 5G network functions execute within operator data centers using Azure Operator Nexus.}
    \label{fig:azureow}
\end{figure}

Unlike AWS, which offers a three-tiered zone hierarchy (AZs, LZs, and WZs) for distributed 5G VNF placement within its public cloud, Azure's telecom strategy centers on a \textit{hybrid cloud} model where the management plane resides in Azure cloud regions while the control and user planes are deployed on operator premises. This architectural choice, illustrated in Figure~\ref{fig:azureow}, reflects Microsoft's bet that carriers would prefer to retain physical control over their network infrastructure while benefiting from cloud-based management and automation.

The cornerstone of this approach is \textit{Azure Operator Nexus}~\cite{azure_operator_nexus}, a managed hybrid cloud platform announced at MWC 2023 and made generally available later that year. Operator Nexus delivers a carrier-grade private cloud on the operator's premises, capable of hosting both containerized and virtualized network functions. The platform is built on Microsoft's Mariner Linux, Hybrid Azure Kubernetes Service (AKS-HCI), and Azure Arc, while providing NFVI-specific features such as CPU pinning, NUMA alignment, huge page support, and Layer~2 networking capabilities---features not available in standard Azure IaaS services~\cite{azure_operator_nexus}. An Analysys Mason study, conducted in collaboration with Microsoft, found that deploying a cloud-native 5G SA network on Operator Nexus could reduce Total Cost of Ownership (TCO) by up to 38\% compared to a do-it-yourself private cloud model over five years, primarily through reduced operational expenditures~\cite{azure_nexus_telecomlead}.

Complementing Operator Nexus, \textit{Azure Stack Edge}~\cite{azure_private5g_overview} serves as the on-premises compute platform for lighter-weight 5G deployments. Azure Stack Edge devices, connected to Azure via Azure Arc, run Kubernetes clusters on which 5G core packet core instances are deployed. Each device supports a complete set of 5G network functions---including the subscriber database, policy control, control plane, and user plane---orchestrated centrally through the Azure portal and Azure Resource Manager (ARM) APIs~\cite{azure_private5g_overview}.

\subsubsection{Evolution of Azure 5G Services}
\label{subsubsec:azure_services}

Microsoft's telecom ambitions were catalyzed by two 2020 acquisitions: \textit{Affirmed Networks}, a provider of cloud-native virtualized Evolved Packet Core (vEPC) software, and \textit{Metaswitch Networks}, which brought unified communications and cloud-native IMS capabilities~\cite{azure_telecom_blog}. These acquisitions, combined with the subsequent acquisition of AT\&T's Network Cloud technology in June 2021~\cite{att_azure_2021}, provided Microsoft with carrier-grade 5G expertise and production-proven software. The resulting portfolio was marketed under the \textit{Azure for Operators} umbrella and comprised several key services:

\textit{Azure Private 5G Core} (AP5GC) was a fully managed service for deploying 5G core VNFs on Azure Stack Edge devices for enterprise private networks. Launched in general availability at MWC 2023~\cite{azure_mwc2023}, AP5GC supported both 4G and 5G SA RANs, integrated with Azure Monitor for real-time observability, and offered High Availability (HA) deployments across paired Azure Stack Edge devices. Partners included AT\&T, BT, Deutsche Telekom, Etisalat, STC, and Telef\'{o}nica for private MEC deployments~\cite{azure_mwc2023}. However, AP5GC was retired on September~30, 2025~\cite{azure_private5g_overview}, with customers directed to migrate to partner solutions available in the Azure Marketplace, such as Nokia 4G/5G Private Wireless and Ericsson Private 5G.

\textit{Azure Operator 5G Core} (AO5GC) was a more ambitious packet-core-as-a-service offering aimed at public mobile network operators, entering private preview in 2023. AO5GC was intended to extend the Affirmed Networks vEPC capabilities into a cloud-managed 5G core for carrier-scale deployments. However, the preview was halted in June 2024 as part of a broader restructuring of the Azure for Operators team~\cite{azure_layoffs_lightreading, azure_layoffs_mobileeurope}.

\textit{Azure Programmable Connectivity} (APC) was previewed at MWC 2023 as a unified API layer enabling developers to interact with operator network capabilities (e.g., QoS on demand, number verification) across multiple carriers, aligned with the GSMA Open Gateway initiative~\cite{azure_mwc2023}. Collaborating partners included AT\&T, Rogers, T-Mobile, Deutsche Telekom, Telef\'{o}nica, and Singtel.

\subsubsection{Major Operator Partnerships}
\label{subsubsec:azure_partnerships}

\textbf{AT\&T.} The AT\&T--Microsoft partnership represents the most consequential telecom-hyperscaler relationship for Azure. In June 2021, AT\&T sold its internally developed Network Cloud technology to Microsoft and committed to migrating its entire 5G mobile network to Azure~\cite{att_azure_2021}. Microsoft acquired the carrier-grade cloud platform---which had been running AT\&T's 5G core at scale since 2018---along with approximately 100 engineering staff, including key architects of AT\&T's containerized core~\cite{att_azure_sdxcentral}. AT\&T subsequently chose Azure Operator Nexus as the platform for its 5G Near Edge network functions~\cite{azure_operator_nexus}. This transaction signaled a broader industry realization: even the largest operators, with a decade of investment in internal cloud development, concluded that hyperscaler platforms offered superior economics and scalability. AT\&T's 5G core migration to Azure planned to start with the 5G SA core and expand to encompass all mobile network traffic managed using Azure technologies.

\textbf{Etisalat (e\&).} At MWC 2024, UAE-based Etisalat emerged as Microsoft's second Tier-1 carrier customer for Azure Operator Nexus~\cite{azure_nexus_fierce}. The deployment enables Etisalat to run 5G and other network workloads on the hybrid cloud platform, leveraging Azure's security, lifecycle management, and AIOps capabilities from the cloud while maintaining physical infrastructure on premises.

\textbf{Deutsche Telekom.} In September 2023, Deutsche Telekom launched its ``Campus Network Smart'' service running on Azure Private MEC with Azure Private 5G Core deployed on Azure Stack Edge~\cite{dt_azure_sdxcentral}. This service targeted enterprise campus environments using a pay-as-you-grow model. DT is also among the partners collaborating on Azure Programmable Connectivity and was identified as a user of the Azure Private MEC ecosystem alongside BT, STC, Tampnet, and Telef\'{o}nica~\cite{azure_mwc2023}.

\textbf{Nokia and stc.} In March 2024, Nokia and Saudi Telecom Company (stc) successfully performed an O-RAN-based 5G Private Wireless Network trial on Azure Operator Nexus~\cite{nokia_stc_nexus}, demonstrating the platform's ability to host disaggregated RAN and core functions in a carrier-grade hybrid cloud environment.

\subsubsection{The 2024 Restructuring and Strategic Pivot}
\label{subsubsec:azure_pivot}

In June 2024, Microsoft undertook a significant restructuring of its telecom-focused operations, cutting up to 1,500 positions across the Azure for Operators and Mission Engineering teams~\cite{azure_layoffs_lightreading, azure_layoffs_mobileeurope}. A leaked internal memo from Executive Vice President Jason Zander attributed the reorganization to Microsoft's intensified investment in artificial intelligence~\cite{azure_layoffs_itpro}. The restructuring halted the Azure Operator 5G Core and Azure Operator Call Protection previews, while remaining Azure Operator Nexus staff were reassigned to the Cloud + AI organization's Azure Edge and Platform product line~\cite{azure_layoffs_itpro}. The subsequent retirement of Azure Private 5G Core in September 2025~\cite{azure_private5g_overview} completed the transition away from first-party 5G network function offerings.

This pivot carries significant implications for the cloud-based cellular deployment landscape. Rather than developing and operating its own 5G core VNFs, Microsoft has shifted toward a \textit{marketplace and AI-enablement model}. Operators seeking 5G solutions on Azure are now directed to partner offerings from Nokia, Ericsson, and others available in the Azure Marketplace~\cite{azure_private5g_overview}. Meanwhile, Microsoft's telecom-adjacent investments have concentrated on AI-driven network operations: at MWC 2024, the company previewed AIOps Copilot for Azure Operator Insights, enabling network engineers to interact with operational data using natural language~\cite{azure_nexus_fierce}. Azure Operator Nexus remains active as the carrier-grade hybrid cloud platform, and AT\&T continues to use it for near-edge 5G workloads, but the breadth of Microsoft's telecom ambition has narrowed considerably.

The Azure experience offers an important counterpoint to the AWS trajectory described in Section~\ref{subsec:aws}. While AWS has steadily deepened its telecom engagement---adding new Wavelength Zones, second-generation Outposts for telco, and the Telco Network Builder---Microsoft attempted to build a comprehensive telecom stack (from packet core to RAN to operator BSS/OSS) and ultimately found the economics unsustainable relative to its AI opportunity cost. The lesson for TSPs evaluating cloud migration is that hyperscaler commitment to telecom-specific services can be volatile, and vendor lock-in to any single provider's proprietary 5G offerings carries non-trivial continuity risk, as evidenced by the forced migration away from both Azure Private 5G Core and Azure Operator 5G Core. As noted in~\cite{5g_map_mobicom2025}, vendor-agnostic measurement platforms such as 5G-MAP become essential in this context, enabling operators to evaluate and compare deployment options across multiple cloud providers without committing to a single ecosystem.


\subsection{Google Cloud Platform}
\label{subsec:gcp}

Google Cloud Platform (GCP) has pursued a distinct strategy for cloud-based 5G deployments compared to AWS and Azure. Rather than acquiring telecom-specific companies or building first-party 5G core VNFs, Google has positioned itself as an open, Kubernetes-native infrastructure provider for telecom, leveraging its heritage as the originator of Kubernetes and its strengths in data analytics and AI. GCP's approach emphasizes an ecosystem-driven model where network equipment providers such as Ericsson and Nokia deploy their 5G solutions on Google's infrastructure, while Google provides the underlying platform, AI capabilities, and edge hardware. In this subsection, we examine the GCP infrastructure for telecom, its flagship partnerships, and its emerging AI-centric telecom strategy.

\subsubsection{GCP Infrastructure for 5G Deployments}
\label{subsubsec:gcp_infra}

\begin{figure}
    \centering

    \begin{subfigure}[t]{\columnwidth}
        \centering
        \includegraphics[width=\columnwidth,trim={0cm 14.5cm 21cm 0cm},clip]{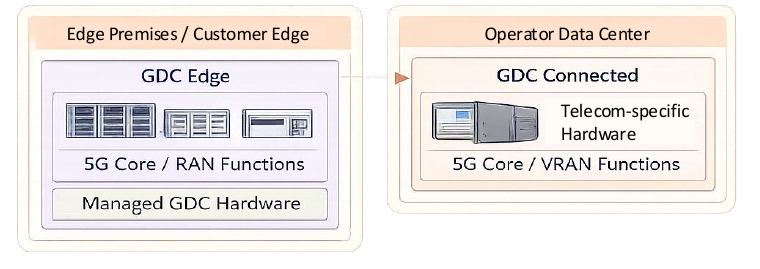}
        \caption{GDC Edge enables managed edge deployments at operator or enterprise premises for hosting 5G core and RAN functions alongside edge applications.}
        \label{fig:gcp_arch_a}
    \end{subfigure}

    \vspace{1.5em}

    \begin{subfigure}[t]{\columnwidth}
        \centering
        \includegraphics[width=\columnwidth,trim={0cm 14.5cm 21.5cm 0cm},clip]{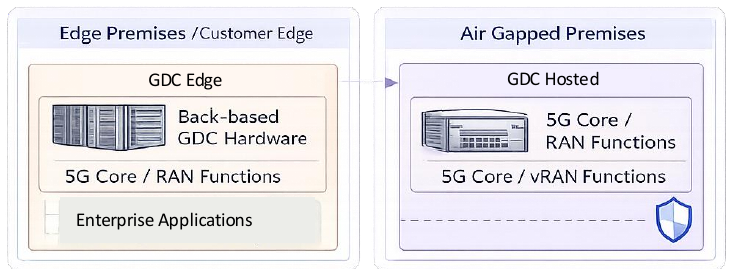}
        \caption{GDC Connected provides telecom-optimized infrastructure within operator data centers for running high-throughput 5G core and vRAN workloads.}
        \label{fig:gcp_arch_b}
    \end{subfigure}

    \vspace{1.5em}

    \begin{subfigure}[t]{\columnwidth}
        \centering
        \includegraphics[width=\columnwidth,trim={0cm 15.5cm 23cm 0cm},clip]{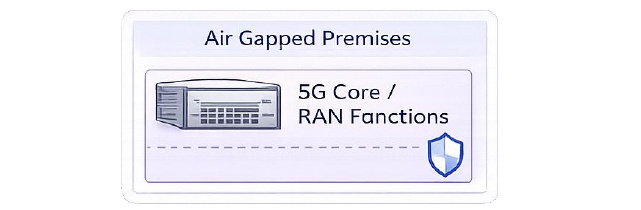}
        \caption{GDC Hosted supports sovereign and air-gapped environments where cloud services must operate entirely on-premises.}
        \label{fig:gcp_arch_c}
    \end{subfigure}
    
    \vspace{1.5em}

    \caption{Google Distributed Cloud (GDC) architecture for telecom deployments. Google's telecom platform is built on a Kubernetes-native foundation using Anthos and Google Kubernetes Engine (GKE) across distributed infrastructure.}
    \label{fig:gcp_arch}
    
\end{figure}

GCP's telecom infrastructure is anchored by the \textit{Google Distributed Cloud} (GDC) product family~\cite{gcp_gdc_edge_ga, gcp_gdc_connected}, which extends Google Cloud services from centralized regions to operator and customer premises. As illustrated in Figure~\ref{fig:gcp_arch}, GDC comprises three deployment modes tailored to different telecom requirements:

\textit{GDC Edge} is a fully managed product that brings Google Cloud infrastructure to operator and customer edge locations~\cite{gcp_gdc_edge_ga}. Made generally available in March 2022, GDC Edge enables operators to run 5G core and RAN functions at the edge alongside enterprise applications such as anomaly detection via computer vision, IoT sensor processing, and local data scrubbing before cloud transfer. GDC Edge hardware is available in two form factors: rack-based configurations (comprising six servers, two top-of-rack switches, cabling, and optics) and compact Edge Appliances. Rack configurations are available across the United States, Canada, and seven European countries~\cite{gcp_gdc_edge_ga}.

\textit{GDC Connected} targets telecom operator data centers specifically, offering customized hardware with an optimized data plane for high-throughput packet processing~\cite{gcp_gdc_connected}. This variant supports telecom use cases including 5G core CNFs and virtual RAN (vRAN) functions, while maintaining an open architecture that avoids locking operators into proprietary technology. GDC Connected transforms the traditional siloed telecom infrastructure model into a service-based architecture that can run core network functions in the cloud~\cite{gcp_gdc_connected}.

\textit{GDC Hosted} addresses sovereign and air-gapped deployment requirements for public-sector and regulated-industry customers. Announced alongside GDC Edge in October 2021~\cite{gcp_gdc_announcement}, GDC Hosted supports customers with strict data residency, security, and privacy requirements who need to modernize on-premises deployments.

All three GDC variants are built on Anthos (now Google Kubernetes Engine Enterprise), Google's open-source-based platform that unifies the management of infrastructure and applications across on-premises, edge, and multiple public clouds~\cite{gcp_gdc_announcement}. This Kubernetes-native foundation differentiates GCP from both AWS (which relies on proprietary zone-based infrastructure) and Azure (which developed its own carrier-grade platform through Operator Nexus). Google's CEO Sundar Pichai has drawn an explicit analogy between Anthos for Telecom and Android: just as Android provided an open platform for mobile applications, Anthos provides an open, Kubernetes-based platform for network-centric applications~\cite{gcp_telecom_strategy}.

\subsubsection{Major Operator Partnerships}
\label{subsubsec:gcp_partnerships}

\textbf{T-Mobile.} The T-Mobile--GCP partnership, announced in 2023, focuses on enhancing 5G capabilities through \textit{5G Advanced Network Solutions} (ANS) and GDC Edge~\cite{TMobilea76online}. This collaboration aims to provide enterprises and government organizations with tools for digital transformation across retail, manufacturing, logistics, and smart cities. By integrating T-Mobile's 5G networks with Google's edge computing and AI technologies, businesses can leverage low latency and high-speed connectivity for applications such as augmented reality (AR) and computer vision. A notable demonstration was the ``magic mirror'' proof of concept, which uses cloud-based processing to create interactive retail experiences~\cite{TMobilea76online}. The partnership represents a hybrid model where T-Mobile provides the 5G radio network and Google provides the edge compute and AI platform.

\textbf{Deutsche Telekom.} In February 2023, Deutsche Telekom, Google Cloud, and Ericsson demonstrated a cloud-native 5G network pilot that represented a significant transformation milestone~\cite{dt_gcp_ericsson_pilot}. The pilot ran Ericsson's 5G core applications on GCP infrastructure, testing the feasibility of hosting carrier-grade network functions in Google's public cloud. Deutsche Telekom has subsequently continued to explore Google's data analytics and AI capabilities for network optimization, and has tested GDC Edge for supporting infrastructure and services closer to end users.

\textbf{Bell Canada and Verizon.} Bell Canada is deploying GDC Edge for its 5G core network functions, making it one of the first North American carriers to run production 5G VNFs on Google's edge infrastructure~\cite{gcp_gdc_edge_ga}. Verizon is similarly leveraging GDC Edge to deliver edge services to enterprises. Additional global partnerships include AT\&T, Reliance Jio, TELUS, and Indosat Ooredoo~\cite{gcp_gdc_edge_ga}.

\textbf{Mavenir.} In September 2022, Mavenir announced a partnership to deliver cloud-based 5G solutions on Google Cloud~\cite{mavenir_gcp}, bringing its Open RAN and cloud-native 5G core capabilities to GCP's infrastructure. This partnership extends Google's 5G ecosystem beyond the traditional equipment vendors to include Open RAN-focused players.

\subsubsection{Ericsson On-Demand: 5G Core as a Service on GCP}
\label{subsubsec:gcp_ericsson}

The most significant recent development in the GCP telecom ecosystem is \textit{Ericsson On-Demand}, launched in June 2025~\cite{ericsson_ondemand_pr}. This offering represents the first true 5G core SaaS product built natively on a public cloud, where both the control plane and user plane run on Google Cloud infrastructure using Google Kubernetes Engine (GKE)~\cite{ericsson_ondemand_sdxcentral}. Managed end-to-end by Ericsson's 24/7 Site Reliability Engineering (SRE) teams with AI-assisted troubleshooting and lifecycle automation, the platform can deploy a full 5G core in minutes, scale elastically based on demand, and operate on a consumption-based billing model with no upfront capital expenditure~\cite{ericsson_ondemand_pr}.

Ericsson On-Demand leverages Google Cloud's full-stack AI infrastructure---spanning 42 cloud regions and over two million miles of terrestrial and subsea fiber---to enable global deployments with built-in compliance and sovereignty options~\cite{ericsson_ondemand_fierce}. The platform is initially targeted at smaller and medium-sized operators who may lack the resources for traditional on-premises 5G core deployments. Use cases include Wide Area Enterprise networks, Fixed Wireless Access (FWA) rollouts, and market entry testing~\cite{ericsson_ondemand_pr}. As noted by analysts, with only 73 5G SA networks launched across 40 countries as of May 2025 (compared to 354 non-standalone deployments), simplified deployment pathways like Ericsson On-Demand could accelerate the adoption of 5G SA architectures~\cite{ericsson_ondemand_telecomtv}.

The significance of Ericsson On-Demand for the GCP telecom strategy cannot be overstated. It effectively positions GCP as the public cloud platform hosting a major equipment vendor's production 5G core, complementing AWS's role as the platform for Nokia's 5G Core SaaS (via the Citymesh deployment discussed in Section~\ref{subsec:aws}). This parallel evolution suggests an emerging paradigm where network equipment providers serve as the integration layer between hyperscale clouds and mobile operators, rather than operators building direct relationships with cloud providers for core network hosting.

\subsubsection{AI-Centric Telecom Strategy}
\label{subsubsec:gcp_ai}

GCP's telecom strategy in 2025--2026 has converged on AI as the primary value proposition. At MWC 2025, Google Cloud showcased several AI-driven telecom solutions~\cite{gcp_mwc2025_blog}: Amdocs launched a \textit{Network AIOps} solution built on GCP to automate complex 5G network operations and enhance service reliability through AI-powered insights. Vodafone Italy announced the completion of its ``Nucleus'' AI-ready data platform on Google Cloud, re-engineering its data pipelines in partnership with Amdocs. Google also demonstrated its \textit{Autonomous Network Operations} framework, which integrates with platforms like Ericsson On-Demand to ingest network telemetry into BigQuery, build digital twins, and deploy AI agents using Vertex AI and Gemini models for root cause analysis and predictive maintenance~\cite{ericsson_ondemand_fierce}.

This AI-centric approach positions GCP uniquely among the three hyperscalers. While AWS leads in direct infrastructure partnerships (Wavelength Zones, Outposts for telco) and Azure pursued---and largely retracted---a full-stack operator platform, GCP has found a differentiated position as the AI and data analytics backbone for telecom. By providing Kubernetes-native infrastructure through GDC, partnering with equipment vendors for 5G core SaaS delivery, and layering Gemini-powered AI agents on top, GCP offers an ecosystem-driven model that avoids the vendor lock-in risks that have affected Azure's telecom customers while providing deeper AI integration than AWS's current portfolio.
\section{Key Investigation Areas} \label{sec:key_investigation}
\subsection{Security, Trust, and Compliance}
\label{subsec:security_trust}

The transition of cellular infrastructure from vertically integrated telecom platforms to cloud-native, distributed computing environments introduces a fundamental shift in the security and trust assumptions underlying mobile networks. Traditional mobile network deployments relied on tightly controlled operator infrastructure, hardware appliances, and relatively static trust boundaries. In contrast, cloud-based 5G and emerging 6G deployments operate on highly dynamic, multi-tenant computing substrates spanning public clouds, operator private clouds, enterprise environments, and edge infrastructure. As a result, the security model of the cellular system must extend beyond protocol-level protections defined by telecommunications standards to incorporate the broader attack surface and trust dependencies of modern cloud platforms.

\begin{figure}[t]
    \centering
    \includegraphics[width=\columnwidth,trim={0cm 0cm 2cm 1.8cm},clip]{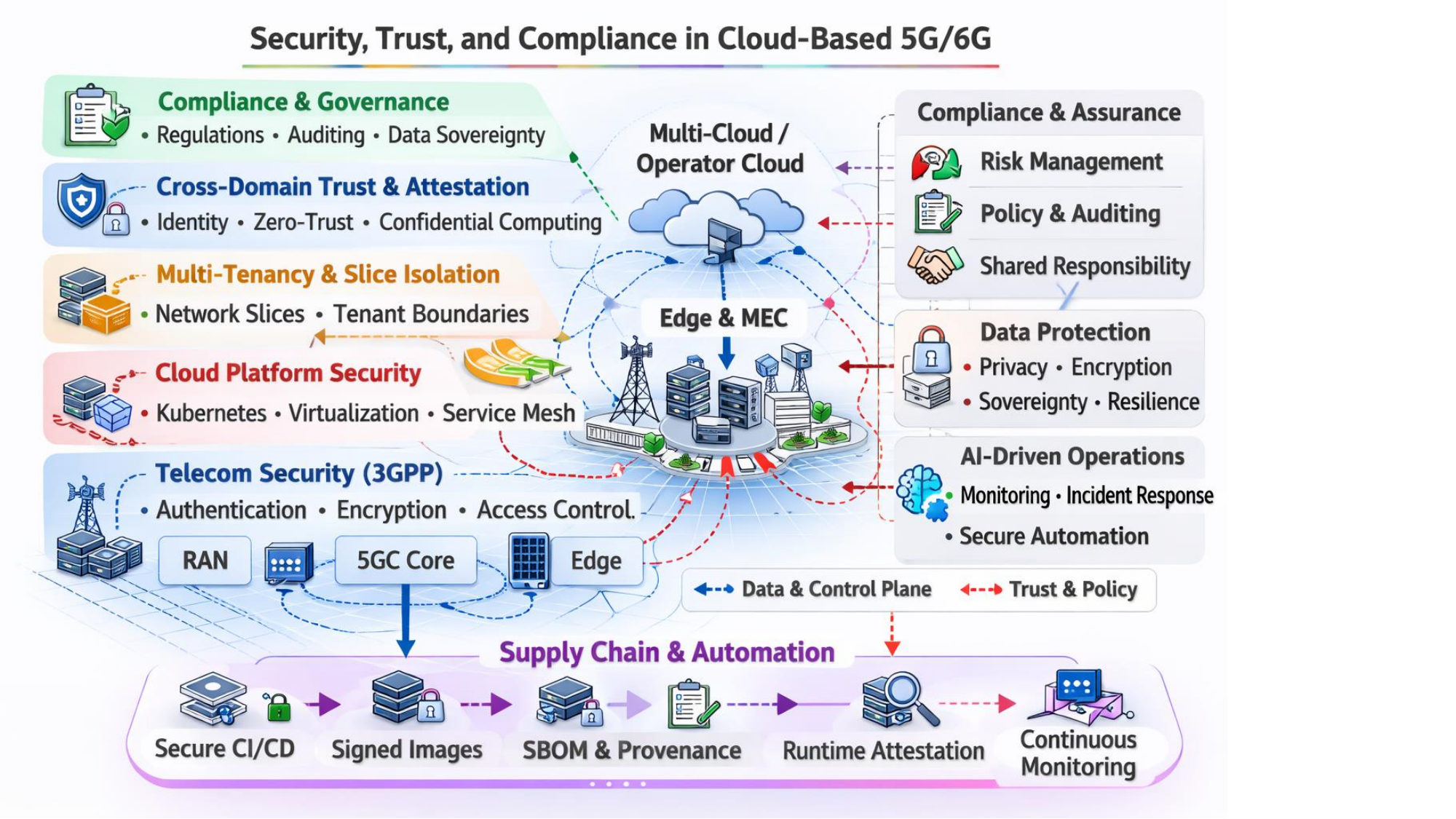}
     \caption{Security overview summarizing the 3GPP security principles, cloud platform security, multi-tenancy and slice isolation, cross-domain trust and attestation, and compliance in a multi-cloud operator network.}
     \label{fig:securityow}
\end{figure}

The security architecture of the 5G system is defined primarily through the 3GPP TS~33-series specifications, which establish the mechanisms for subscriber authentication, key management, network function authorization, and service-based architecture protection. In particular, TS~33.501~\cite{3gpp33501} introduces security procedures for the SBA, including mutual authentication between network functions, transport security based on TLS~\cite{3gpp33210}, and authorization frameworks implemented through the NRF~\cite{3gpp29510}. These mechanisms were designed to secure a logically disaggregated core network architecture in which network functions interact through standardized service interfaces. While these mechanisms provide strong protocol-level protections, their design assumes that the underlying infrastructure is operated within a trusted administrative domain controlled by the mobile network operator. Cloud-native deployments challenge this assumption. When network functions are deployed as virtual machines or containerized workloads on shared cloud infrastructure, the operator must rely on external orchestration platforms, hypervisors, container runtimes, and infrastructure management systems that lie outside the scope of traditional telecom security specifications.

This introduces an important conceptual distinction between \emph{network-level security}, as defined by 3GPP specifications, and \emph{platform-level security}, which governs the execution environment in which network functions operate. Compromise of the cloud orchestration layer, container runtime, or host operating system can undermine the security guarantees provided by telecom protocols, regardless of the strength of the protocol design itself. 3GPP has recognized this expansion through a series of study items: TR~33.848~\cite{3gpp33848} on the security impacts of virtualization identifies 27 Key Issues spanning confidentiality of sensitive data, function isolation, memory introspection, container security, and attestation requirements. Additional studies on SBA security enhancement, edge computing security (TR~33.839~\cite{3gpp33839}, TR~33.749~\cite{3gpp33749}), and network slicing security (TR~33.813~\cite{3gpp33813}, TR~33.874~\cite{3gpp33874}, TR~33.886~\cite{3gpp33886}) further reflect the broadening scope of the 3GPP security agenda. However, a significant gap remains between the \emph{procedural} security specified by these standards and the \emph{operational} security required when those procedures execute on infrastructure the operator does not fully control. As a result, understanding the security posture of cloud-based cellular deployments requires analyzing the interaction between 3GPP-defined mechanisms and the broader cloud computing security stack. 

From the cloud security perspective, NIST SP~800-190~\cite{nist_sp800190} provides foundational guidance on application container security across five tiers: images, registries, orchestrators, containers, and host OS. NISTIR~8320~\cite{nistir8320} and its companion NISTIR~8320A~\cite{nistir8320a} extend this to hardware-enabled security for multi-tenant container environments, directly applicable to telecom CNF deployments where isolation guarantees must exceed typical enterprise requirements. NIST SP~1800-33~\cite{nist_sp180033} bridges these domains by demonstrating a 5G standalone network with security capabilities mapped to both 3GPP and NIST frameworks. Government bodies have also recognized the urgency of this intersection: CISA and NSA published a four-part ``Security Guidance for 5G Cloud Infrastructures''~\cite{cisa_5g_cloud} addressing lateral movement prevention, network resource isolation, data protection, and infrastructure integrity. The GSMA 5G Security Guide (FS.40)~\cite{gsma_fs40} and Baseline Security Controls (FS.31)~\cite{gsma_fs31} provide industry-level security frameworks that span NFVI, container security, VNF lifecycle management, and MEC platform controls. At the European level, ENISA's 5G Security Controls Matrix~\cite{enisa_5g_controls} maps controls to ISO/IEC 27002, ISO/IEC 27005, and NIST SP~800-53, while ENISA's NIS2 implementation guidance~\cite{enisa_nis2} establishes mandatory risk management, incident reporting, and business continuity requirements for telecom operators classified as Essential Entities.

Given these considerations, the security implications of cloud-based cellular deployments cannot be understood solely through the lens of telecom standards or cloud security frameworks independently. Instead, a systematic investigation is required to analyze how the two domains intersect and where new vulnerabilities, trust assumptions, and compliance requirements emerge. In this section, we examine the security, trust, and compliance implications of cloud-native 5G and future 6G deployments across several key investigative dimensions: the expanded attack surface introduced by cloud-native network functions, multi-tenancy and slice isolation challenges, trust establishment across heterogeneous infrastructure domains, cross-domain trust in distributed edge deployments, software supply chain and automation security, and the compliance and assurance requirements associated with operating telecom workloads on shared cloud platforms. Our investigation is summarized in Figure~\ref{fig:securityow}.

\subsubsection{Cloud-Native Attack Surface for Cellular Network Functions}
\label{subsubsec:cnf_attack_surface}

Cloud-native deployments introduce a substantially expanded attack surface compared to traditional telecom infrastructure. Network functions that were previously implemented as tightly integrated hardware appliances are now realized as software workloads deployed through container orchestration platforms such as Kubernetes. While this architectural shift provides significant benefits in scalability and operational flexibility, it also exposes cellular infrastructure to the security risks associated with modern distributed computing environments. 3GPP TR~33.848~\cite{3gpp33848} explicitly acknowledges this expansion, noting that containers do not provide the same level of isolation as VMs (Key Issue~25) and that container breakout vulnerabilities (Key Issue~26) pose risks to colocated network functions. NIST SP~800-190~\cite{nist_sp800190} systematically categorizes these risks across image, registry, orchestrator, container runtime, and host OS layers, while NISTIR~8320A~\cite{nistir8320a} demonstrates a proof-of-concept using Intel SGX and TPM-based attestation to establish hardware roots of trust in container platforms.

In containerized deployments, network functions are packaged as container images stored in registries, deployed through orchestration frameworks, and interconnected through service meshes and virtual networking overlays. Each of these components represents a potential attack vector. Minna et al.~\cite{minna_k8s_security2021} demonstrate that Kubernetes network abstractions enable unexpected attacks when approached with traditional network-security mental models, presenting three practical attacks: FirewallHole, which bypasses overlay firewalls via VXLAN mimicry; Hit\&Spread, which exploits API server access to move laterally from a compromised container; and Replace\&Propagate, which leverages supply-chain vulnerabilities to inject malicious container images. Nam et al.~\cite{bastion_atc2020} identify critical container-network threats including eavesdropping, ARP spoofing, traffic injection, and unauthorized host access, and present BASTION, a container-aware communication sandbox that provides per-container network visibility and traffic isolation. Song et al.~\cite{song_5g_container2024} directly analyze container security threats against 5G core NFs built on SBA, demonstrating how denial-of-service, privilege escalation, and container escape attacks translate to containerized 5G network function deployments. The broader NFV security landscape is surveyed by Madi et al.~\cite{madi_nfv_survey2021}, who derive a three-dimensional threat taxonomy for 5G NFV environments, and by Marku et al.~\cite{marku_outsourced_vnf2020}, who provide a comprehensive overview of cryptographic and trusted-hardware methods for protecting outsourced VNFs.

East-west traffic between CNFs within a cluster presents particular security challenges. In legacy deployments, inter-NF traffic traversed controlled network segments. In Kubernetes, pod-to-pod communication defaults to unencrypted traffic on a flat network. Service meshes such as Istio and Linkerd can enforce mutual TLS and fine-grained authorization policies. Budigiri et al.~\cite{budigiri_k8s_netpol2021} evaluate eBPF-based Kubernetes network policies (Calico, Cilium) for 5G URLLC edge use cases, reporting 0.7--0.8$\times$ latency reduction and 1.2--3.5$\times$ throughput improvement over iptables-based enforcement while analyzing remaining security vulnerabilities. Kulkarni and Fahmy~\cite{kulkarni_ztx2024} propose ZTX-SEM, a zero-trust security module for cloud-native 5G control planes on Kubernetes with protocol-agnostic packet interception, achieving a 75\% reduction in resource utilization and 28\% decrease in session setup times compared to Istio while encrypting all inter-NF east-west traffic. The 5G-STREAM framework~\cite{5g_stream_dsn2025} demonstrates how a purpose-built two-tier service mesh can provide authorization and reliability for 5G core microservices in distributed cloud environments, reducing control plane latency by up to 2$\times$ per HTTP transaction at minor resource cost. Similarly, the 5G-WAVE framework~\cite{5g_wave_infocom2024} proposes decentralized inter-VNF authorization using WAVE attestations rather than OAuth~2.0 tokens, addressing key vulnerabilities of the centralized NRF-as-authorization-server model including authorization code injection, access token leakage, credential phishing, and authorization flooding.

The O-RAN architecture further extends the cloud-native attack surface to the RAN domain. Polese et al.~\cite{polese_oran_survey2023} provide the most comprehensive O-RAN survey to date, with a dedicated security section covering open interface vulnerabilities, AI/ML model threats in the RIC, and cloud-platform attack surface. Groen et al.~\cite{groen_oran_impl2025} present the first holistic O-RAN security analysis with experimental evidence, covering the security of open interfaces, AI/ML intelligence, and platforms with cost-benefit tradeoffs. Their companion work~\cite{groen_oran_tmc2024} provides the first investigation of encryption impact on O-RAN E2 and Open Fronthaul interfaces. The XRF framework~\cite{xrf_infocom2023} addresses the fundamental system security requirements of authentication, authorization, and discovery for xApps in the O-RAN near-RT RIC, demonstrating scalable OAuth~2.0-based token distribution in a Kubernetes-deployed microservice environment.

\subsubsection{Multi-Tenancy and Slice Isolation}
\label{subsubsec:slice_isolation_security}

Network slicing enables the logical partitioning of cellular infrastructure to support multiple services with distinct performance and security requirements. However, when slices are deployed on shared cloud infrastructure, the notion of isolation must be reconsidered across multiple layers of the computing stack. De Alwis et al.~\cite{dealwis_slicing_survey2024} provide the most comprehensive survey on network slicing security, covering lifecycle security, inter-slice and intra-slice threats, slice broker security, ZTM security, and blockchain-based approaches. 3GPP has conducted multiple studies on this topic: TR~33.813~\cite{3gpp33813} addresses network slicing security enhancement including slice-specific authorization, TR~33.874~\cite{3gpp33874} examines enhanced security for network slicing phase~2 with AF authorization, and TR~33.886~\cite{3gpp33886} continues with phase~3 on temporary slice authorization. Salahdine et al.~\cite{salahdine_intelligent_slicing2022} propose AI/ML-based approaches to securing network slicing, addressing both inter-slice and intra-slice challenges with emphasis on anomaly detection in multi-tenant cloud-native deployments.

In traditional telecom environments, slice isolation is typically enforced through logical separation mechanisms within the core network and radio access network. Cloud-native deployments introduce additional layers of resource sharing, including shared compute nodes, container orchestration clusters, storage systems, and virtual networking overlays. Gonzalez et al.~\cite{gonzalez_isolation2020} systematically define isolation levels ranging from logical to bare-metal/air-gap across RAN, transport, and core domains, highlighting that strong isolation in shared cloud infrastructure remains an open challenge with a significant gap between standardized concepts and actual compute-level enforcement. Consequently, the security of network slices becomes dependent not only on telecom control-plane mechanisms but also on the isolation guarantees provided by the underlying cloud infrastructure.

Potential isolation failures may arise through side-channel attacks, which represent a particularly acute threat in shared cloud environments. Liu et al.~\cite{liu_sca_slice2020} directly address how logical (vs. physical) isolation makes co-located slices vulnerable to side-channel attacks, proposing an SCA-aware resource allocation algorithm that balances utilization against security for URLLC and eMBB slices. More recently, Shao et al.~\cite{shao_rl_sca2024} demonstrate an RL-based cache timing attack on co-located 5G slices, achieving 95--98\% accuracy in extracting authentication keys and registration data from a victim slice through shared memory and cache resources. These findings suggest that namespace-based isolation in shared Kubernetes clusters is insufficient for security-critical slices.

Stronger isolation mechanisms have been explored through microVM-based approaches. Firecracker~\cite{firecracker_nsdi2020} provides hardware-enforced isolation with container-like performance using lightweight microVMs, deployed at scale in AWS Lambda and Fargate, and offers an archetype for per-tenant isolation applicable to 5G workloads. Anjali et al.~\cite{anjali_containers_vee2020} provide the first fine-grained comparison of LXC, gVisor, and Firecracker, measuring host kernel code coverage and performance tradeoffs. However, even these stronger isolation mechanisms face challenges. Xiao et al.~\cite{xiao_microvm_sec2023} demonstrate that container operations forwarded to the host kernel can break Kata Container and Firecracker isolation, causing privilege escalation, 93.4\% I/O degradation, and 60\% packet loss. Weissman et al.~\cite{weissman_firecracker_sidechannel2023} further show that Firecracker's virtualization has little effect on microarchitectural side-channel attacks such as Spectre and MDS/Zombieload, concluding that the same host-level mitigations remain required for multi-tenant protection. These findings indicate that even hardware-assisted isolation mechanisms do not fully resolve the multi-tenancy security problem for cloud-based cellular deployments.

Ensuring strong slice isolation in cloud-based deployments therefore requires coordinated security mechanisms spanning the telecom architecture, virtualization layer, and orchestration platform. The 5G-WAVE framework~\cite{5g_wave_infocom2024} addresses one dimension of this problem by proposing decentralized inter-VNF authorization that removes reliance on the NRF as a central OAuth~2.0 server, thereby reducing the blast radius of a single-point compromise and addressing the inter-slice fraud scenarios identified in the slicing security literature.

\subsubsection{Trust Establishment, Attestation, and Confidential Computing}
\label{subsubsec:trust_attestation}

Traditional mobile network trust is anchored in well-established cryptographic primitives: the USIM/eSIM credential, the 5G-AKA protocol, and the operator's control over its domain. Cloud-native deployment introduces a second, parallel trust stack: hardware roots of trust, measured/secure boot, platform attestation, Trusted Execution Environments (TEEs), workload identity, and verifiable infrastructure state. 3GPP has recognized the importance of this second stack through its treatment of Hardware-Mediated Execution Enclaves (HMEEs) in TR~33.848~\cite{3gpp33848}, which identifies HMEE as a solution for multiple Key Issues including confidentiality of sensitive data (KI~2), data location and lifecycle (KI~5), function isolation (KI~6), memory introspection protection (KI~7, 15), key storage trustworthiness (KI~11), NF migration security (KI~12), and attestation at the NF level (KI~13).

A body of systems research has established the paradigm of shielding network functions inside hardware enclaves. SafeBricks~\cite{safebricks_nsdi2018} combines Intel SGX enclaves with Rust-based enforcement to shield generic NFs from untrusted cloud operators, achieving 0--15\% overhead with a 20$\times$ TCB reduction. LightBox~\cite{lightbox_ccs2019} provides the first SGX-enabled system achieving near-native middlebox speed with stateful processing, tracking 1.5 million concurrent flows at 10~Gbps within the enclave. ShieldBox~\cite{shieldbox_sosr2018} deploys Click-based NFs inside SGX enclaves via the SCONE framework with Docker-based deployment and remote attestation. TrustedClick~\cite{trustedclick_sdnnfvsec2017} established the foundational SGX+Click paradigm for secure NFs, while EndBox~\cite{endbox_dsn2018} executes middlebox functions at the network edge inside SGX enclaves for scalable deployment. SafeLib~\cite{safelib_tnsm2025} provides the most comprehensive open-source SGX-based framework to date, supporting stateful and stateless VNFs with kernel-bypass (DPDK) networking integrated with libVNF for streamlined development. Wang et al.~\cite{wang_sgx_vnf2018} analyze the remaining challenges for VNF protection with SGX, including secure data processing, state protection, and single-enclave scalability limitations.

In the specific context of 5G control plane functions, the P-AKA study~\cite{paka_dsn2024} provides the first empirical characterization of deploying 5G-AKA functions inside Intel SGX enclaves. The results show 1.2--1.5$\times$ overhead in function execution time and 2.2--2.9$\times$ in response latency, but this overhead constitutes only 5.58\% of end-to-end UE session setup delay, suggesting that HMEE is feasible for security-critical control plane functions. Beyond SGX, the TEE landscape includes competing alternatives. Al Atiiq and Risdianto~\cite{atiiq_sev_nfv2024} evaluate AMD SEV-SNP for NFV deployment, finding approximately 20\% average performance penalty but with a significant practical advantage: unlike SGX, SEV requires no NF codebase modification, offering a more practical deployment path for existing VNF implementations. Vomvas et al.~\cite{vomvas_zte_b5g2024} propose Zero Trust Execution (ZTE), a vertical extension of zero trust to model untrusted execution environments, demonstrating TEE integration with Open5GS and UERANSIM with minimal performance overhead and no changes to the 5G standard. Valero et al.~\cite{valero_tee_5g2021} propose TEE-as-a-Service (TEEaaS) using ARM TrustZone for 5G edge nodes with remote attestation and location-aware VM deployment. Baldoni et al.~\cite{trustedvim_eucnc2019} present TrustedVIM, an architecture using ARM TrustZone with VOSYSMonitor for mixed-critical edge infrastructure, with only approximately 4\% impact on VM boot time.

The zero-trust paradigm provides the architectural framework for this trust model. NIST SP~800-207~\cite{nist_sp800207} establishes the foundational principles, tenets, and deployment models. Ramezanpour and Jagannath~\cite{ramezanpour_izta2022} present the first architectural concept for intelligent ZTA (i-ZTA) in 5G/6G, introducing real-time Monitoring, Evaluating risk, and Deciding (MED) components within an O-RAN-aligned SBA design. Kholidy et al.~\cite{kholidy_zt_milcom2022} propose a dynamic data-driven Zero Trust Security Framework using Vulnerability, Exploitability, and Attackability (VEA-bility) metrics to quantify end-to-end trust across 5G domains and tenants. Bello et al.~\cite{bello_zt_5g2022} address sustained zero-trust principles for the 5G mobile core, considering the dynamic SBA and cloud-native deployment model. In the O-RAN context, the XRF framework~\cite{xrf_infocom2023} provides concrete zero-trust implementation by establishing authentication, authorization, and discovery mechanisms for xApps that remove reliance on the near-RT RIC as a trusted entity, while the 5G-WAVE~\cite{5g_wave_infocom2024} and 5G-STREAM~\cite{5g_stream_dsn2025} frameworks represent concrete steps toward zero-trust inter-VNF communication in the core network.

\subsubsection{Cross-Domain Trust in Distributed Edge Deployments}
\label{subsubsec:edge_trust}

Cloud-native cellular deployments frequently span multiple administrative domains, including public cloud providers, operator private clouds, enterprise premises infrastructure, and edge computing platforms. In such environments, establishing trust across heterogeneous infrastructure domains becomes a central security challenge. Edge computing platforms such as multi-access edge computing (MEC) enable latency-sensitive services by placing network functions and application workloads closer to end users. However, these deployments often involve infrastructure operated by different entities, each with distinct security policies and operational procedures.

3GPP has studied security enhancements for edge computing support through TR~33.839~\cite{3gpp33839}, which analyzes security threats for edge application deployment in 5G including authentication and authorization of edge application servers, secure EAS discovery, and protection of EES/ECS interfaces. The normative TS~33.558~\cite{3gpp33558} captures the resulting security requirements for enabling edge applications. The continued evolution is reflected in TR~33.749~\cite{3gpp33749}, which addresses enhanced federation scenarios and advanced trust models for multi-stakeholder edge deployments in Release~19. ETSI White Paper No.~46~\cite{etsi_mec_wp46} analyzes security use cases where edge computing makes typical cloud security approaches insufficient, adding MEC federation threat analysis and GSMA OPG alignment in its second edition.

The MEC security landscape has been comprehensively surveyed. Ranaweera et al.~\cite{ranaweera_mec_survey2021} investigate threat vectors in the ETSI-standardized MEC architecture covering authentication, authorization, access control, data privacy, and virtual platform isolation. Their companion work~\cite{ranaweera_mec_usecases2021} maps security vulnerabilities to specific 5G use cases (critical infrastructure, eMBB, mMTC, autonomous driving, AR/VR, UAVs) deployed in MEC contexts. Nencioni et al.~\cite{nencioni_mec_survey2023} present a MEC taxonomy across security, dependability, and performance dimensions, including discussion of quantum-resistant cryptography for MEC. Nowak et al.~\cite{nowak_mec_verticals2021} study MEC security across twelve representative vertical industries, identifying the most sensitive use cases for prioritizing protection mechanisms.

The INSPIRE-5Gplus project~\cite{inspire5gplus_ares2020} proposes an integrated architecture combining AI-driven security management, DLT-based trust, and TEEs for closed-loop end-to-end zero-touch security management across multi-stakeholder 5G environments. Benzaid et al.~\cite{benzaid_trust_5g2021} discuss trust management spanning the entire 5G ecosystem including software and hardware supply chain, NFV/cloud infrastructure, and multi-stakeholder environments. The broader 5G security landscape is comprehensively surveyed by Khan et al.~\cite{khan_5g_security2020}, who cover SDN, NFV, slicing, MEC, physical-layer security, and the 3GPP SA3 security framework including edge and local breakout scenarios, and by Ahmad et al.~\cite{ahmad_security_beyond5g2019}, who identify gaps between telecom-specific and IT/cloud security frameworks.

The 5G-MAP framework~\cite{5g_map_mobicom2025} provides empirical evidence of how VNF placement across different AWS cloud zones (Availability Zones, Local Zones, Wavelength Zones) impacts both performance and security. The strategic placement of side-car proxies adjacent to each VNF within the same pod ensures that network domain security requirements as defined by 3GPP TS~33.210~\cite{3gpp33210} remain intact even when SCPs manage end-to-end TLS sessions, demonstrating that security-preserving observability is achievable in distributed cloud deployments. Data sovereignty at the edge is an additional dimension: edge-resident user data may be subject to jurisdictional requirements that constrain where it can be processed and stored. Cloud elasticity features such as workload migration and autoscaling may inadvertently move data across jurisdictional boundaries unless carefully constrained by policy.

\subsubsection{Software Supply Chain and Automation Security}
\label{subsubsec:supply_chain}

The adoption of cloud-native architectures introduces a complex software supply chain that spans container images, orchestration templates, third-party libraries, and automated deployment pipelines. Unlike appliance-based deployments where software was tightly coupled with validated hardware, cloud-native CNFs are composed from heterogeneous, frequently updated components sourced from diverse supply chains. Each element---container images, Helm charts, Kubernetes operators, CI/CD pipelines, base OS artifacts, third-party CNFs, and model artifacts for AI-driven orchestration---represents a potential attack vector. Minna et al.~\cite{minna_k8s_security2021} demonstrate one vector through their Replace\&Propagate attack, which leverages supply-chain vulnerabilities to inject malicious container images into Kubernetes deployments.

The zero-touch management and closed-loop automation paradigm central to cloud-native 5G/6G operations creates additional security dimensions. Benzaid and Taleb~\cite{benzaid_zsm_security2020} introduce the ETSI ZSM framework's attack surface, analyzing threats from SDN/NFV, AI/ML, and open APIs that enable fully autonomous network management. They recommend mitigations including protection against injection attacks on common data services and closed-loop manipulation. Coronado et al.~\cite{coronado_ztm_survey2022} provide a comprehensive survey bridging ZTM and mobile network research, taxonomizing management solutions with security challenges across ETSI ZSM, 3GPP, and MEF standardization efforts. Carrozzo et al.~\cite{carrozzo_ztops_noms2020} propose a zero-touch security architecture combining distributed AI for cognitive orchestration with DLT and smart contracts for cross-domain trust among non-trusted parties.

The growing reliance on AI/ML for network management introduces adversarial machine learning as an emerging threat. Benzaid and Taleb~\cite{benzaid_ai_b5g2020} examine AI as a double-edged sword---defender, offender, and victim---in beyond-5G security, discussing adversarial ML, data poisoning, and model evasion that can undermine autonomous network management. Apruzzese et al.~\cite{apruzzese_wild_networks2022} propose a ``myopic'' adversarial threat model for 5G requiring no target system compromise, demonstrating that appending junk data to packets disrupts ML-based traffic classification and QoS across six ML applications envisioned for 5G. Sagduyu et al.~\cite{sagduyu_adv_ml_5g2021} identify adversarial ML attack surfaces against 5G wireless including attacks on CBRS spectrum sharing classifiers and GAN-based spoofing of physical-layer UE authentication.

Federated learning, increasingly proposed for privacy-preserving mobile network management, introduces its own security challenges. Lim et al.~\cite{lim_fl_survey2020} provide a foundational survey covering FL fundamentals, communication efficiency, resource allocation, privacy, and security in heterogeneous mobile edge environments, including analysis of adversarial participants. Isaksson and Norrman~\cite{isaksson_fl_5g2020} integrate FL into the 3GPP NWDA architecture with multi-party computation protecting local model update confidentiality, demonstrating significantly lower communication overhead than prior work.

\subsubsection{Compliance, Assurance, and Regulatory Implications}
\label{subsubsec:compliance}

Operating cellular infrastructure within cloud environments raises important compliance and regulatory considerations that go beyond simply listing applicable regulations. The more fundamental challenge is that cloud deployment changes \emph{how compliance must be demonstrated}. Telecom operators used to certify relatively static appliances; cloud-native 5G/6G requires evidence for dynamic infrastructure, ephemeral workloads, rolling updates, policy-as-code, and shared-responsibility boundaries.

The primary telecom security assurance ecosystem is built on the GSMA Network Equipment Security Assurance Scheme (NESAS)~\cite{gsma_nesas} and 3GPP Security Assurance Specifications (SCAS). The SCAS framework includes TS~33.117~\cite{3gpp33117} for general security assurance requirements, with NF-specific specifications (TS~33.511--33.527) covering gNB, AMF, SMF, UPF, UDM, AUSF, SEPP, NRF, NWDAF, SCP, MnF, and virtualized network products. Regional regulatory bodies including ENISA have adopted this framework as a basis for EU 5G certification. The ENISA 5G Security Controls Matrix~\cite{enisa_5g_controls} provides a dynamic matrix of controls supporting EU Member States in implementing the EU 5G Cybersecurity Toolbox, mapped to ISO/IEC 27002, ISO/IEC 27005, and NIST SP~800-53.

Cloud-native deployments introduce shared-responsibility models that create potential gaps in compliance coverage. The 5G Americas white paper ``Evolving 5G Security for the Cloud''~\cite{5gamericas_cloud_security2022} defines the cloud shared-responsibility model for 5G deployments across IaaS, PaaS, and SaaS tiers, identifying that hybrid cloud and MEC deployments pose additional risks from responsibilities retained by the MNO. The CISA/NSA ``Security Guidance for 5G Cloud Infrastructures''~\cite{cisa_5g_cloud} explicitly identifies that operators must take responsibility for securing their tenancy in the cloud when shared-responsibility gaps exist. The GSMA Baseline Security Controls (FS.31)~\cite{gsma_fs31} provide comprehensive voluntary controls covering NFVI/container security, VNF lifecycle management, NFV orchestration, security operations, MEC platform controls, and roaming/interconnect security, aligned to CIS Controls v8.1.

Lawful interception compliance in cloud-hosted environments presents particular challenges. 3GPP TS~33.127~\cite{3gpp33127} and TS~33.128~\cite{3gpp33128} specify LI architecture supporting network-layer and service-based interception in 5G, defining LI\_X1, LI\_X2, and LI\_X3 interfaces for interception at AMF, SMF, UPF, and other NFs. ETSI TS~104~007~\cite{etsi_li_104007} provides a comprehensive LI system blueprint for virtualized and cloud-native 5G, notably referencing NIST SP~800-207 (Zero Trust Architecture) and IETF RFC~9334 (RATS Architecture) for hardware root of trust and remote attestation in LI systems. When the infrastructure hosting these LI interfaces is operated by a third-party cloud provider, the jurisdictional and operational complexities multiply.

At the European regulatory level, ENISA's NIS2 implementation guidance~\cite{enisa_nis2} establishes mandatory requirements for telecom operators classified as Essential Entities, including 24-hour early warning, 72-hour full incident report, and 1-month final report timelines, along with business continuity planning obligations. For regulated sectors deploying private 5G/6G, compliance requirements compound. Chen et al.~\cite{chen_zt_healthcare2021} address this for healthcare, presenting a zero-trust security system for 5G smart healthcare achieving real-time security situational awareness, continuous authentication, and fine-grained access control using 3GPP SUCI/AKA mechanisms for patient data protection.

\textbf{Synthesis.} The investigation areas above reveal that cloud-based cellular security is not a simple extension of either telecom security or cloud security, but a \emph{composition} of both threat models. 3GPP provides the procedural security baseline for mobility, signaling, and subscriber protection; cloud-native deployment reopens the problem at the substrate, orchestration, and supply-chain layers; multi-tenancy and geographic distribution complicate trust establishment and assurance; and the trajectory toward 6G---with AI-native operations and increased automation---further expands the trusted computing base. Compliance, accordingly, must evolve from static certification toward continuous evidence generation, runtime attestation, and policy-driven enforcement across shared-responsibility boundaries.
\subsection{Performance, Latency, and Reliability}
\label{subsec:performance_latency}

The performance characteristics of cellular networks have historically been achieved through tightly integrated hardware appliances deployed within carefully engineered operator infrastructure. Traditional mobile core and radio access network elements were implemented using specialized networking hardware, deterministic packet processing pipelines, and dedicated transport infrastructure designed to satisfy stringent latency and reliability requirements. The transition toward cloud-based deployments introduces a fundamental shift in this operational model. In cloud-native cellular systems, network functions are executed as virtualized or containerized software workloads operating on shared computing infrastructure. While this architectural shift enables elasticity, scalability, and faster service innovation, it also raises important questions regarding the ability of general-purpose cloud platforms to satisfy the strict performance and reliability requirements of cellular networks.

\begin{figure}[t]
    \centering
    \includegraphics[width=\columnwidth,trim={0cm 0cm 21cm 2cm},clip]{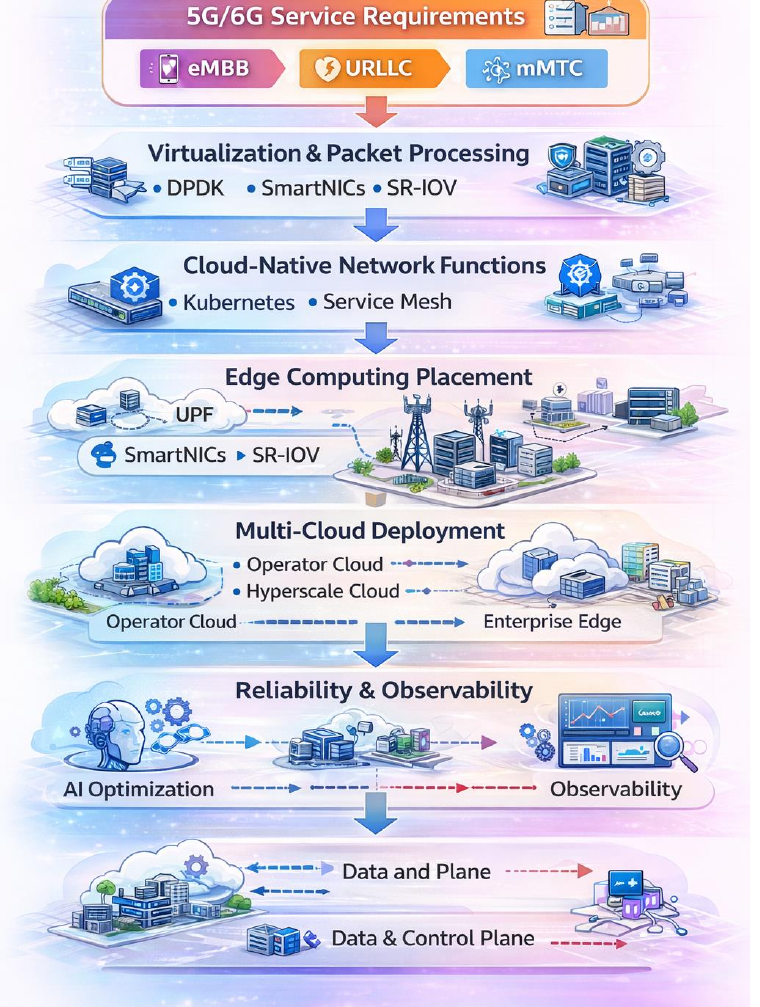}
     \caption{Performance and reliability overview summarizing service requirements over virtualization, cloud-native deployments, edge computing, multi-cloud with high reliability and observability.}
     \label{fig:perfrelow}
\end{figure}

The service categories introduced in the 5G system impose diverse and demanding performance requirements. Enhanced Mobile Broadband services emphasize high throughput and large-scale traffic handling, whereas URLLC require extremely low end-to-end latency and reliability levels approaching five-nines availability. mIoT prioritize scalability and the ability to support extremely large device populations. Achieving these requirements within distributed cloud environments presents significant technical challenges due to virtualization overhead, shared infrastructure resources, multi-tenant orchestration environments, and geographically distributed deployments. Consequently, understanding the performance implications of cloud-native cellular deployments requires a systematic investigation of how cloud computing paradigms interact with telecom performance requirements. The summary of this section is provided in Figure~\ref{fig:perfrelow}.

\subsubsection{Virtualization Overhead and Data Plane Acceleration}
\label{subsubsec:perf_requirements}

Cellular networks are engineered to satisfy stringent service-level performance targets that differ significantly from those typically assumed in cloud computing environments. In traditional telecom deployments, deterministic performance was achieved through specialized packet processing hardware, optimized network interfaces, and tightly controlled resource allocation mechanisms. In contrast, cloud infrastructure is primarily designed for flexible resource sharing and elastic scaling, often prioritizing throughput and resource utilization over deterministic latency guarantees. Virtualization introduces several sources of performance overhead that can affect packet processing latency and throughput. Hypervisors, virtual switches, and software-defined networking layers introduce additional processing stages that may increase packet forwarding delays.

The magnitude of this overhead has been quantified across multiple studies. Kourtis et al.~\cite{kourtis_sriov_dpdk2015} provide foundational measurements comparing LibPCAP, SR-IOV, and DPDK for VNF packet processing, finding that SR-IOV combined with DPDK achieves approximately 81\% of native physical DPDK throughput in a VM, while LibPCAP saturates at 1~Gbps---an 87.5\% throughput reduction compared to DPDK. Perez et al.~\cite{perez_virt_comparison2021} compare bare metal, KVM VMs, Docker containers, Kubernetes-orchestrated containers, Firecracker micro-VMs, and FaaS for a 5G monitoring platform, finding that containers offer superior throughput and lower latency than VMs while maintaining deployment agility. More recently, a study on low-latency container packet processing~\cite{container_lowlatency2024} evaluates DPDK, real-time kernels, and CPU pinning for URLLC workloads, demonstrating that cache architecture selection (shared vs.\ non-shared caches) impacts container latencies more than virtualization overhead itself when proper optimizations are applied.

For the data plane specifically, the 5G User Plane Function (UPF) has been the focus of intensive acceleration efforts. Chen et al.~\cite{chen_dpdk_upf2020} virtualize the UPF using Docker containers with Intel DPDK on x86 platforms, finding that only two physical cores are needed to handle 40~Gbps traffic, achieving 100\% throughput for packets of 256~bytes or larger and 60.69\% for 64-byte packets. Christakis et al.~\cite{christakis_upf_eval2024} deploy and evaluate four UPF implementations (SPGWU-UPF, P4-Switch-UPF, VPP-UPF, SmartNIC-P4-UPF) using OpenAirInterface in a real 5G environment, finding that P4-based UPFs outperform in throughput and packet loss while VPP-UPF (DPDK-based) shows competitive latency at lower cost. Beyond DPDK, emerging kernel-bypass frameworks offer alternative acceleration paths. Vieira et al.~\cite{vieira_ebpf_xdp_survey2020} provide a comprehensive survey of eBPF and XDP for fast in-kernel packet processing, covering architecture, program types, hardware offloading to SmartNICs, and comparisons with DPDK. An XDP-accelerated UPF running inside Docker containers~\cite{xdp_upf_docker2023} demonstrates that XDP acceleration in containers significantly improves packet processing while retaining containerization benefits. Hardware offloading takes this further: Synergy~\cite{synergy_icnp2022} designs a 5G UPF running on SmartNICs with a two-level flow-state access mechanism, achieving 2.32$\times$ lower handover latency and at least 2.04$\times$ lower packet loss compared to host-based buffering. Borromeo et al.~\cite{borromeo_fpga_smartnic2022} implement 5G DU Low-PHY functions on FPGA-based SmartNICs using OpenCL, demonstrating lower processing time and power consumption for URLLC-grade latency requirements.

Shared compute infrastructure can also lead to resource contention between workloads, commonly referred to as the ``noisy neighbor'' effect, which may impact latency-sensitive network functions. Margolin et al.~\cite{margolin_noisy_neighbor2016} demonstrate that SVM and random forest classifiers can detect noisy-neighbor degradation in NFV infrastructure with greater than 90\% accuracy. More recent work~\cite{noisy_neighbor_b5g2025} uses eBPF-based kernel instrumentation of the 5G UPF to measure per-TEID flow latency under multi-tenant contention, finding that even prioritized network slices are susceptible to noisy-neighbor degradation, with measurable per-packet latency impacts that challenge QoS isolation guarantees.

\subsubsection{Cloud-Native Network Function Performance}
\label{subsubsec:cnf_performance}

The transition from VNFs to CNFs further changes the performance characteristics of cellular infrastructure. CNFs are typically implemented as containerized microservices orchestrated by platforms such as Kubernetes. While microservice-based architectures enable modular service development and independent scaling of network functions, they also introduce new sources of performance overhead. In the 5G service-based architecture, control plane functions interact through HTTP/2-based service interfaces, generating significant east-west traffic within the core network. The overhead of this communication model has been systematically investigated.

Jain et al.~\cite{l25gc_sigcomm2022} present L$^2$5GC, which re-architects the 5G core using shared-memory IPC to eliminate HTTP serialization overhead while remaining 3GPP compliant, reducing control plane event completion time by approximately 50\% and data plane latency by approximately 2$\times$ compared to free5GC, with integrated transparent failure recovery. Goshi et al.~\cite{pp5gs_tnsm2023} introduce PP5GS, a procedure-based functional decomposition of NFs evaluated on Kubernetes with free5GC as baseline. PP5GS requires up to 34\% fewer computing resources, generates at least 40\% less signaling traffic, and completes complex procedures up to 50\% faster than standard SBA decomposition. Ahmad et al.~\cite{ahmad_sigcomm2020} propose a low-latency and consistent cellular control plane that reduces control plane latency through optimized state management, while their follow-up work~\cite{ahmad_conext2022} demonstrates control plane interventions enabling emerging edge applications through a 5G framework.

The performance of open-source 5G core implementations has been extensively benchmarked. Mukute et al.~\cite{mukute_benchmark2024} introduce both macro-benchmarking (registration/deregistration under varying UE loads) and micro-benchmarking (Linux kernel system call profiling) for OpenAirInterface, Open5GS, and free5GC, finding that Open5GS leads in control-plane performance. Lando and Schierholt~\cite{lando_oss_5g2023} compare free5GC and Open5GS in Docker-based deployment with UERANSIM, reporting that Open5GS uses 4.1$\times$ less CPU than free5GC during control plane operations. Barbosa et al.~\cite{barbosa_oss_5g2025} evaluate free5GC, OAI, Open5GS, and SD-Core on COTS hardware, finding that Open5GS provides best control-plane latencies, OAI achieves highest data-plane throughput, and free5GC has lowest resource consumption. The 5GC-Bench framework~\cite{5gc_bench2025} profiles per-VNF CPU and memory under increasing loads with OAI 5GC, identifying that AMF and SMF are CPU-bound, UDR is memory-bound, and UDM incurs high CPU usage from cryptographic operations. These per-VNF characterizations are essential for capacity planning in cloud deployments.

Service meshes used for mutual authentication, traffic management, and observability introduce additional latency. Aldas and Babakian~\cite{aldas_servicemesh2023} propose enhancing Envoy proxy within Istio to support 5G-specific semantics including NRF integration and SCP functionality, comparing communication latencies across three SCP deployment scenarios. A Red Hat-supported study~\cite{mtls_comparison2024} evaluates mTLS overhead in Istio, Istio Ambient, Linkerd, and Cilium on Kubernetes, finding that mTLS enforcement increases P99 latency by 166\% for Istio but only 8\% for Istio Ambient and 33\% for Linkerd---directly applicable to 3GPP-mandated mTLS in 5G SBA. The 5G-STREAM framework~\cite{5g_stream_dsn2025} demonstrates how a purpose-built two-tier service mesh can reduce control plane latency by up to 2$\times$ the inter VNF-NRF latency per HTTP transaction at a minor resource cost of 0.1~USD/hr on AWS for a VNF managing 50,000 requests per minute. The 5G-NECTAR framework~\cite{atalay_nfvsdn2024} specifically measures service mesh encapsulation overhead on cloud-deployed 5G core VNFs, addressing overlay networking mechanisms used for inter-VNF communication on AWS.

Container networking implementations also significantly affect performance. Qi et al.~\cite{qi_cni_icdcs2020} provide a comprehensive evaluation of CNI plugins (Flannel, Weave, Cilium, Calico, Kube-router) with fine-grained CPU cycle-per-packet measurements, finding that Cilium performs best for intra-host communication through eBPF-optimized routing, while Kube-router and Calico excel for inter-host traffic. FlexCore~\cite{flexcore_apnet2023} proposes an XDP/eBPF-based SCTP load balancer for 5G core supporting slice-aware, UE-aware, and procedure-aware load balancing, achieving up to 79\% latency reduction on stateless architectures and 63\% for latency-critical slices. These findings underscore that the choice of networking stack, service mesh, and load balancing strategy can dominate overall control plane latency in cloud-native 5G deployments.

\subsubsection{Edge Computing and Latency Optimization}
\label{subsubsec:edge_latency}

Edge computing has emerged as a key architectural mechanism for addressing latency constraints in cloud-based cellular networks. By deploying compute and storage resources closer to end users, edge platforms enable latency-sensitive services to operate with reduced round-trip delays compared to centralized cloud deployments. The magnitude of achievable latency reductions has been characterized through both simulation and real-world measurement.

Fezeu et al.~\cite{fezeu_mmwave_pam2023} provide the first in-depth measurement of commercial 5G mmWave PHY latency on AT\&T and Verizon networks, finding a best achievable PHY latency of 0.85~ms occurring only approximately 2.27\% of the time. They explore the latency benefits of deploying on AWS Wavelength, AWS Local Zone, and AWS Regional Zone, demonstrating that server placement significantly impacts end-to-end delay. Narayanan et al.~\cite{narayanan_5g_sigcomm2021} present a comprehensive measurement study of commercial 5G across multiple carriers (mmWave, sub-6~GHz), deployment schemes (NSA vs.\ SA), and applications, revealing key 5G latency characteristics and handover behaviors. Lumos5G~\cite{lumos5g_imc2020} provides a composable ML framework for context-aware 5G throughput prediction achieving 1.37$\times$ to 4.84$\times$ reduction in prediction error, establishing baseline performance data for edge computing latency analysis. The 5G-MAP framework~\cite{5g_map_mobicom2025} conducts large-scale deployment experiments across 8 AWS regions and 18 edge zones spanning 7 countries, identifying topologies that can considerably reduce session setup time as inter-VNF latency is reduced by up to five times through strategic VNF placement.

One of the primary mechanisms used to reduce latency in distributed 5G deployments is the flexible placement of the User Plane Function. Leyva-Pupo et al.~\cite{leyva_upf_placement2019} propose an ILP and heuristic framework for jointly optimizing edge node and UPF placement to minimize deployment costs under sub-1~ms latency constraints, achieving greater than 20\% cost savings with placement heuristics within one UPF of optimal. Their subsequent work~\cite{leyva_dynamic_upf2022} addresses dynamic UPF placement and chaining reconfiguration to handle user mobility while ensuring QoS, using an optimal stopping theory scheduling mechanism to determine reconfiguration timing based on latency violations. Jin et al.~\cite{jin_vnf_infocom2020} jointly optimize edge server and physical link resources under latency constraints with maximum resource reuse and a proven worst-case performance bound. Cziva et al.~\cite{cziva_vnf_infocom2018} investigate dynamic latency-optimal VNF placement at the network edge, proposing approaches that adapt to changing network conditions under limited edge resources. Zhang et al.~\cite{zhang_vnf_infocom2019} address VNF placement for 5G network slices accounting for co-location interference effects, jointly considering cloud and transport KPIs.

Simulation-based evaluation further corroborates these findings. Virdis et al.~\cite{virdis_mec_eval2020} evaluate MEC deployment options using Simu5G~\cite{simu5g_2020} for in-vehicle infotainment and remote driving, finding that 4G radio access is the bottleneck preventing MEC service scaling while 5G enables considerably higher MEC service penetration. Alawe et al.~\cite{alawe_amf_ccnc2018} propose a control-theory-based algorithm for AMF load balancing and dynamic scaling using NFV, confirming fair load distribution while scaling dynamically with massive IoT/MTC traffic. Barrachina-Mu\~{n}oz et al.~\cite{barrachina_cn5g2022} deploy Open5GS as CNFs in Kubernetes with Prometheus monitoring and MEC support, demonstrating UPF re-selection and mobility scenarios with over-the-air transmissions and providing a Helm-chart-based blueprint for cloud-native 5G testbed deployments.

\subsubsection{Multi-Cloud and Cross-Domain Performance}
\label{subsubsec:multicloud_performance}

Cloud-based cellular deployments frequently span multiple infrastructure domains, including hyperscale cloud providers, operator private clouds, regional edge facilities, and enterprise environments. While this multi-domain architecture enables flexible deployment models and improved geographic coverage, it also introduces additional latency sources and performance variability.

The 5G-MAP framework~\cite{5g_map_mobicom2025} provides the most comprehensive cross-region measurement study to date, deploying the OAI 5G core across 8 AWS regions and 18 edge zones spanning 7 countries. The results demonstrate that moving control plane functions (AMF, NRF) to edge zones significantly reduces 5G-AKA and session setup latencies, while identifying that inter-site hop count is the dominant factor in control plane performance degradation. ECHO~\cite{echo_mobicom2018} presents a cellular core designed for public cloud deployment (Azure) with stateless processing frontends paired with high-availability storage backends, deployed across 3 data centers (2 in Europe, 1 in the US), demonstrating efficient state synchronization and component replacement across distributed sites while handling VM failures inherent to public clouds.

Multi-cluster Kubernetes orchestration introduces additional complexity. Syrigos et al.~\cite{syrigos_multicluster2023} propose multi-domain orchestration using SUSE Rancher and Submariner for encrypted L3 cross-cluster connectivity, evaluating throughput under different Submariner configurations (encrypted vs.\ unencrypted tunnels) for OAI 5G core deployment over a 25~Gbps fabric. Dumitru-Guzu et al.~\cite{dumitru_crosscluster2024} use the Liqo operator for on-demand scaling of Open5GS functions across distributed clusters, demonstrating significant latency and throughput improvements for in-band peering via VPN tunnels and validating three end-to-end slicing use cases (eMBB, URLLC, mMTC). The 5G-NECTAR framework~\cite{atalay_nfvsdn2024} provides specific measurements of service mesh encapsulation overhead for inter-VNF communication across AWS deployment zones.

At the architectural level, Santos et al.~\cite{santos_hyperstrator2020} propose a ``hyperstrator'' architecture coordinating per-segment orchestrators for RAN, transport, and core, demonstrating negligible overhead for cross-segment slice provisioning while confirming the necessity of coordinated resource management for consistent end-to-end QoS. Taleb et al.~\cite{taleb_multidomain2019} introduce a technology-agnostic multi-domain orchestration architecture with four strata, addressing cross-domain latency management and federated network slice instantiation. Woo et al.~\cite{woo_elastic_nsdi2018} address elastic scaling of stateful network functions without packet drops or state inconsistency, demonstrating efficient state transfer during scale-out and scale-in operations---a foundational capability for distributed 5G NF state management across cloud sites. Overlay networking mechanisms commonly used in container orchestration platforms introduce encapsulation overhead and routing complexity. Encrypted tunnels used to secure inter-domain communication further increase packet processing latency, as characterized by the multi-cluster studies above. These factors complicate the design of distributed cellular control planes and require careful placement of latency-sensitive functions to minimize cross-domain communication delays.

\subsubsection{Reliability and Fault Tolerance}
\label{subsubsec:reliability}

Reliability has historically been a defining characteristic of telecommunications infrastructure. Traditional mobile networks were engineered using hardware redundancy, geographically distributed core sites, and deterministic failover mechanisms designed to achieve extremely high availability targets. In cloud-native environments, reliability is achieved through different mechanisms, including microservice replication, automated orchestration, and distributed system design principles. A central design question is whether network functions should be stateful, procedurally stateless, or fully stateless, as each approach offers different reliability-performance tradeoffs.

The foundational work on stateless NFs by Kablan et al.~\cite{statelessnf_nsdi2017} proposes StatelessNF, decoupling state from processing using DPDK, Docker containers, and RAMCloud, with a network-wide orchestrator managing scaling and failure recovery via OpenFlow. This directly influenced stateless NF design patterns adopted in 5G core specifications. Kulkarni and Fahmy~\cite{kulkarni_stateless2021} quantify the performance cost of statelessness by comparing procedural versus transactional paradigms in 5G, proposing optimizations including shared UE state via a common database and non-blocking API calls. State sharing reduces latency by an average of 10\%, clarifying the performance-resilience tradeoff.

Paul et al.~\cite{paul_k8s_5g2025} provide the most comprehensive evaluation to date, comparing stateful, procedurally-stateless (Aether SD-Core), and fully-stateless (UDSF-based) 5G core designs on Kubernetes. Fine-grained state checkpointing reduces throughput by up to 75\% but achieves 100\% registration success under NF failure, compared to 80\% failure for the stateful design. Importantly, Kubernetes orchestration imposes only approximately 1.5\% overhead, demonstrating that the platform itself is not the bottleneck. L$^2$5GC~\cite{l25gc_sigcomm2022} takes a different approach, integrating transparent failure recovery into its shared-memory architecture to achieve both low latency and high reliability simultaneously.

For network slice-level reliability, Vittal and Franklin~\cite{vittal_harness2022} propose HARNESS, an intelligent scheduling system using XDP and eBPF for high-availability 5G slicing, achieving 60\% reduction in dropped control plane requests, 50\% reduction in average response time, and 3.2\% improvement in slice service availability. Their subsequent work~\cite{vittal_resilience_tnsm2024} builds a self-resilient 5G core combining large-scale ILP-based Column Generation and AI-based deep learning in a closed-loop SON paradigm, validated on a real 5G testbed with significant improvements in resilience across multiple slicing scenarios. Upadhya et al.~\cite{upadhya_noisy2020} propose dynamic CPU pinning combined with load balancing via dynamic network slicing as a lightweight alternative to computationally expensive VM/container migration for addressing noisy-neighbor-induced reliability degradation.

The scaling and resource consumption dimensions of reliability have been characterized in our prior work. Atalay et al.~\cite{atalay_wcnc2022} present a resource consumption study for scaling network slices in a 5G testbed, while the companion cost assessment study~\cite{atalay_globecom2022} quantifies the deployment costs associated with network-slice-as-a-service offerings.

\subsubsection{Performance Observability and Autonomous Optimization}
\label{subsubsec:observability}

The dynamic nature of cloud-native cellular infrastructure necessitates advanced observability mechanisms capable of monitoring performance across distributed computing environments. Traditional telecom network management systems relied on centralized monitoring frameworks and relatively static infrastructure assumptions. In contrast, cloud-native deployments involve highly dynamic workloads that may be instantiated, migrated, or terminated in response to changing traffic conditions.

Distributed tracing has emerged as a critical observability mechanism for understanding control plane behavior in microservice-based 5G cores. Zhao et al.~\cite{zhao_5gc_tracer2025} propose 5GC-Tracer, an eBPF-based application-centric distributed tracing system for 5G core networks that requires no code instrumentation. Validated during the Paris 2024 Paralympic Games on a real 5GC deployment, it achieves high trace collection rates with low overhead and enables statistical anomaly localization for QoS degradation. Soldani et al.~\cite{soldani_ebpf2023} introduce the ``Sauron'' eBPF platform deployed at Rakuten Mobile, estimating energy consumption, deriving performance counters, parsing NAS protocols, and detecting unauthorized access---all with less than 1\% overhead. The 5G-MAP framework~\cite{5g_map_mobicom2025} employs side-car proxies adjacent to each VNF within the same pod to provide detailed control plane telemetry on individual HTTP transactions, enabling operators to evaluate performance and operation of cloud-based 5G deployments without requiring direct access to VNF internals. Barrachina-Mu\~{n}oz et al.~\cite{barrachina_cn5g2022} demonstrate a cloud-native 5G experimental platform with end-to-end monitoring using Prometheus and Grafana dashboards integrated with a Kubernetes-deployed 5G core.

AI-driven optimization represents an increasingly important mechanism for maintaining performance in cloud-native cellular deployments. Solozabal et al.~\cite{solozabal_drl_vnf2020} extend Neural Combinatorial Optimization to VNF placement using sequence-to-sequence networks with reinforcement learning, achieving near-optimal results orders of magnitude faster than CPLEX solvers. Dalgkitsis et al.~\cite{dalgkitsis_drl_globecom2020} leverage DDPG reinforcement learning to automate VNF deployment between edge and cloud nodes, demonstrating superior dynamic resource allocation enabling zero-touch service management. Liu et al.~\cite{onslicing_conext2021} implement online end-to-end network slicing with constraint-aware DRL on a real testbed using OAI LTE and 5G NR, achieving 61.3\% resource usage reduction compared to rule-based solutions while maintaining near-zero SLA violation (0.06\%) throughout online learning. Polese et al.~\cite{polese_coloran_tmc2023} present ColO-RAN, the first publicly available large-scale O-RAN testing framework, designing three DRL-based xApps for closed-loop RAN slicing and scheduling control across 7 base stations and 42 users. Passas et al.~\cite{passas_ai_autoscaling2022} develop AI-driven dynamic scaling for cloud-native 5G AMF, demonstrating that deep learning approaches enable higher user admissions and improved scalability compared to threshold-based autoscaling.

As cellular networks evolve toward 6G architectures, the integration of autonomous network management systems and AI-driven optimization mechanisms is expected to play an increasingly important role. Coronado et al.~\cite{coronado_ztm_survey2022} provide a comprehensive survey bridging zero-touch management and mobile network research, covering ETSI ZSM architecture, autonomic management, SON, intent-based networking, and AI/ML-driven automation for 5G and 6G. The trajectory toward closed-loop automation is clear: systems like OnSlicing~\cite{onslicing_conext2021} demonstrate that DRL-based approaches can achieve resource savings exceeding 60\% with near-zero SLA violations on live infrastructure, while 5GC-Tracer~\cite{zhao_5gc_tracer2025} shows that eBPF-based instrumentation can provide the real-time observability data needed to feed such closed-loop systems without performance degradation.

\textbf{Synthesis.} The investigation above reveals a consistent pattern: cloud-native cellular deployments can meet telecom-grade performance requirements, but only through careful co-design of the software stack and infrastructure. Data plane acceleration (DPDK, SmartNIC, XDP) restores near-line-rate performance, but the control plane remains the primary bottleneck---standard HTTP/2 SBA introduces substantial latency that shared-memory alternatives and procedure-based decomposition can halve. Service mesh overhead is significant (up to 166\% P99 latency increase for sidecar-based Istio) but sidecarless architectures dramatically reduce this penalty. Edge placement of UPF and AMF/NRF is the single most impactful latency optimization, with real-world measurements confirming up to 5$\times$ reduction in inter-VNF latency. Reliability requires explicit architectural choices: stateless designs achieve near-perfect failure recovery but impose 10--75\% throughput penalties depending on checkpointing granularity. AI-driven optimization, particularly DRL for VNF placement and network slicing, has matured from simulation to real-testbed validation with resource savings exceeding 60\% at near-zero SLA violation rates.
\subsection{Scalability, Elasticity, and Resource Efficiency}
\label{subsec:scalability}

Cloud-native architectures promise significant improvements in scalability and infrastructure utilization compared to traditional telecom deployments. Historically, mobile networks were engineered using static capacity planning models in which infrastructure was provisioned to accommodate projected peak traffic demand. While this approach ensured reliable service delivery during periods of heavy network usage, it often resulted in substantial resource overprovisioning and inefficient infrastructure utilization during off-peak periods. The adoption of cloud computing paradigms introduces the possibility of dynamically scaling network resources in response to real-time demand fluctuations, thereby improving both scalability and resource efficiency.

\begin{figure}[t]
    \centering
    \includegraphics[width=\columnwidth,trim={0cm 0cm 20.5cm 1cm},clip]{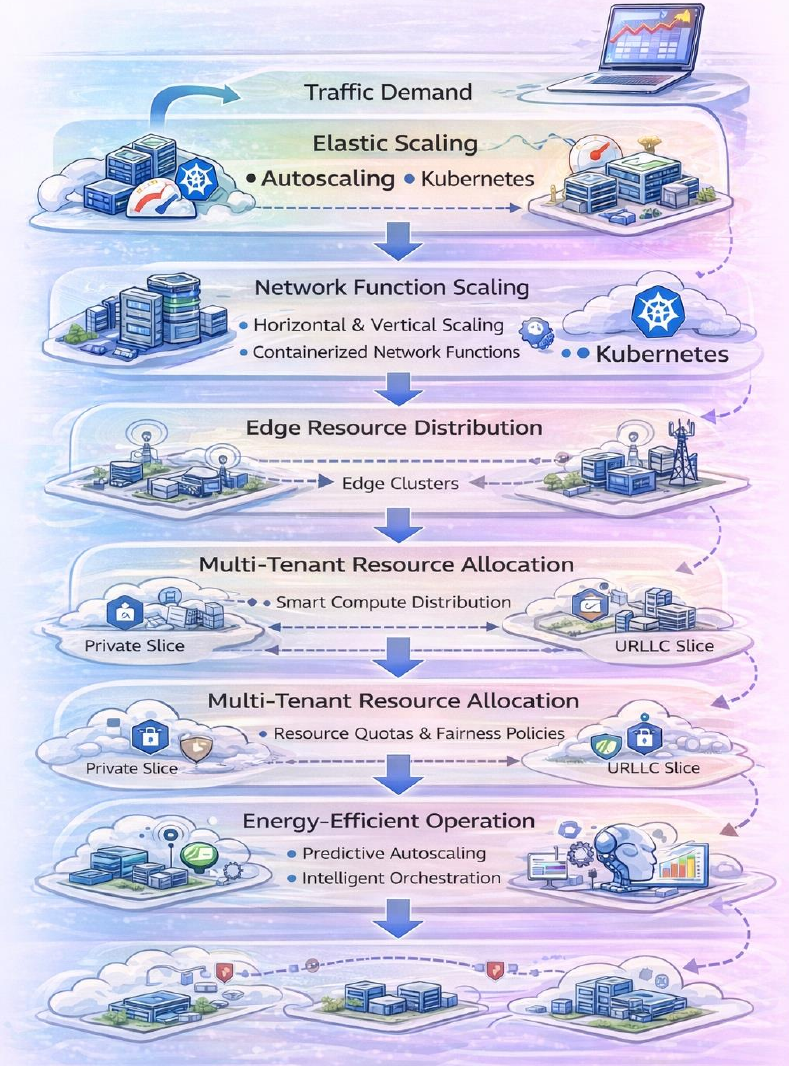}
     \caption{Scalability and resource efficiency framework for cloud-based cellular networks showing the interplay between elastic scaling, distributed edge management, multi-tenant slice orchestration, energy-aware operation, and AI-driven optimization}
     \label{fig:scalow}
\end{figure}

Despite these potential benefits, applying cloud elasticity mechanisms to cellular networks presents unique challenges. Telecom workloads differ significantly from typical cloud applications in that many network functions maintain persistent session state, handle latency-sensitive traffic, and operate under strict service-level agreements. In addition, the geographically distributed nature of cellular infrastructure introduces constraints related to workload placement, traffic locality, and mobility management. Consequently, achieving scalable and efficient operation in cloud-based cellular networks requires careful consideration of network function scaling strategies, distributed resource allocation mechanisms, and intelligent orchestration frameworks. Figure~\ref{fig:scalow} summarizes the investigation sub-categories in this section.

\subsubsection{Elastic Scaling of Cloud-Native Network Functions}
\label{subsubsec:cnf_scaling}

Cloud-native network functions are commonly implemented as containerized microservices orchestrated by platforms such as Kubernetes. These orchestration frameworks provide mechanisms for both horizontal scaling (deploying additional instances) and vertical scaling (increasing resources per instance). While such mechanisms are well established in cloud computing environments, applying them to telecom network functions introduces additional complexities because many control-plane components maintain persistent context information related to user equipment sessions, mobility management, and policy enforcement.

Several systems have addressed this challenge through fundamentally different architectural approaches. CoreKube~\cite{corekube_mobicom2023} presents a message-focused, cloud-native mobile core with truly stateless workers orchestrated via Kubernetes. By externalizing all state to a shared data store, CoreKube achieves dynamic scaling with minimal compute overhead and seamless failure recovery through Kubernetes autoscaling and self-healing primitives. TEGRA~\cite{tegra2025} proposes soft-state microservice design patterns with Kubernetes HPA and custom UE-based sticky load balancing, processing requests 20$\times$ faster than free5GC, 11$\times$ faster than Open5GS, and 1.75$\times$ faster than Aether, with autoscaling achieving 2$\times$ lower latencies than Aether under dynamic load. Paul et al.~\cite{paul_k8s_5g2025} provide the most comprehensive comparison of stateful, procedurally-stateless (Aether SD-Core), and fully-stateless (UDSF-based) 5G core designs on Kubernetes, finding that Kubernetes orchestration imposes only approximately 1.5\% overhead, but fine-grained state checkpointing reduces throughput by up to 75\%. L$^2$5GC~\cite{l25gc_sigcomm2022} takes a different approach, replacing HTTP-based SBA communication with shared-memory IPC to eliminate serialization overhead while remaining 3GPP compliant, reducing control plane event completion time by approximately 50\%.

The tradeoff between statefulness and scalability has been systematically quantified. Goshi et al.~\cite{pp5gs_tnsm2023} introduce PP5GS, a procedure-based functional decomposition requiring up to 34\% fewer computing resources, generating at least 40\% less signaling traffic, and completing complex procedures up to 50\% faster than standard SBA decomposition. Kulkarni et al.~\cite{kulkarni_stateless2021} quantify the cost of statelessness in 5G, finding that sharing UE state among AMF and SMF via a common database reduces latency by an average of 10\%, and their follow-up work~\cite{kulkarni_lowcost2022} proposes optimizations avoiding redundant database reads that reduce statelessness cost by 33\% on average. Goshi et al.~\cite{goshi_procedure_aware2024} further propose Piggyback and Proactive-Push approaches for procedure-aware stateless systems, reducing synchronous procedure completion time by 44--70\% without additional data-store overhead. For session continuity during horizontal scaling, Du et al.~\cite{mlsld_dcan2022} introduce ML-SLD, a non-3GPP-defined middleware NF that servitizes the N2 interface to avoid service interruptions during large-scale AMF scaling events, while Sthawarmath et al.~\cite{sthawarmath_stateless2022} demonstrate that stateless control-plane functions with external state stores enable seamless horizontal scaling where any NF instance can serve any subscriber request by retrieving state on demand.

Predictive autoscaling using machine learning has emerged as a key mechanism for proactive resource provisioning. Alawe et al.~\cite{alawe_traffic_forecast2018} compare DNN and LSTM neural networks for forecasting 5G core traffic to drive scaling decisions, finding that LSTM-based forecast-driven scaling significantly outperforms threshold-based reactive solutions. Subramanya and Riggio~\cite{subramanya_fl_autoscaling2021} model autoscaling as a time-series forecasting problem and demonstrate that federated learning addresses data privacy concerns in multi-domain scenarios while maintaining effective scaling with minimal QoS violations. Passas et al.~\cite{passas_ai_autoscaling2022} apply deep learning to a customized orchestrator for proactively scaling AMF replicas in a real testbed, enabling significantly higher user admission rates than default Kubernetes autoscaling. Alawe et al.~\cite{alawe_amf_ccnc2018} provide a control-theory-based algorithm for dynamic AMF load balancing and scaling, confirming fair load distribution under massive IoT/MTC traffic.

\subsubsection{Network Slicing Resource Management}
\label{subsubsec:multitenant_scaling}

Cloud-native cellular infrastructure often supports multiple tenants and services operating on shared computing platforms. Network slicing enables operators to create logically isolated virtual networks tailored to the requirements of different applications. In such multi-tenant environments, resource allocation mechanisms must ensure both efficiency and strong isolation guarantees while supporting the diverse QoS profiles of eMBB, URLLC, and mMTC service categories.

The foundational resource allocation challenge for network slicing is formalized by Popovski et al.~\cite{popovski_slicing2018}, who study orthogonal versus non-orthogonal RAN resource slicing across three generic 5G service types, introducing Heterogeneous NOMA and demonstrating that non-orthogonal slicing yields significant performance gains. Alsenwi et al.~\cite{alsenwi_drl_slicing2021} address eMBB-URLLC coexistence through deep reinforcement learning, maximizing eMBB throughput while satisfying stringent URLLC latency and reliability constraints via risk-sensitive optimization. Liu et al.~\cite{onslicing_conext2021} implement online end-to-end network slicing with constraint-aware DRL on a real testbed using OAI, achieving 61.3\% resource usage reduction compared to rule-based solutions while maintaining near-zero SLA violation (0.06\%) throughout online learning. MicroOpt~\cite{microopt2024} leverages differentiable neural network-based slice models with gradient descent for dynamic resource scaling, achieving up to 21.9\% improvement in resource allocation compared to state-of-the-art approaches.

Slice admission control determines which slice requests are accepted and how resources are reserved. Ojijo and Falowo~\cite{ojijo_admission_survey2020} provide a comprehensive survey covering admission control objectives, strategies, and optimization algorithms. Villota-Jacome et al.~\cite{villota_drl_admission2022} propose SARA and DSARA mechanisms using RL and deep RL for admission control and resource allocation in 5G core network slicing, differentiating core from edge nodes and processing requests across eMBB, URLLC, and mMTC use cases. Bega et al.~\cite{bega_deepcog2020} design DeepCog, a custom loss function ($\alpha$-OMC) that accounts for the monetary cost of overprovisioning versus underprovisioning, demonstrating over 50\% reduction in resource management costs using real-world metropolitan mobile network data. Sciancalepore et al.~\cite{sciancalepore_traffic2017} address traffic forecasting per slice with admission control based on SLAs and adaptive correction of forecasted load, demonstrating how predictive methods optimize network utilization while meeting service guarantees.

Managing shared infrastructure across multiple slices introduces fairness and isolation challenges. Yarkina et al.~\cite{yarkina_multitenant2022} combine features of complete partitioning and complete sharing policies, achieving up to an order of magnitude improvement in session loss probability compared to static slicing while maintaining comparable user data rates. Samdanis et al.~\cite{samdanis_slice_broker2016} introduce the 5G Network Slice Broker concept enabling MVNOs and vertical market players to dynamically request and lease resources from infrastructure providers with SLA-based admission control. Our prior work characterizes slice scaling from both resource consumption~\cite{atalay_wcnc2022} and cost assessment~\cite{atalay_globecom2022} perspectives, quantifying how CPU and memory usage grow with increasing slice instances in containerized 5G testbed deployments and the associated deployment costs for network-slice-as-a-service offerings.

\subsubsection{Distributed Resource Management in Edge Environments}
\label{subsubsec:edge_scaling}

Edge computing architectures introduce additional challenges for scalability and resource efficiency. Rather than concentrating compute resources within centralized data centers, edge deployments distribute infrastructure across numerous geographically dispersed locations with more limited capacity. The distributed nature of this infrastructure can lead to resource fragmentation, where available capacity is unevenly distributed across locations and traffic demand fluctuates significantly across regions and time periods.

Kim et al.~\cite{kim_marl_slicing2021} develop a hierarchical MEC architecture with two cooperative PPO-based algorithms that maximize resource efficiency for end-to-end network slicing while recognizing slice request characteristics, demonstrating high QoS satisfaction across heterogeneous requirements. D'Oro et al.~\cite{doro_sledge2020} prove that optimal joint network-MEC slice instantiation is NP-hard and propose near-optimal algorithms that instantiate slices 7.5$\times$ faster and within 25\% of the optimum, validated on a 24-radio testbed with 9 smartphones. Yu et al.~\cite{yu_drl_fl_mec2021} propose the I-UDEC framework combining two-timescale DRL with federated learning for privacy-preserving distributed resource management in 5G ultra-dense networks, achieving up to 31.87\% reduction in task execution time.

VNF and CNF placement optimization has been comprehensively surveyed by Attaoui et al.~\cite{attaoui_vnf_cnf_survey2023}, who classify 294 articles on placement approaches by metrics, methods (heuristic, meta-heuristic, ML), and environment (cloud, edge, fog). Agarwal et al.~\cite{agarwal_vnf_placement2019} formulate joint VNF placement, resource assignment, and traffic routing using a queuing-based model, proposing MaxZ, a fast strategy that jointly accounts for mutual interactions among placement, assignment, and routing decisions. Kianpisheh et al.~\cite{kianpisheh_edge_cloud2020} propose Markov-based approximation and node-ranking heuristics for VNF placement across edge-cloud tiers, achieving up to 21\% cost improvement over state-of-the-art schemes. Behravesh et al.~\cite{behravesh_mec_tnsm2021} tackle joint time-sensitive user association and service function chain placement in MEC-enabled 5G considering dynamic user mobility and time-varying traffic demands. The 5G-MAP framework~\cite{5g_map_mobicom2025} provides empirical evidence across 8 AWS regions and 18 edge zones showing that strategic placement of control plane functions at edge locations reduces inter-VNF latency by up to 5$\times$, offering actionable guidance for operators making cloud placement decisions.

Network-aware container scheduling further improves edge resource utilization. Wojciechowski et al.~\cite{wojciechowski_netmarks2021} propose NetMARKS, which uses Istio service mesh metrics for network-aware Kubernetes pod scheduling, reducing application response time by up to 37\% and saving up to 50\% of inter-node bandwidth. These results demonstrate that default Kubernetes scheduling, which considers only CPU and memory, is insufficient for latency-sensitive telecom workloads at the edge.

\subsubsection{Resource Consumption Characterization and Cost Efficiency}
\label{subsubsec:resource_char}

Understanding the resource consumption profiles of individual 5G core network functions is essential for capacity planning and cost optimization in cloud deployments. The 5GC-Bench framework~\cite{5gc_bench2025} profiles per-VNF CPU and memory under increasing loads with OAI 5GC, identifying that AMF, SMF, and UDM are CPU-bound while UDR is memory-bound, with UDM incurring particularly high CPU usage from cryptographic operations. Mukute et al.~\cite{mukute_benchmark2024} introduce both macro-benchmarking and micro-benchmarking (Linux kernel system call profiling) for three open-source 5G cores, finding that Open5GS achieves lowest control-plane latencies. Lando et al.~\cite{lando_oss_5g2023} report that free5GC uses only 0.17\% CPU for data plane operations versus 22.1\% for Open5GS due to single-threaded user-plane handling, while free5GC scales to 1001.1~Mbps with 8 UEs.

Our prior work characterizes resource consumption from a scaling perspective: Atalay et al.~\cite{atalay_wcnc2022} measure the limits of 5G microservice virtualization using lightweight containers for different slicing models, quantifying how CPU and memory usage grow with increasing slice instances and UEs. The companion cost assessment study~\cite{atalay_globecom2022} provides deployment cost modeling for network-slice-as-a-service offerings. A comparative evaluation of horizontal, vertical, and hybrid VNF scaling strategies~\cite{cost_placement_im2021} finds that throughput and event capacity depend approximately 80\% on CPU and 20\% on memory, with hybrid scaling providing the optimal balance between reliability and resource utilization. Studies comparing VMs and containers for 5G MEC~\cite{vm_container_mec2019} confirm that containers provide significantly reduced migration time, faster service recovery, and inherent advantages for control plane latency.

\subsubsection{Energy Efficiency and Sustainable Infrastructure}
\label{subsubsec:energy}

As cellular networks evolve toward large-scale distributed cloud infrastructures, energy efficiency becomes an increasingly important design consideration. Lopez-Perez et al.~\cite{lopezperez_energy_survey2022} provide a comprehensive survey reviewing power consumption models for 5G RANs, covering detailed breakdowns for DRAN, CRAN, and H-CRAN architectures alongside massive MIMO, lean carrier design, sleep modes, and ML-based energy optimization strategies.

Energy-aware VNF placement has been investigated through both optimization and learning approaches. Al-Quzweeni et al.~\cite{alquzweeni_energy2019} develop MILP optimization for VM location and server utilization in optical-network-supported 5G NFV architecture, achieving up to 38\% energy savings (average 34\%). Kar et al.~\cite{kar_energy_tnsm2018} jointly optimize VNF chain placement and traffic routing to minimize total energy costs through dynamic consolidation and migration. GreenNFV~\cite{greennfv_sc2023} tunes CPU sharing, frequency scaling, LLC allocation, DMA buffer size, and packet batch size via DRL, achieving 4.4$\times$ higher throughput with 1.5$\times$ better energy efficiency in throughput SLA mode, or 3$\times$ higher throughput with 50\% energy reduction in energy SLA mode. Tipantuna and Hesselbach~\cite{tipantuna_energy_slicing2020} analyze requirements for energy-aware 5G network slicing per 3GPP specifications, integrating NFV, SDN, and IoT connectivity for slice creation that adapts to renewable energy availability.

Sleep mode strategies offer substantial energy savings at the RAN level. Renga et al.~\cite{renga_sleep_twc2023} study Advanced Sleep Modes where base stations gradually deactivate components during inactivity, achieving up to 90\% energy reduction with throughput loss of 3--19\% and latency under 5~ms. El Amine et al.~\cite{elamine_sleep_rl2022} develop an RL agent that learns optimal multi-level sleep mode policies (micro-sleep to deep sleep) across heterogeneous base stations based on traffic load patterns, balancing energy savings with QoS requirements.

\subsubsection{AI-Driven Resource Optimization and Orchestration}
\label{subsubsec:ai_scaling}

Future cellular networks are expected to increasingly rely on intelligent automation to manage the complexity of large-scale distributed infrastructure. Deep reinforcement learning has emerged as a dominant approach for VNF scaling and placement. Solozabal et al.~\cite{solozabal_drl_vnf2020} extend Neural Combinatorial Optimization to VNF placement using sequence-to-sequence networks with RL, achieving near-optimal results orders of magnitude faster than CPLEX solvers. Pei et al.~\cite{pei_drl_jsac2020} formulate VNF placement as Binary Integer Programming and propose DDQN-VNFPA using Double Deep Q-Networks, demonstrating improved SFC acceptance ratio, throughput, delay, and load balancing on real-world network topology. Pujol Roig et al.~\cite{pujol_drl_jsac2020} develop a DRL agent that manages VNF horizontal and vertical scaling across multiple resource types, achieving significant cost reduction while maintaining QoS. Saha et al.~\cite{saha_drl_survey2023} survey DRL-based approaches for slice scaling and VNF placement, categorizing state/action/reward design patterns and identifying open challenges including scalability and sim-to-real transfer.

Traffic prediction for proactive provisioning has been addressed by several systems. DeepCog~\cite{bega_deepcog2020} introduces a custom loss function accounting for the monetary asymmetry of overprovisioning versus underprovisioning, demonstrating over 50\% reduction in resource management costs. Sevgican et al.~\cite{sevgican_nwdaf2020} present the first comprehensive NWDAF study with ML implementation per 3GPP standards, evaluating linear regression, LSTM, and RNN for load prediction and anomaly detection. Lee et al.~\cite{lee_fl_comst2024} provide a comprehensive survey of federated learning-empowered mobile network management covering access-to-core functions, reviewing FL for traffic prediction, radio resource management, slicing, and 3GPP NWDAF standardization.

The orchestration landscape is evolving from standards-based frameworks toward cloud-native tooling. Yilma et al.~\cite{yilma_osm_onap2020} benchmark OSM and ONAP across onboarding delay, runtime orchestration delay, and deployment process delay, revealing that ONAP requires significantly more resources while OSM offers better documentation and lighter footprint. Kalim et al.~\cite{kalim_k8s_tnsm2020} provide a comprehensive performance evaluation of Kubernetes for 5G NF deployments, establishing the feasibility of container orchestration for telecom workloads. Kube5G~\cite{kube5g_globecom2020} introduces a Kubernetes Operator achieving 5G provisioning in under 2 minutes and reconfiguration in under 1 minute. Scotece et al.~\cite{scotece_5gkube2023} argue that IT-world DevOps tools can directly deploy softwarized 5G cores without complex ETSI MANO overhead, demonstrating significant OPEX reduction. Goshi et al.~\cite{goshi_interdep2021} study inter-NF dependency effects on resource utilization, showing that a single bottleneck NF can constrain the entire 5G core if dependencies are not properly modeled. The shift toward GitOps-based automation is exemplified by 5G-CT~\cite{bonati_5gct2025}, which uses Red Hat OpenShift with ArgoCD and Tekton for automated end-to-end 5G and O-RAN deployment and testing, validated with months of automated over-the-air testing. For multi-cloud scenarios, Osmani et al.~\cite{osmani_multicloud_k8s2021} integrate Federated Kubernetes with Network Service Mesh for seamless multi-cloud workload connectivity, while Li et al.~\cite{li_5growth2021} present the 5Growth platform integrating ETSI NFV MANO with Kubernetes deployments across core and edge, validated with real vertical use cases across European 5G testbeds. Coronado et al.~\cite{coronado_ztm_survey2022} provide a comprehensive survey bridging zero-touch management and mobile network research, covering ETSI ZSM architecture, autonomic management, SON, and AI/ML-driven automation for 5G and 6G.

\textbf{Synthesis.} The investigation above reveals that cloud-native scalability for cellular networks is achievable but requires deliberate architectural choices. Stateless designs (CoreKube, TEGRA, PP5GS) reduce resource consumption by 34--55\% and enable near-instantaneous horizontal scaling, though the cost of statelessness ranges from 10\% to 75\% throughput overhead depending on checkpointing granularity. Kubernetes imposes less than 2\% orchestration overhead but requires network-aware scheduling (NetMARKS achieves 37\% response time reduction) and custom autoscaling beyond default HPA. AI-driven resource optimization has matured from simulation to real-testbed validation: DeepCog's cost-aware forecasting achieves 50\% resource management cost reductions, OnSlicing demonstrates 61\% resource savings with near-zero SLA violations on live infrastructure, and federated DRL addresses multi-domain privacy. Energy efficiency gains of 34--90\% are achievable through VNF consolidation and advanced sleep modes, while the orchestration landscape is shifting from ETSI MANO toward GitOps-based approaches that reduce deployment time from minutes to seconds.
\subsection{Orchestration and Automation}
\label{subsec:orchestration}

The large-scale deployment of cloud-native cellular infrastructure introduces significant complexity in the management and operation of network resources. Traditional mobile networks relied on relatively static infrastructure environments in which network elements were manually provisioned, configured, and upgraded through centralized management systems. In contrast, cloud-based cellular architectures consist of highly dynamic software workloads distributed across multiple cloud environments, edge sites, and administrative domains. Managing such environments requires sophisticated orchestration and automation mechanisms capable of coordinating network functions, infrastructure resources, and service policies across heterogeneous computing platforms.

\begin{figure}[t]
    \centering
    \includegraphics[width=\columnwidth,trim={0cm 0cm 20cm 0cm},clip]{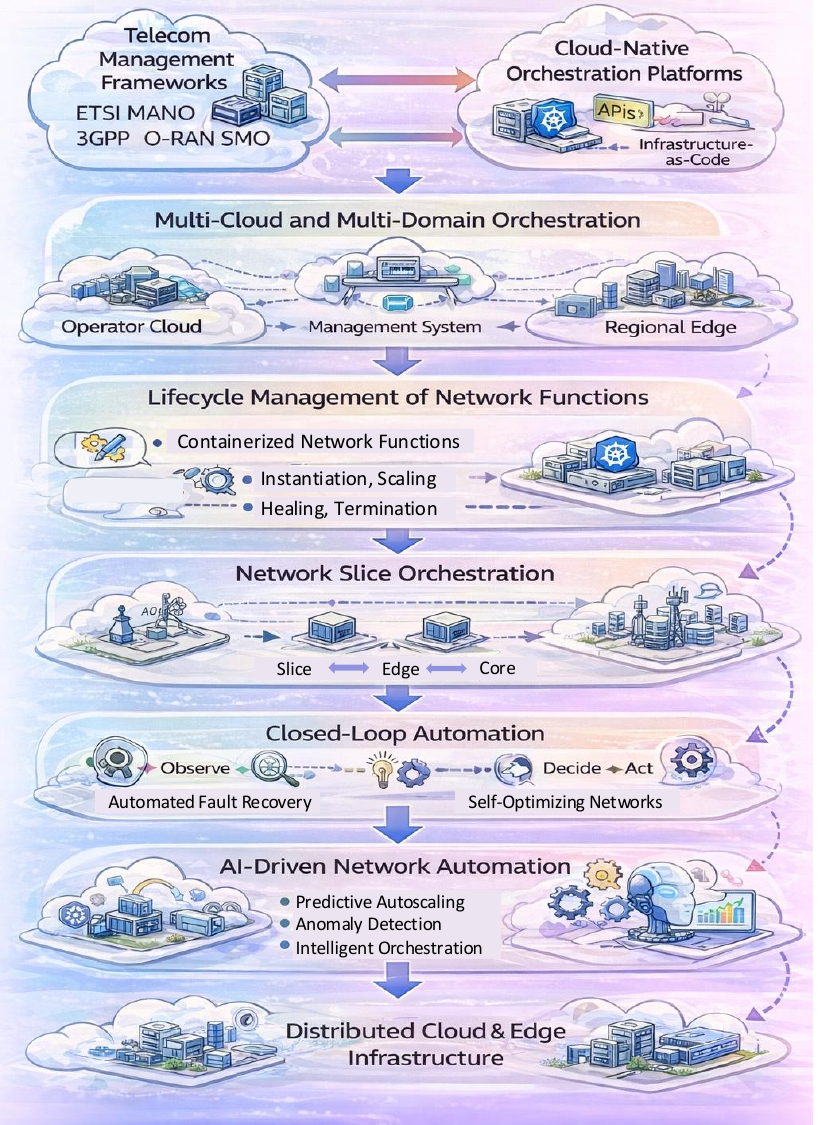}
    \caption{Overview of orchestration and automation mechanisms in cloud-based cellular networks spanning telecom management frameworks, cloud-native lifecycle management, multi-domain orchestration, network slice lifecycle, closed-loop automation, and AI-driven network intelligence.}
    \label{fig:orchautoov}
\end{figure}

Cloud-native orchestration frameworks enable automated deployment and lifecycle management of network functions, allowing operators to dynamically instantiate, scale, and recover services in response to changing network conditions. However, integrating telecom management architectures with modern cloud orchestration systems presents substantial challenges. Cellular networks must maintain strict performance guarantees, service-level agreements, and security policies while operating within distributed infrastructure environments that may span multiple cloud providers and edge locations. Consequently, understanding the orchestration and automation requirements of cloud-based cellular networks requires investigating the interaction between telecom management frameworks, cloud-native orchestration platforms, and emerging AI-driven automation techniques. Figure~\ref{fig:orchautoov} summarizes the investigation categories of this section.

\subsubsection{Telecom Management Frameworks and Cloud Orchestration}
\label{subsubsec:mano_cloud}

Traditional telecom network management has been guided by the ETSI Management and Orchestration (MANO) architecture and the 3GPP network management specifications (TS~28-series). These frameworks define mechanisms for network function virtualization management, infrastructure resource allocation, and service lifecycle management. In parallel, the O-RAN Service Management and Orchestration (SMO) framework coordinates RAN resources and network services. Cloud-native deployments introduce a complementary orchestration layer centered around Kubernetes and cloud provider infrastructure management systems. Integrating these telecom and cloud paradigms remains a central challenge.

Yilma et al.~\cite{yilma_osm_onap2020} provide the first formal benchmarking of OSM and ONAP using a vCPE VNF, introducing KPIs for onboarding delay, deployment delay, runtime orchestration delay, and quality of decision. Their results show that OSM-4 uses only 2.27\% of the vCPU resources required by ONAP-B, though both platforms lacked runtime scaling support at the time of evaluation. Trakadas et al.~\cite{trakadas_mano_comparison2020} extend this comparative analysis to include the SONATA MANO framework from the EU 5GTANGO project, evaluating all three against common functional and operational KPIs. A fundamental tension has emerged between ETSI MANO and Kubernetes-native orchestration. Gawel and Zielinski~\cite{gawel_k8s_mano2019} systematically evaluate the degree to which Kubernetes satisfies ETSI MANO specifications through stress and chaos testing on a virtual IMS deployment, identifying specific gaps. The proposal for ``true cloud-native'' MANO~\cite{true_cn_mano2021} argues that cloud-nativeness requires more than replacing VMs with containers, advocating for architectures that exploit Kubernetes' declarative model, operator pattern, and native lifecycle management while addressing network service composition and dynamic dependency management. Scotece et al.~\cite{scotece_5gkube2023} present 5G-Kube, demonstrating that IT-world DevOps tools (containers, Kubernetes, Helm) can directly deploy softwarized 5G cores without complex ETSI MANO overhead, reducing both complexity and cost.

For the O-RAN management plane, Polese et al.~\cite{polese_oran_survey2023} provide the most comprehensive deep dive into the SMO framework, covering the Non-RT RIC, O-RAN interfaces (E2, A1, O1, O2), and AI/ML workflows for data-driven closed-loop control. Polese et al.~\cite{polese_6g_oran2024} extend this to 6G, discussing CI/CD/CT pipelines and GitOps methodology for managing O-RAN deployments, while Marinova and Leon-Garcia~\cite{marinova_oran2024} describe how the SMO serves as an automation platform for orchestrating O-RAN NFs, radio resources, and network slices at scale. Coronado et al.~\cite{coronado_ztm_survey2022} provide the most comprehensive survey bridging zero-touch management and mobile network research, covering the convergence of ETSI ZSM, ETSI NFV MANO, 3GPP management, and cloud-native platforms. Liyanage et al.~\cite{liyanage_zsm_survey2022} complement this by mapping how ETSI ZSM, ETSI NFV, 3GPP SA5, and TM Forum standards converge toward automated hybrid telecom/cloud management.

\subsubsection{Lifecycle Management of Cloud-Native Network Functions}
\label{subsubsec:lifecycle}

Cloud-native network functions must support a complete lifecycle that includes instantiation, configuration, scaling, upgrading, and termination. Container orchestration platforms provide mechanisms for automated workload scheduling, service discovery, and health monitoring, enabling a declarative approach where network functions are defined through configuration models and the orchestration platform automatically maintains the desired system state.

Arouk and Nikaein~\cite{kube5g_globecom2020} present Kube5G, a cloud-native 5G service platform with a novel CNF design as nested reusable layers for building, packaging, and upgrading multi-version NFs during CI/CD. Their benchmarks show 4G/5G network provisioning in under 2 minutes and updates in under 1 minute via the Kube5G-Operator. Their companion work~\cite{arouk_noms2020} demonstrates network automation using Kubernetes and OpenShift Operators covering deployment phases from basic install to auto-pilot. Chun et al.~\cite{chun_k8s_5g2019} identify enhancements needed in Kubernetes for 5G NFV infrastructure, proposing NUMA-aware scheduling, hugepage isolation, enhanced pod specifications, and resource advertising for latency-sensitive CNFs requiring near-line-rate packet processing.

The shift toward GitOps-based automation represents a significant paradigm transition in telecom operations. Leiter et al.~\cite{leiter_gitops2023} demonstrate how Kubernetes Operators enable NETCONF-based NF configuration combined with GitOps principles, with ArgoCD syncing UPF custom resources from Git repositories to Kubernetes clusters. Bonati et al.~\cite{bonati_5gct2025} present 5G-CT, an OpenShift- and GitOps-based automation framework deploying a softwarized end-to-end 5G and O-RAN system in seconds without human intervention, validated with months of automated over-the-air testing using software-defined radios and OpenAirInterface. Apostolakis et al.~\cite{apostolakis_cn_mobile2022} implement a cloud-native mobile network using Open5GS on Kubernetes with Docker containers, bridging the gap between monolithic open-source 5G implementations and cloud-native deployment frameworks.

Observability during the CNF lifecycle is equally critical. Khichane et al.~\cite{khichane_5gc_observer2023} propose 5GC-Observer, a non-intrusive observability framework using eBPF technology to monitor cloud-native 5G core NFs on Kubernetes without modifying NF source code, enabling real-time QoS degradation detection through statistical anomaly methods. Barrachina-Mu\~{n}oz et al.~\cite{barrachina_cn5g2022} deploy Open5GS and Prometheus-based monitoring as CNFs in a multi-tier Kubernetes cluster, demonstrating end-to-end monitoring via Grafana in UPF re-selection and mobility scenarios. Their extended work~\cite{barrachina_zt_iet2023} adds a zero-touch orchestrator with a decision engine interacting with the Kubernetes scheduler for automated workload management. The 5G-STREAM framework~\cite{5g_stream_dsn2025} provides a purpose-built two-tier service mesh for 5G core microservices that dynamically maintains VNF routing and authorization information using distributed configuration management, demonstrating that lifecycle management must encompass not only the NFs themselves but also their communication and security infrastructure.

\subsubsection{Multi-Domain and Multi-Cloud Orchestration}
\label{subsubsec:multicloud_orchestration}

Cloud-based cellular deployments frequently span multiple infrastructure domains, including hyperscale public clouds, operator private clouds, and geographically distributed edge environments. Orchestrating network services across such heterogeneous infrastructure requires mechanisms for workload portability, policy enforcement, and cross-domain coordination.

Multi-cluster Kubernetes orchestration has emerged as the dominant approach for distributed 5G deployments. Osmani et al.~\cite{osmani_multicloud_k8s2021} present the first solution integrating Federated Kubernetes (KubeFed) with Network Service Mesh for seamless multi-cloud workload connectivity tailored to 5G telco requirements. Syrigos et al.~\cite{syrigos_multicluster2024} employ SUSE Rancher for multi-cluster management and Submariner for L3 cross-cluster connectivity, evaluating the framework over 25~Gbps fabric connecting distributed clusters. Iorio et al.~\cite{iorio_liqo2023} propose the ``liquid computing'' vision with Liqo, an open-source framework using Virtual Kubelet to collapse remote clusters into virtual nodes, demonstrating minimal overhead (few milliseconds) compared to vanilla Kubernetes. Dumitru-Guzu et al.~\cite{dumitru_crosscluster2024} use Liqo to peer multiple Kubernetes clusters across different cloud tenants for deploying Open5GS, validating end-to-end network slicing across eMBB, URLLC, and mMTC use cases. The 5G-MAP framework~\cite{5g_map_mobicom2025} provides the most comprehensive cross-region measurement study to date, deploying the OAI 5G core across 8 AWS regions and 18 edge zones spanning 7 countries and demonstrating that strategic VNF placement across cloud zones can reduce inter-VNF latency by up to 5$\times$.

At the architectural level, Taleb et al.~\cite{taleb_multidomain2019} introduce a management and orchestration architecture for multi-domain network slices with four strata spanning service conductor, domain-specific orchestration, sub-domain MANO, and logical multi-domain slice instances. Li et al.~\cite{li_5growth2021} present 5Growth, an end-to-end service platform integrating ETSI NFV MANO with Kubernetes deployments across core and edge, validated with real vertical use cases across European 5G testbeds. The 5G-TRANSFORMER project~\cite{5gtransformer_nfvsdn2019} demonstrates federated network service deployment across administrative domains in under 5 minutes, with federation operations consuming minimal additional time. The Network Service Federation vision~\cite{nsfv_federation2020} enables automated multi-domain orchestration of composite NFV network services, addressing the operational complexity of cross-provider deployments.

\subsubsection{Network Slice Orchestration}
\label{subsubsec:slice_orchestration}

Network slicing represents a fundamental capability requiring coordination across multiple network domains and infrastructure layers. Slice lifecycle management encompasses instantiation, scaling, monitoring, reconfiguration, and termination while preserving service-level guarantees and inter-slice isolation.

Afolabi et al.~\cite{afolabi_e2e_tmc2020} propose an end-to-end Network Slicing Orchestration System with a Dynamic Auto-Scaling Algorithm (DASA) using hierarchical architecture with dedicated entities per domain, enabling autonomous resource adaptation through proactive and reactive provisioning validated via queuing models. Chiu et al.~\cite{chiu_cn_mano2022} present a cloud-native MANO framework for automating end-to-end network slicing following 3GPP management specifications using Kubernetes, Helm, and bandwidth management techniques. Ebrahimi et al.~\cite{ebrahimi_slicing_comst2024} provide a comprehensive survey analyzing network slicing resource management across RAN, transport, and core domains, examining interdependencies and cross-domain issues for end-to-end 6G contexts. Wyszkowski et al.~\cite{wyszkowski_slice_design2024} present a systematic tutorial on organizing a standards-aligned network slice and subnet design process, addressing 3GPP's gap in design-time aspects and proposing automation opportunities.

Dynamic slice modification is addressed through AI-driven approaches. Wei et al.~\cite{wei_slice_reconfig2020} develop the Intelligent Network Slice Reconfiguration Algorithm (INSRA) based on the Branching Dueling Q-network to handle intractable multi-dimensional discrete action spaces for core network slice reconfiguration under dynamic traffic. Bega et al.~\cite{bega_deepcog2020} present DeepCog, which uses a custom loss function accounting for overprovisioning and underprovisioning costs, demonstrating over 50\% reduction in resource management costs with real-world metropolitan data. Their AZTEC system~\cite{bega_aztec2020} combines per-slice traffic forecasting with admission control formulated as a geometric knapsack problem, enabling autonomous slice resource management. Abbas et al.~\cite{abbas_ibn_access2021} propose an intent-based networking platform automating lifecycle management of multi-domain network slices, where users provide high-level intentions that are automatically translated into slice configurations, eliminating the need for expert-level MANO knowledge. Our prior work provides concrete measurements of the resource consumption~\cite{atalay_wcnc2022} and deployment costs~\cite{atalay_globecom2022} associated with scaling network slices in cloud-native testbed deployments.

\subsubsection{Closed-Loop Automation and Zero-Touch Management}
\label{subsubsec:closedloop}

Closed-loop automation frameworks continuously monitor network conditions, analyze operational metrics, and automatically adjust configurations without human intervention. These systems follow control loop structures such as OODA (Observe-Orient-Decide-Act), MAPE-K (Monitor-Analyze-Plan-Execute-Knowledge), and COMPA (Collect-Organize-Monitor-Plan-Act), adapted for telecom-specific requirements.

Benzaid and Taleb~\cite{benzaid_ai_ztm2020} provide one of the earliest treatments of AI-driven ZSM, analyzing how the ETSI ZSM framework leverages AI/ML for self-managing capabilities while identifying adversarial ML risks, explainability challenges, and security concerns for autonomous management. Gallego-Madrid et al.~\cite{gallego_ml_zsm2022} survey ML adoption in ZSM frameworks, analyzing supervised, unsupervised, and reinforcement learning methods applied to different ZSM management domains. Gomes et al.~\cite{gomes_intent_loops2021} propose intents as a mechanism for coordinating hierarchies of closed loops within ETSI ZSM, supporting OODA, MAPE-K, and COMPA patterns with hierarchical and peer interactions for delegation and escalation. Boutaba et al.~\cite{boutaba_ai_cla2021} propose an AI-driven closed-loop automation architecture using the monitor-analyze-plan-execute paradigm integrated with O-RAN's RIC, covering data analytics, anomaly detection, root cause analysis, and remediation for autonomous slice orchestration.

Monitoring provides the data foundation for closed-loop systems. Saha et al.~\cite{saha_monarch_noms2023} present MonArch for end-to-end network slice monitoring and per-slice KPI computation, validated on a Free5GC/Kubernetes testbed with up to 50 network slices. Their extended journal version~\cite{saha_monarch_tnsm2024} adds adaptive monitoring algorithms dynamically adjusting collection intervals, demonstrating consistent ingestion times of 2.25--2.75~ms across varying slice counts. Subramanya and Riggio~\cite{subramanya_fl_autoscaling2021} demonstrate that deep learning models using both centralized and federated approaches achieve effective predictive horizontal and vertical autoscaling of VNFs in multi-domain settings while preserving domain isolation. Digital twins are emerging as a complementary mechanism: Apostolakis et al.~\cite{apostolakis_dt2023} propose three digital twin application scenarios for 6G including twinning network appliances for configuration optimization, model training for RL-based algorithms, and what-if analysis for network planning.

\subsubsection{AI-Driven Network Automation}
\label{subsubsec:ai_automation}

The increasing scale and complexity of cloud-native cellular infrastructure has motivated AI techniques that go beyond traditional closed-loop automation toward AI-native architectures where intelligence is embedded throughout the network management stack.

The 3GPP Network Data Analytics Function (NWDAF) provides the standardized interface for AI/ML integration in 5G core networks. Chouman et al.~\cite{chouman_nwdaf2022} implement a functional NWDAF prototype integrated with Open5GS, collecting data via standardized N34 and N23 interfaces and applying unsupervised learning to analyze NF interactions. Manias et al.~\cite{manias_nwdaf_globecom2022} extend this prototype to apply $k$-means clustering for characterizing 5G core signaling patterns for proactive NF scaling. Shafiee et al.~\cite{shafiee_nwdaf2025} present the first implementation of UPF Event Exposure Service for standardized real-time data collection and proof-of-concept closed-loop automation involving UPF, NWDAF, and extended SMF. Bega et al.~\cite{bega_ai_standards2020} bridge the gap between standards (ETSI ZSM, ENI, 3GPP NWDAF) and practical AI/ML algorithms, demonstrating AI-based closed-loop automation for network slicing admission control.

Intent-based networking using large language models represents an emerging frontier. Manias et al.~\cite{manias_llm_intent2024} develop a custom LLM for 5G intent-based networking that extracts and interprets user intents into actionable network policies, reducing human intervention toward zero-touch management. Mekrache et al.~\cite{mekrache_llm_ibn2024} propose an LLM-centric architecture spanning the complete intent lifecycle---decomposition, translation, negotiation, activation, and assurance---validated at the EURECOM 5G facility. Dzeparoska et al.~\cite{dzeparoska_llm_policy2023} leverage LLMs to process user intents into structured policy abstractions linked with APIs for automated execution, establishing the concept of intent drift through KPI monitoring.

DRL-based approaches continue to advance toward practical deployment. Liu et al.~\cite{onslicing_conext2021} implement online end-to-end network slicing with constraint-aware DRL on a real testbed, achieving 61.3\% resource usage reduction with near-zero SLA violation. Polese et al.~\cite{polese_coloran_tmc2023} present ColO-RAN, the first publicly available large-scale O-RAN testing framework with DRL-based xApps for closed-loop RAN slicing and scheduling control. The XRF framework~\cite{xrf_infocom2023} addresses the authentication and authorization dimension of O-RAN automation, providing scalable mechanisms for xApp management. Explainable AI is gaining attention as a prerequisite for operator trust: Brik et al.~\cite{brik_xai_oran2024} present a comprehensive tutorial on XAI in 6G O-RAN covering methods, metrics, and an automation pipeline for XAI model training and deployment. As Letaief et al.~\cite{letaief_6g_roadmap2019} envision in their highly influential roadmap paper, 6G will support ubiquitous AI from core to edge, with edge AI, federated learning, and intelligent radio as key technologies where AI plays a central role in designing and optimizing network architectures, protocols, and operations.

\textbf{Synthesis.} The orchestration landscape for cloud-based cellular networks is converging from two directions: the telecom standards world (ETSI NFV MANO, 3GPP SA5, ETSI ZSM) and the cloud-native world (Kubernetes, Helm, GitOps). OSM requires only 2.27\% of the resources of ONAP, yet both lag behind Kubernetes-native approaches that achieve provisioning in under 2 minutes and GitOps-based deployment in seconds. Multi-cluster Kubernetes tools (Liqo, Submariner, KubeFed) enable cross-domain orchestration with minimal overhead, though federation across independent operators remains demonstrated primarily through EU-funded prototypes. Network slice lifecycle management has matured through AI-driven approaches: DeepCog achieves 50\% cost reduction, AZTEC enables zero-touch slice capacity allocation, and DRL-based reconfiguration handles intractable action spaces. The 2023--2025 period shows a notable shift toward LLM-based intent processing for network management, moving beyond earlier pattern-matching IBN approaches. Closed-loop automation and NWDAF implementations are transitioning from prototypes to deployed systems, while digital twins and XAI remain nascent but represent critical enablers for trustworthy autonomous 6G operations.
\section{Future Directions} \label{sec:future_trends}
\begin{figure}[t]
    \centering
    \includegraphics[width=\columnwidth,trim={0cm 0cm 5cm 0cm},clip]{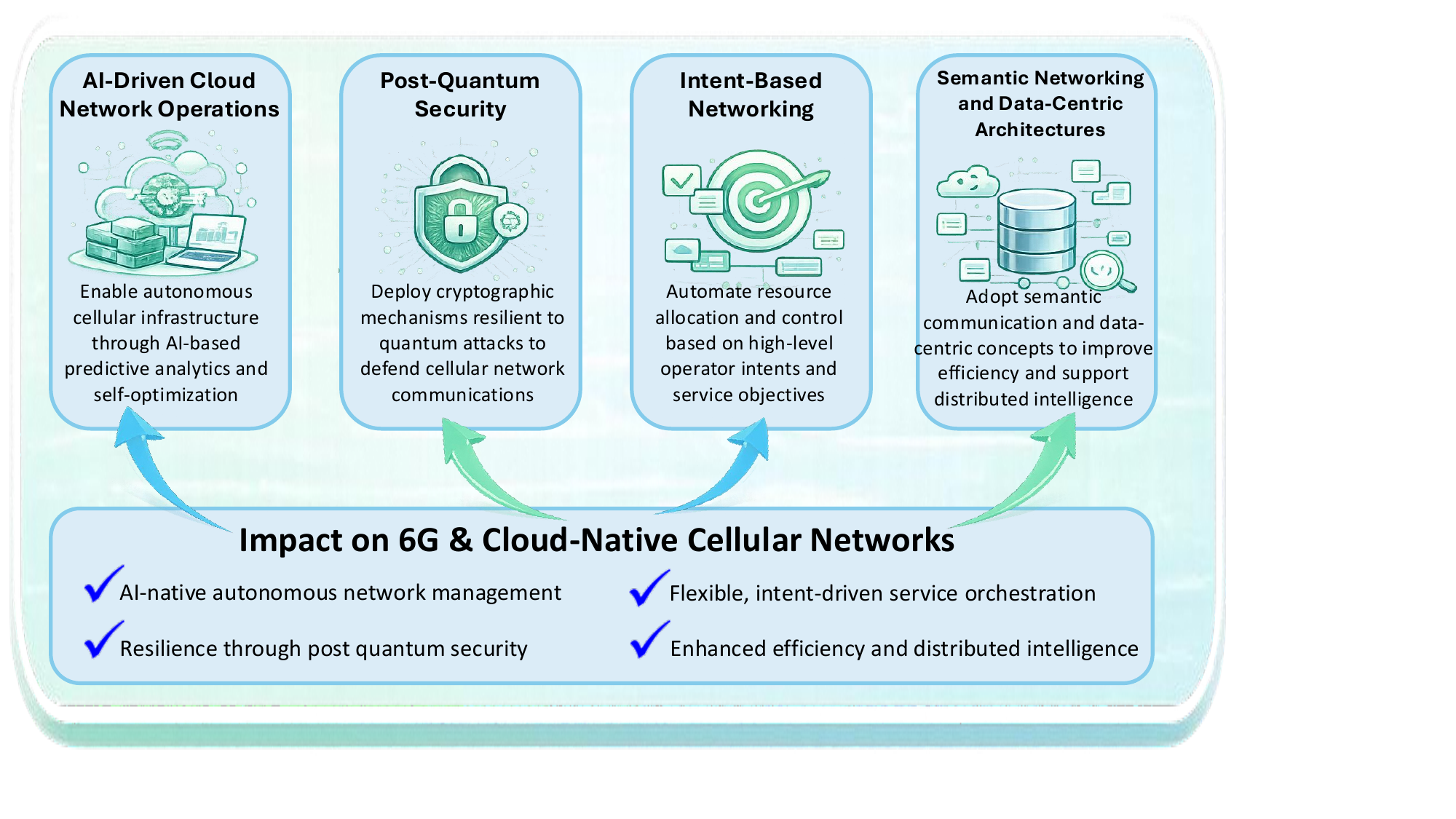}
    \caption{Future directions to enable a more robust and cloud-native 6G deployment with AI-native orchestration, post-quantum secure communication, intent-based networking and semantic principles}
    \label{fig:futuredirov}
\end{figure}

Last but not least, this subsection explores the future directions for enabling 6G cloud-native cellular deployments. The summary of the discussion points is presented in Figure~\ref{fig:futuredirov}.

\subsection{AI-Driven Cloud Network Operations}
Future 6G networks will be AI-native, embedding intelligence throughout the stack[cite: 6422, 7279]. Seminal vision papers argue that 6G will support ubiquitous AI services beyond mobile Internet \cite{A1}. Scalable and trustworthy edge AI systems with integrated wireless communication and decentralized ML models are essential for a holistic system architecture \cite{A2}. 6G wireless systems will be driven by autonomous systems and extended reality, requiring a comprehensive research agenda for pervasive AI \cite{A3}. 

Beyond performance, 6G must be human-centric with high security and AI-driven intelligence as core requirements \cite{A4}. Zero-touch network and service management (ZSM) frameworks leverage AI as a key enabler for fully autonomous closed-loop management, though they introduce risks regarding data quality and model interpretability \cite{A5}. Federated learning (FL) offers a path toward communication efficiency and privacy preservation at the wireless edge \cite{A6}. A holistic 6G architecture will likely build on pervasive intelligence throughout the stack, from resource management to service orchestration \cite{A7, A8}. Furthermore, integrating network digital twins allows for high-fidelity virtual replicas to support AI training and predictive optimization \cite{A9}. Finally, the emergence of Large Language Models (LLMs) provides a path toward AGI-enabled autonomous network management \cite{A10}.

\subsection{Post-Quantum Security for Cloud-Based 5G/6G}
The emergence of quantum computing threatens existing cryptographic frameworks \cite{B5, B8}. Comprehensive surveys highlight the integration of Quantum Key Distribution (QKD) and Post-Quantum Cryptography (PQC) into VPN security frameworks and 5G demonstrations \cite{B1}. Performance studies on TLS 1.3 show that post-quantum signature candidates are viable for securing Service-Based Architecture (SBA) interfaces \cite{B2, B7}. KEM-based key exchanges (KEMTLS) can further reduce bandwidth overhead and CPU cycles compared to standard PQ TLS 1.3 \cite{B3}. 

Specific protocol elements like the Subscription Concealed Identifier (SUCI) \cite{B4} and the AKA protocol \cite{B6} have been successfully hardened using post-quantum KEMs like CRYSTALS-Kyber. Current research also examines NIST-shortlisted PQC algorithms for securing IPsec and TLS in 5G/6G contexts \cite{B9}. These efforts aim to advance 3GPP and NIST standards toward a "quantum-ready" core \cite{B10}.

\subsection{Intent-Based Networking (IBN)}
IBN simplifies network management by allowing operators to express high-level goals instead of low-level commands \cite{C2, C3}. Comprehensive surveys define the IBN closed-loop expression, translation, resolution, activation, and assurance \cite{C1}. ML-powered frameworks support proactive service assurance across radio and cloud domains \cite{C4}, while one-touch platforms automate the lifecycle of end-to-end network slices \cite{C5}. 

Natural language interfaces, such as Lumi, enable operators to express intents which are then translated via sequence-to-sequence models \cite{C6, C7}. Recently, LLM-centric approaches have been proposed for configuring network services using natural language, validated on 5G facilities \cite{C8}. Advanced models like NetLM use transformers to understand network dynamics and capture packet data sequences for autonomous operation \cite{C9}.

\subsection{Semantic Networking and Named Data Networking}
Semantic communication fundamentally challenges the Shannon theorem by prioritizing meaning and task-relevance over bit-level accuracy \cite{D1, D4}. Foundational deep learning systems like DeepSC minimize semantic errors in text \cite{D1}, speech \cite{D2}, and multi-user multimodal transmission \cite{D3}. Reviews categorize these into semantic-oriented, goal-oriented, and semantic-aware types \cite{D5}. New paradigms like WePCN introduce a semantic base representation to minimize data while maximizing knowledge \cite{D6, D7}.

Named Data Networking (NDN) provides a data-centric alternative to endpoint-based routing \cite{D8, D9, atalay2025envisioning}. Integrating it with MEC offers mutual benefits for connected vehicles and IoT \cite{D10, atalay2025towards}. This shift also necessitates data-centric security and access control mechanisms, where security is attached to data objects rather than session channels \cite{D11}.
\section{Conclusion} \label{sec:conc}
The transition from hardware-bound cellular infrastructures to cloud-native deployments represents one of the most significant architectural shifts in the evolution of mobile networks. 
This survey examined the emerging landscape of cloud-based cellular deployments by presenting a structured taxonomy that captures deployment architectures, multi-tenancy and isolation mechanisms, orchestration models, and economic ownership structures. 
We further analyzed key investigation areas, including security and trust, performance and reliability, scalability and resource efficiency, and automation and orchestration, that determine the feasibility and robustness of operating cellular networks on distributed cloud platforms. Through an examination of hyperscaler ecosystems and telecom and cloud collaborations, we showed how public cloud providers are increasingly shaping the deployment and operational models of modern cellular systems. 
Finally, we discussed several promising future directions, including AI-driven network operations, post-quantum security, intent-based networking, and semantic communication paradigms, which collectively point toward more autonomous, secure, and intelligent 6G era infrastructures. As cellular systems continue to converge with cloud computing and edge platforms, addressing the interoperability, trust, and operational challenges identified in this survey will be essential for realizing scalable and resilient next-generation mobile networks.

\bibliographystyle{IEEEtran}
\bibliography{references}

\end{document}